\documentclass[12pt]{article}

\ifx\pdfoutput\undefined
\usepackage[dvips,bookmarks]{hyperref}
\else
\usepackage{hyperref}
\fi
\hypersetup{colorlinks=false,bookmarksopen,bookmarksnumbered,citecolor=blue,
   pdfstartview=FitH}

\usepackage{amsmath,amssymb,amsthm}
\usepackage{bbm,dsfont,mathrsfs,slashed,cancel,tensor} 
\numberwithin{equation}{section}
\usepackage{graphicx} 
\usepackage{booktabs} %for tables
\usepackage{multirow} %for tables
\usepackage{float}
\usepackage{cite} %[1,2,3] is [1-3]
\usepackage{ytableau} %Young tableau
\ytableausetup{centertableaux} %Generally have this on
\DeclareRobustCommand{\svdots}{% s for `scaling'
  \vbox{%
    \baselineskip=0.4\normalbaselineskip
    \lineskiplimit=0pt
    \hbox{.}\hbox{.}\hbox{.}%
    \kern-0.2\baselineskip
  }%
}
\usepackage{tikz}
\newcommand{\WS}{W(S)}
\newcommand{\WK}{W(K)}

\newcommand{\GK}{G(K)}

\newcommand {\cD}{{\cal D}}

\newcommand {\cH}{{\cal H}}

\newcommand {\cM}{{\cal M}}
\newcommand {\cN}{{\cal N}}

\newcommand {\cR}{{\cal R}}

\newcommand {\cV}{{\cal V}}

\def\a{\alpha}
\def\b{\beta}

\def\d{\delta}
\def\e{\epsilon}

\def\g{\gamma}
\def\G{\Gamma}

\def\k{\kappa}
\def\l{\lambda}

\def\n{\nabla}

\def\q{\theta}
\def\r{\rho}
\def\s{\sigma}
\def\t{\tau}

\def\x{\xi}
\def\z{\zeta}

\def\L{\Lambda}
\def\O{\Omega}

\def\ri{{\rm i}}
\newcommand{\ua}{\underline{a}}
\newcommand{\ub}{\underline{b}}
\newcommand{\uc}{\underline{c}}
\newcommand{\ud}{\underline{d}}
\newcommand{\ue}{\underline{e}}

\newcommand{\Sp}{\mathrm{Sp}}
\newcommand{\USp}{\mathrm{USp}}
\newcommand{\SU}{\mathrm{SU}}
\newcommand{\SL}{\mathrm{SL}}

\newcommand{\SO}{\mathrm{SO}}

\newcommand{\OSp}{\mathrm{OSp}}
\newcommand{\Spin}{\mathrm{Spin}}
\newcommand{\bbR}{{\mathbb R}}
\newcommand{\bbC}{{\mathbb C}}
\newcommand{\bbH}{{\mathbb H}}

\newcommand{\bbZ}{{\mathbb Z}}

\newcommand{\hnabla}{{\hat{\nabla}}}
\newcommand{\tr}{{\mathrm{tr}}}
\newcommand{\ve}{\varepsilon}

\newcommand{\bbD}{{\mathbb{D}}}

\newcommand{\be}{\begin{equation}}
\newcommand{\ee}{\end{equation}}
\newcommand{\bea}{\begin{eqnarray}}
\newcommand{\eea}{\end{eqnarray}}

\newcommand{\ba}{\begin{array}}
\newcommand{\ea}{\end{array}}
\newcommand{\bsubeq}{\begin{subequations}}
\newcommand{\esubeq}{\end{subequations}}

\newcommand{\hf}{\frac{1}{2}}
\newcommand{\eps}{\varepsilon}

\DeclareMathOperator{\ad}{ad}
\DeclareMathOperator{\sgn}{sgn}
\DeclareMathOperator{\Hom}{Hom}

\parskip=\medskipamount

\definecolor{myblack}{rgb}{0.1,0.1,0.1}
\newcommand{\blackalpha}[1]{*(myblack)\textcolor{white}{#1}}
\theoremstyle{plain}

\newtheorem{theorem}{Theorem}[section]

\newtheorem{proposition}[theorem]{Proposition}

\theoremstyle{definition}

\definecolor{ck}{rgb}{0.858, 0, 0.478}

\begin{document}
\begin{titlepage}
\begin{center}
{\Large \bf Six-dimensional ${\mathcal{N}}=(2,0)$ Conformal Superspace}
\\ 
\end{center}

\begin{center}
{\bf
Christian Kennedy, Gabriele Tartaglino-Mazzucchelli
} \\
\vspace{5mm}
\footnotesize{
{\it School of Mathematics and Physics, University of Queensland,
\\
 St Lucia, Brisbane, Queensland 4072, Australia}}
\vspace{2mm}

 \texttt{christian.kennedy@uq.edu.au}\\
 \texttt{g.tartaglino-mazzucchelli@uq.edu.au} 
\end{center}

\begin{abstract}
\baselineskip=14pt
We develop a new off-shell formulation for six-dimensional conformal supergravity obtained by gauging the 6D $\cN = (2, 0)$ superconformal algebra in superspace. 
We provide the complete gauged algebra, which proves to be considerably constrained compared to other conformal superspaces constructed in the past. This formulation is employed to obtain the unique 6D $\cN = (2, 0)$ Bach tensor superfield, which describes the multiplet of equations of motions for conformal supergravity, by using a general ansatz fixed by truncation to 6D $\cN = (1, 0)$ results. We also translate some results into components for precise matching against the literature.

\end{abstract}
\vfill

\vfill
\end{titlepage}

\newpage
\renewcommand{\thefootnote}{\arabic{footnote}}
\setcounter{footnote}{0}

\tableofcontents

%%%%%%%%%%%%%%%%%%%%%%%%%%%%%%%%%%%%%%%%%%%%%%%%%%%%%%
%%%%%%%%%%%%%%%%%%%%%%%%%%%%%%%%%%%%%%%%%%%%%%%%%%%%%%

\allowdisplaybreaks

\section{Introduction}

A powerful method for engineering off-shell supergravity–matter couplings efficiently relies on the observation that Poincar\'e gravity can be realised as conformal gravity coupled to a compensating scalar field \cite{Deser:1970hs,Zumino:1970tu}. This perspective plays a fundamental role in both the superconformal tensor calculus and superspace formulations of supergravity --- see \cite{Freedman:2012zz,Lauria:2020rhc} and \cite{Gates:1983nr,Buchbinder:1998twe,Kuzenko:2022skv,Kuzenko:2022ajd} for reviews and a list of relevant references. In the supersymmetric case, conformal gravity provides an off-shell representation of the local superconformal algebra, with the vielbein appearing among its independent fields. This representation is referred to as the \emph{Weyl multiplet} of conformal supergravity. Classifying and constructing Weyl multiplets has a long history and is an interesting mathematical problem in its own right  ---  see \cite{deWit:1979dzm,Bergshoeff:1980is,Siegel:1978mj,Howe:1981gz,Fradkin:1982xc,Bergshoeff:1985mz,Bergshoeff:1999,Kugo:2000hn,Bergshoeff:2001hc,Butter:2009cp,Butter:2011sr,Kuzenko:2011xg,Butter:2013goa,Butter:2013rba,Butter:2014xxa,Butter:2016,Butter:2017pbp,Butter:2019edc,Howe:2020xrg,Howe:2020hxi,Hutomo:2022hdi,Kuzenko:2022qnb,Kuzenko:2023qkg,Adhikari:2023tzi,Adhikari:2024qxg,Adhikari:2024esl,Adhikari:2025wwb,Kuzenko:2025bud} for an incomplete list of relevant references for extended supergravity theories in various space-time dimensions. Moreover, these multiplets become starting points for a wide range of physical applications.

For example, among the interesting applications of off-shell superconformal techniques, localization approaches are worth mentioning for the study of several observables in supersymmetric quantum field theories \cite{Pestun:2016zxk}, as well as applications to holography. In particular, in recent years, new efforts have been invested in pushing the AdS/CFT correspondence to new precision tests where, on the gravity side, higher-derivative supergravity characterises subleading corrections in observables such as the entropy of black holes. For a list of references on new precision tests of holography based on higher-derivative supergravity see the recent works
\cite{Baggio:2014hua,Butter:2018wss,Bobev:2020egg,Bobev:2021oku,Bobev:2021qxx,Liu:2022sew,Hristov:2022lcw,Cassani:2022lrk,Gold:2023ymc,Cassani:2024tvk,Casarin:2024qdn,Ma:2024ynp,Saskowski:2024otc,Hristov:2024cgj,Cassani:2024kjn} as well as the recent review \cite{Ozkan:2024euj}.

Among the classes of (supersymmetric) higher-derivative gravity invariants, a special role is played by (supersymmetric) conformal anomalies. Conformal anomalies are one of the key data of a conformal field theory. For this reason, they have been the subject of substantial investigations; see, for example, \cite{Osborn:1993cr,Duff:1980qv,Duff:1993wm,Bonora:1985cq,Buchbinder:1986im,Kuzenko:2013gva,Gomis:2015yaa,Cordova:2015fha,Cordova:2015vwa,Herzog:2013ed,Bastianelli:2000hi} for the case of two, four, and six space-time dimensions.
These quantum anomalies are associated to invariants for conformal gravity and arise when studying conformal field theories in curved manifolds. Conformal anomalies were classified by Deser and Schwimmer \cite{Deser:1993yx} and are divided into two families denoted as type A and type B. Type A anomalies are defined by the topological Euler invariant of a curved background, while type B anomalies are defined by the independent conformal gravity invariants existing in a given space-time dimension. For example, in ${\rm D}=4$ only one type B anomaly exists, which is associated with the square of the Weyl tensor, $C^{abcd} C_{abcd}$. In six dimensions, which is the main focus of our paper, there are three independent contributions to conformal gravity describing type B anomalies. They are given by the following terms
\begin{equation}\label{eq:ConfInvs}
c_1\, C_{abcd} C^{aefd} C_e{}^{bc}{}_f
+c_2\, C_{ab}{}^{cd} C_{cd}{}^{ef} C_{ef}{}^{ab}
+c_3 \,C_{abcd} (\d_e^a \Box - 4 \mathcal{R}_e{}^a + \tfrac{6}{5} \d_e^a \mathcal{R}) C^{ebcd}
\,,
\end{equation}
where $C_{abcd}$ is the Weyl tensor, $\cR_{ab}$ is the Ricci tensor, and $c_i$ represent the three type B anomaly coefficients.

It is well known that superconformal algebras exist in Lorentzian space-time dimensions less than or equal to six \cite{Nahm:1977tg}. In the six-dimensional case, rigid superconformal field theories with no spin higher than one in the spectrum exist only with chiral $\mathcal{N}=(1,0)$ and $\mathcal{N}=(2,0)$ supersymmetries. The associated (standard) Weyl multiplets for 6D $\mathcal{N}=(1,0)$ and $\mathcal{N}=(2,0)$ conformal supergravity were constructed within the component fields superconformal tensor calculus in 1985 \cite{Bergshoeff:1985mz} and 1999 \cite{Bergshoeff:1999}, respectively. Obtaining the supersymmetric extension of the 6D conformal anomalies \eqref{eq:ConfInvs} has represented a challenging open problem for decades. Various constraints on the 6D conformal supergravity invariants were obtained by using indirect analysis in \cite{Bastianelli:2000hi,Kulaxizi:2009pz,Beccaria:2015uta}. 
In particular, in these works, it was argued that only two independent 6D conformal supergravity actions should exist with $\mathcal{N}=(1,0)$ and one for $\mathcal{N}=(2,0)$ supersymmetry, hence relating the anomaly coefficients in \eqref{eq:ConfInvs}. It was only in 2016 \cite{Butter:2016} that for $\mathcal{N}=(1,0)$ this claim was proven directly by using supersymmetry. In particular, a key ingredient was to employ \emph{conformal superspace} techniques to formulate two $(1,0)$ locally superconformal invariants, as well as to prove that only two independent supercurrents can exist to describe the multiplets of equations of motion for $(1,0)$ conformal supergravity. Successively, the results of \cite{Butter:2016} were reduced to component fields in \cite{Butter:2017} and further elaborated in \cite{Casarin:2024qdn}.\footnote{In \cite{Casarin:2024qdn}, results for the 6D $(2,0)$ action obtained as an uplift of the $(1,0)$ ones were extended to include up to fermions squared contributions to compute the anomaly for the 6D $(2,0)$ conformal supergravity theory. See also \cite{Casarin:2023ifl} and reference therein for heat-kernel calculation of the conformal anomaly of various 6D theories, including the challenging case of higher-derivative ones.} In these last two papers, the complexity of the 6D conformal supergravity actions was fully unfolded, showing that these invariants comprise thousands of terms. Moreover, a careful analysis of some of the structures of the two $\cN=(1,0)$ conformal supergravity actions made it possible to prove in \cite{Butter:2017} that only a linear combination of the two $(1,0)$ invariants can possess $\cN=(2,0)$ completion. One of the main results of our paper will be to show that there exists only one 6D $\cN=(2,0)$ supercurrent (the supersymmetric Bach tensor) that can describe the entire multiplet of equations of motion of $(2,0)$ conformal supergravity.
The uniqueness results in \cite{Butter:2017} and in our present work are then in agreement with the previous alternative analysis mentioned above.

The main motivation for our paper is to take another step towards obtaining the complete calculation for the unique action for 6D $\cN=(2,0)$ conformal supergravity. Given the success of the techniques employed in \cite{Butter:2016}, a natural attempt could follow two steps: 1) engineer a conformal superspace formulation of the $(2,0)$ Weyl multiplet of conformal supergravity; 2) use superspace techniques, including the cohomological superform approach \cite{Castellani:1991eu,Gates:1997ag,Gates:1998hy}, to define the $(2,0)$ invariant in terms of a compact building block that could then be reduced to components using straightforward (though computationally challenging) techniques as in \cite{Butter:2017}.
The aim of this paper is to present the solution to the first step, leaving the second step for future research. As we will see, the construction of 6D $\cN=(2,0)$ conformal superspace proves to be an involved exercise in its own right, presenting interesting features compared to other dimensions and amount of supersymmetries.

Before summarising in more detail the analysis in our paper, it is worth pausing to talk more about conformal superspaces. 
Conformal superspace was originally introduced
by D.~Butter for 4D $\mathcal{N} = 1$ supergravity in a seminal work \cite{Butter:2009cp} and then extended to
other space-time dimensions $2 \leq {\rm D} \leq 6$ for various amount of supersymmetry  in 
\cite{Butter:2013goa,Butter:2011sr,Kuzenko:2023qkg,Butter:2014xxa,Butter:2016,Butter:2019edc,Kuzenko:2022qnb,Kuzenko:2025bud}  ---  see \cite{Kuzenko:2022skv,Kuzenko:2022ajd} for recent reviews and \cite{Howe:2020xrg,Howe:2020hxi} for a related approach based on local supertwistors. 
The main idea of this approach is to merge the advantages of the curved superspace techniques with the standard superconformal tensor calculus. This is achieved by gauging the entire superconformal algebra in superspace and by imposing suitable conventional constraints that make the superspace geometry dependent only in terms of a single primary superfield that describes the supersymmetric analogue of the Weyl tensor (Cotton tensor in 3D). In the last 15 years, conformal superspaces have been instrumental in pushing forward the construction of higher-derivative invariants. such as
the $\cN$-extended conformal supergravity 
actions in three dimensions for $3\leq \cN \leq 6$ \cite{Butter:2013goa,Butter:2013rba,Kuzenko:2013vha}, the $\cN=(1,0)$ conformal supergravity invariants \cite{Butter:2016, Butter:2017}, the construction of the complete 4D $\cN=4$ conformal supergravity action \cite{Butter:2016mtk,Butter:2019edc}, and recently the completion of all curvature squared invariants in 5D $\cN=1$, and 6D $\cN=(1,0)$ supergravity \cite{Butter:2014xxa,Novak:2017wqc,Butter:2018wss,Gold:2023ymc,Gold:2023dfe,Gold:2023ykx}  ---  see \cite{Butter:2013lta,Kuzenko:2014jra,Kuzenko:2015jda,Kuzenko:2015jxa,Kuzenko:2016tfz} for more references on the application of conformal superspace to study higher-derivative supergravity invariants. An advantage of conformal superspace is that at the component level, it recovers the Weyl multiplet and transformation rules of conformal supergravity as formulated within the superconformal tensor calculus  ---  in fact, constructing a conformal superspace geometry uniquely leads to the complete algebra for a gauging of the superconformal algebra. Completing the classification of these constructions to the 6D $\cN=(2,0)$ case is then an important step in finalising the development of this formalism as well as a stepping stone towards studies that could shed new light on the structure of the mysterious 6D $\cN=(2,0)$ superconformal field theories, as well as finding applications to holography and more.

Coming back to our paper, our construction of 6D $\cN=(2,0)$ conformal superspace  yields multiple results about 6D $\cN=(2,0)$ conformal supergravity. 
First, conformal superspace itself provides an off-shell covariant formalism to work with conformal supergravity. This formalism is quite desirable because it inherently encodes superconformal symmetry within the superspace geometry --- essentially packing some of the structure into the definitions of the objects themselves. 
Second, we provide explicitly the entire structure of the gauged superconformal algebra and the supersymmetry transformations of the fields as seen in Section~\ref{sect:ConformalSuperspace}. More details of the derivation of the 6D $\cN=(2,0)$ conformal superspace geometry are given in Section \ref{sect:DerivationOf6DConformalSuperspace}. As we mention in Section~\ref{sect:component-reduction}, it is relatively easy to translate these results directly to components. Some explicit formulas are also given. 
Third, in Section~\ref{sect:ConventionalConstraintsAndDeformations} we have proved a rigidity result about the gauged algebra  ---  its generators $\nabla_a,\nabla_\a^i,S_i^\a,K_{a}$ mostly cannot be deformed. In particular, only $\nabla_a$ admits a deformation and it is governed by one real parameter. The discussion in Section~\ref{sect:ConventionalConstraintsAndDeformations} shows that this means the conventional constraints in components are also equally rigid. We find that the \emph{traceless frame} conventional constraints used in 6D $\cN=(1,0)$ components in \cite{Butter:2017} can also be implemented in 6D $\cN=(2,0)$. This discussion is continued in Section~\ref{sect:truncationto6DN=(1,0)2} where we examine the need for deformations when truncating the 6D $\cN=(2,0)$ standard Weyl multiplet to the 6D $\cN=(1,0)$ standard Weyl multiplet.
Fourth, in Section~\ref{sect:BachTensor} we are able to completely determine the 6D $\cN=(2,0)$ Bach tensor, which is the primary superfield describing the whole multiplet of equations of motions for conformal supergravity, up to an irrelevant overall scaling. We make a general ansatz with seven real parameters and fix five of them using our $\cN=(2,0)$ conformal superspace. We then truncate our result and compare against the $\cN=(1,0)$ results of \cite{Butter:2017}. This shows agreement and fixes the last degree of freedom, leaving one independent parameter associated with overall scaling.

For the reader’s convenience, we have added two Appendices to our paper.
In Appendix~\ref{appendix:notationsandconventions} we give our notations and conventions. First, we give some detail on our $\USp(4)$ and gamma matrix conventions and also on reality conditions of the superconformal algebra and fields in our theory. Second, we give some details on superform conventions with discussion on exterior derivatives and interior products. Third, we give details on converting our conventions to those of \cite{Bergshoeff:1999} since their results are foundational for this paper and we directly compare back to them in Section~\ref{sect:component-reduction}. 
In Appendix \ref{appendix:reptheory} we provide some detail on the representation theory of $\SL(n+1,\bbC)$ and $\Sp(2n,\bbC)$ since we heavily leverage it in Sections~\ref{sect:DerivationOf6DConformalSuperspace} and \ref{sect:BachTensor}. First, we give a method to take a $\SL(n+1,\bbC)$ irrep represented by a Young diagram and its Young symmetriser, and determine the corresponding $\SL(n+1,\bbC)$ highest weight and dimension. Second, we give a short derivation of the scaling factor associated with Young symmetrisers when they are used in index notation on $\SL(n+1,\bbC)$ tensors, this allows us define the idempotents --- see equation \eqref{YoungSymmetrisersExample} and \eqref{scalingfactor}. Third, we discuss the multi-term symmetries, or Garnir relations of tensors in a $\SL(n+1,\bbC)$ irrep and how these characterise all the index symmetries of the tensor. Fourth, we give a method to take a $\Sp(2n,\bbC)$ irrep represented by a traceless Young diagram and determine the corresponding $\Sp(2n,\bbC)$ highest weight. We also give the Weyl dimension formula for $\Sp(4,\bbC)$. Fourth, we give some $\SL(4,\bbC) \times \Sp(4,\bbC)$ irrep decompositions of some fields in our theory.

\section{Conformal superspace} \label{sect:ConformalSuperspace}

Conformal superspace in dimensions two \cite{Kuzenko:2022qnb}, three \cite{Butter:2013goa}, four \cite{Butter:2009cp,Butter:2011sr,Kuzenko:2023qkg,Butter:2019edc}, five \cite{Butter:2014xxa}, and six \cite{Butter:2016} mostly possesses the following key properties: (i) the gauging of the entire superconformal algebra; (ii) the curvature and torsion tensors may be expressed in terms of a single primary superfield that describe the supersymmetric analogue of the Weyl tensor (Cotton tensor in 3D);
and (iii) the gauged algebra obeys the same basic constraints as those of a super Yang--Mills
theory. 

An exception to (ii) and (iii) occurs in the maximal 4D $\cN=4$ case \cite{Butter:2019edc}. In our maximal 6D $\cN=(2,0)$ case, (ii) is still obeyed as the whole algebra is described in terms of a single primary super-Weyl tensor superfield, but we also break (iii), so, similarly to the 4D $\cN=4$ case, our gauged algebra is more complex than in all other cases listed in the references above where (iii) holds.

In this section, we will describe how the gauging of simple superconformal algebras can be engineered in general dimensions from first principles. 
In Section~\ref{sect:6DN20conformalsuperspace} we will then specialise to the 6D $\cN=(2,0)$ case, introduce the super-Weyl tensor, and show how to constrain the gauged algebra to describe conformal supergravity. We mainly give all the final results and postpone the details until Section~\ref{sect:DerivationOf6DConformalSuperspace}.

\subsection{\texorpdfstring{The 6D $\cN=(2,0)$ superconformal algebra}{The 6D N=(2,0) superconformal algebra}}

The 6D $\cN=(2,0)$ superconformal group is $\OSp(6,2|4)$ and contains the generators of
translation ($P_{a}$), Lorentz  ($M_{ab}$), 
special conformal ($K_{a}$), dilatation ($\mathbb{D}$), $\USp(4)$ ($J^{ij}$),
$Q$-supersymmetry ($Q_\a^i$) and $S$-supersymmetry ($S^\a_i$) 
transformations.\footnote{
For our spinor conventions and notations we refer to Appendix A of \cite{Butter:2016}. The only difference is the replacement of the $\bbC^2$ symplectic indices of $\SU(2)$ with those of $\bbC^4$ of $\USp(4)$. We discuss the minor effects of this in our Appendix~\ref{appendix:gammamatrixandsymplecticalgebra} and give some basic formulae.} 
Their algebra is
\begin{subequations}\label{SuperconformalAlgebra}
\begin{eqnarray}
&[M_{a b} , M_{c d}] = 2 \eta_{c [a} M_{b] d} - 2 \eta_{d [ a} M_{b ] c} 
\,,\quad
[J^{ij} , J^{kl}] = \O^{k(i} J^{j) l} + \O^{l (i} J^{j) k} \,,
~~~~~~
 %%%
 \\
&[M_{a b}, P_c] = 2 \eta_{c [a} P_{b]} 
\,, \quad 
[M_{a b} , K_c] = 2 \eta_{c [a} K_{b]} 
\,, \quad
[\mathbb{D} , P_a] = P_a 
\,, \quad
 [\mathbb{D} , K_a] = - K_{a} 
 \,,~~~~~~ 
 %%%
 \\
  &{[}M_{ab} , Q_\g^k {]} = - \hf (\g_{ab})_\g{}^\d Q_\d^k 
\,, \quad 
[{\mathbb D} , Q_\a^i ] = \hf Q_\a^i 
\,,\quad
[J^{ij} , Q_\a^k ] = \O^{k (i} Q_\a^{j)} \,,~~~~~~~~
 \\
&{[}M_{ab} , S^\g_k {]} =- \hf (\tilde{\g}_{ab})^\g{}_\d S^\d_k  
\,, \quad 
[{\mathbb D} , S^\a_i ] = - \hf S^\a_i
\,, \quad [J^{ij} , S^\a_k ] = \d_k^{(i} S^{\a j)} 
\,,~~~~~~~~
%%%
\\
&\{ Q_\a^i , Q_\b^j \} 
=
- 2 \ri \Omega^{ij} (\g^{c})_{\a\b} P_{c} 
\,, \quad
\{ S^\a_i , S^\b_j \} = - 2 \ri \Omega_{ij} (\tilde{\g}^{c})^{\a\b} K_{c} 
\,,~~~~~~
%%%
\\
&\{ S^\a_i , Q_\b^j \} = 2 \d^\a_\b \d^j_i \mathbb D - 4 \d^j_i M_\b{}^\a + 8 \d^\a_\b J_i{}^j 
\,,
\quad
[K_a , P_b] = 2 \eta_{a b} {\mathbb D} + 2 M_{ab} 
\,, ~~~~~~
%%%
\\
& [ K_a , Q_\a^i ] = - \ri (\g_a)_{\a\b} S^{\b i} \,, \quad [S^\a_i , P_a ] = - \ri (\tilde{\g}_a)^{\a\b} Q_{\b i} 
 \,, ~~~~~~
\end{eqnarray}
\end{subequations}
with all other (anti)commutators vanishing.
It is useful to represent vectors and antisymmetric tensors in terms of spinorial bilinears. According to the definitions given in Appendix \ref{appendix:gammamatrixandsymplecticalgebra}, the generator $P_a$ can be expressed as $P_{\a\b}=(\gamma^a)_{\a\b}P_a$,
$M_{ab}$ can be expressed as  $M_\a{}^\b = - \frac{1}{4} (\g^{ab})_\a{}^\b M_{ab}$, and so on.
The following alternative representation of some commutation relations is useful
\begin{subequations}
\bea
&[ M_\a{}^\b , Q_\g^k ] = - \d_\g^\b Q_\a^k + \frac{1}{4} \d^\b_\a Q_\g^k 
\,,\quad 
[ M_\a{}^\b , S^\g_k ] = \d_\a^\g S^\b_k - \frac{1}{4} \d^\b_\a S^\g_k
\,,
\\
&[P_{\d\ve},M_\k{}^\l] = -2\d_{[\d}^\l P_{\ve]\k} - \frac{1}{2}\d_\k^\l P_{\d\ve}
\,,\quad
[K^{\z\eta},P_{\d\ve}] = 8 \d_{[\d}^\z \d_{\ve]}^\eta \mathbb{D} - 16\d_{[\d}^\z M_{\ve]}{}^\eta
\,.
\eea
\end{subequations}
We can group together the translation and $Q$-supersymmetry as 
generators of supertranslations, $P_A = (Q_\a^i,P_a )$.
Similarly, we group together the special conformal and $S$-supersymmetry transformations
by denoting $K^A = (S^\a_i,K^a )$ and the closed subset of generators that do not
contain $P_{{A}}$ by $X_{\underline{a}} = (M_{ab} , J_{ij}, \mathbb D , K^A)$. The superconformal algebra then takes the form 
\begin{subequations}
\begin{align}
[X_{\underline{a}} , X_{\underline{b}}] &= -f_{\underline{a} \underline{b}}{}^{\underline{c}} X_{\underline{c}} 
\,,\\
[X_{\underline{a}} , P_{{B}}] &=
- f_{\underline{a} {{B}}}{}^{{C}} P_{{C}}
 -f_{\underline{a} {{B}}}{}^{\underline{c}} X_{\underline{c}}
\,,\\
[P_{{A}} , P_{{B}}] &= -f_{{{A}} {{B}}}{}^{{{C}}} P_{{C}}
\,,
\end{align}
\end{subequations}
where one understands $[-,-]$ to be the $\bbZ_2$-graded commutator, reducing to the commutator and anticommutator depending on its arguments.
The subgroup generated by exponentiating $X_{\ua}$ is denoted by $\cH$.

\subsection{Gauging a superconformal algebra in general}

We follow closely the approaches in previous conformal superspace papers \cite{Butter:2013goa,Butter:2009cp,Butter:2011sr,Kuzenko:2023qkg,Butter:2019edc,Butter:2014xxa,Butter:2016,Kuzenko:2022qnb}, but stress on using first principles in building conformal superspace.
We occasionally leverage details about 6D $\cN =(2,0)$ for concreteness, but it is all generalisable to any case where the superconformal algebra is a simple superalgebra.

We note that notation-wise, we opt to use $[-,-]$ to denote the $\bbZ_2$-graded commutator, which reduces to the commutator and anticommutator depending on its arguments. Similarly, we denote graded symmetrisation and graded antisymmetrisation of indices as $(\dots)$ and $[\dots]$ respectively. For example
\begin{equation}
    T_{[AB]} = \frac{1}{2!} ( T_{AB} - (-1)^{AB} T_{BA})
    \,,
\end{equation}
where we write use conventions $A \equiv \e(A) = \pm 1$ for the grading associated to an index. Moreover, we enforce that all objects have grading determined by their indices.
This is a similar notation to two original conformal superspace papers \cite{Butter:2009cp,Butter:2011sr}, but all later conformal superspace papers \cite{Kuzenko:2022qnb,Butter:2013goa,Kuzenko:2023qkg,Butter:2014xxa,Butter:2016} instead use the notation $[-,-\}$ for graded commutators and $[...)$ for graded antisymmetrisation --- graded symmetrisation does not cleanly fit into these notations. We use graded symmetrisation only briefly, so we fix our notation to allow for it more easily. 

In 6D $\cN =(2,0)$ we use a supermanifold $\cM^{6|16}$ described by local coordinates $z^M = (x^m, \theta_i^\mu)$ where $m=0,1,2,3,4,5$, $i=1,2,3,4$, $\mu=1,2,3,4$ and $\theta_i^\mu$ is a symplectic Majorana-Weyl spinor. All other cases of conformal superspaces are analogous, but the number of independent coordinates, amount of R-symmetry and reality conditions on spinors will vary.

To begin gauging, we associate to each generator $X_{\underline{a}} = (M_{ab} , J_{ij}, \mathbb D , S_i^\a, K^a)$ of $\mathcal{H}$ a connection 1-form $\omega^{\ua} = (\Omega^{ab},\Phi^{ij},B,\mathfrak{F}_\a^i, \mathfrak{F}_a)$, the resulting covariant derivative is
\begin{equation}
\nabla_M = \partial_M - \omega_M = \partial_M - \omega_M{}^{\ua} X_{\ua}
\,.
\end{equation}
Now, drawing inspiration from CFT, we impose that our covariant derivative $\nabla_M$ does not commute with the special conformal generators $S_i^\a,K_a$. The other generators $M_{ab},\bbD,J_{ij}$ are enforced to commute with the covariant derivative $\nabla_M$. In particular, to have a consistent structure we impose that $[K^A,\nabla_M]$ is valued in $\nabla_M,X_{\ua}$ up to some superfield coefficients. This is the distinguishing feature of gauging the (super)conformal algebra versus what is typically seen in Riemmanian geometry or Yang--Mills.

We introduce a local frame superfield (supervielbein) $E_M{}^A$ and its inverse $E_A{}^M$. It is helpful to focus on the 1-form $E^A = dz^M E_M{}^A$. First, we impose that $E^A$ is a $\mathcal{H}$-covariant vector-valued 1-form. We now take a step that seems abstract, but we will discuss it in more detail shortly. We would like to fix $E^A$'s representation by imposing that it is vector-valued in the sense of $E = E^A P_A$. Hence, we have the (minus) adjoint action
\begin{equation}
X_{\ub} E = - \ad(X_{\ub})E = -(-1)^{A\ub} E^A[X_{\ub},P_A]
\,.
\end{equation}
This is not quite right since conformal symmetry means $[X_{\ub},P_A]$ is $P_A,X_{\ua}$-valued. So we modify the above action by projecting on the $P_A$-valued part.  Leveraging that $E_M{}^A E_A{}^N = \d_M^N$ is a trivial representation ($X_{\ua} \d_M^N = 0$) and using the product rule, we get the structure constant versions
\begin{subequations}\label{XactionsE}
\begin{align}
X_{\ub} E_M{}^A &= -(-1)^{\ub M} E_M{}^C f_{C\ub}{}^A
\,,\\
X_{\ub} E_A{}^M &= - f_{\ub A}{}^C E_C{}^M
\,.
\end{align}
\end{subequations}
The underlying bosonic and fermionic parts of $E^A=(E^a,E^\a_i)$  then obey
\begin{subequations}
\begin{alignat}{4}
M_{ab} E^c &= 2\d_{[a}^c E_{b]}
\,,&\quad
J_{ij} E^c &= 0
\,,&\quad
\bbD E^c &= -E^c
\,,&\quad
K^A E^c &= 0
\,,\\
M_{ab} E^\g_k &= - \frac{1}{2} (\tilde{\g}_{ab})^\g{}_\d E^\d_k
\,,&\quad
J_{ij} E^\g_k &= -\O_{k(i} E^\g_{j)}
\,,&\quad
\bbD E^\g_k &= -\frac{1}{2} E^\g_k
\,,\\
S^\b_j E^\g_k &= \ri \O_{jk} (\tilde{\g}_c)^{\b\g} E^c 
\,,&\quad
K^a E^\g_k &= 0
\,.
\end{alignat}
\end{subequations}
These relations are precisely what one expects of the vielbein and gravitino in component descriptions of conformal supergravity.
Indeed, in our theory the vielbein and gravitino arise as pullbacks\footnote{The double bar denotes the pull back to the underlying manifold through its inclusion $\iota: \mathcal{M}^6 \to \mathcal{M}^{6|16}$. Practically, for a differential form this amounts to setting all fermionic coordinate $\theta_i^\mu$'s and their 1-forms $d\theta_i^\mu$'s to zero.} $e^a = E^a ||$ and $\psi_i^\a = 2 E^\a_i ||$. 

Another way to justify \eqref{XactionsE} is to compare to the role of a vielbein in other geometric theories. On any (pseudo-)Riemannian manifold, a vielbein $e^a$ is, geometrically, a local choice of basis on the tangent space such that the Riemannian metric becomes the Euclidean metric. Then all other vielbeins are related to $e^a$ by $\SO(n)$ rotations. This is precisely what the rotation generator $M_{ab}$ does when acting on $E^c \sim e^c$ above. That is to say, the Euclidean algebra $\mathfrak{so}(n) \times \bbR^n$ has an adjoint representation where its rotation generators $M_{ab}$ act on its translation generators $P_{a}$ via rotations. Hence, imposing the adjoint representation in our superconformal algebra case is just a generalisation.

Combining the $\mathcal{H}$-actions on $E^A$ and the fact that $M_{ab},J_{ij},\bbD$ commute with $\nabla_M$, the in-frame covariant derivatives $\nabla_A = E_A{}^M \nabla_M$ satisfy relations identical to the superconformal algebra's translation generator
\begin{subequations}
\begin{alignat}{3}
[M_{ab},\nabla_c] &= 2\eta_{c[a} \nabla_{b]}
\,,&\quad
[J^{ij},\nabla_c] &= 0
\,,&\quad
[\bbD,\nabla_c] &= \nabla_c
\,,\\
[M_{ab},\nabla_\g^k] &= - \frac{1}{2} (\g_{ab})_\g{}^\d \nabla_\d^k
\,,&\quad
[J^{ij},\nabla_\g^k] &= \O^{k(i} \nabla_\g^{k)}
\,,&\quad
[\bbD,\nabla_\g^k] &= \frac{1}{2} \nabla_\g^k
\,.
\end{alignat}
\end{subequations}

The next conceptual step is to define infinitesimal $\mathcal{H}$-gauge transformations of the connection 1-form. To do this, we need to understand covariant fields. A covariant field $\phi$, which can be vector-valued (perhaps through possessing $A,\ua$ indices), has a $\mathcal{H}$-gauge transformation defined by the representation of $X_{\ua}$ on $\phi$. Allow the gauge transformations to be parameterised by $\Lambda = \Lambda^{\ua} X_{\ua}$ with $\Lambda^{\ua}$ being a collection of superfields. The gauge transformation of $\phi$ is then simply
\begin{align}
    \d_{\mathcal{H}}\phi =\Lambda \phi = \Lambda^{\ua}(X_{\ua}\phi)
    \,.
\end{align}
The finite $\mathcal{H}$-gauge transformation is given by exponentiation as $\phi \to U\phi = e^\Lambda \phi$. We now act a local group element $U$ on the covariant derivative of $\phi$. Treating $\nabla_M$ and $U$ as linear operators, we find
\begin{equation}
U(\nabla_M\phi) = U(\nabla_M(U^{-1}U\phi)) = \left(\nabla_M + U[\nabla_M,U^{-1}] \right) U\phi
\,,
\end{equation}
where, unlike in Riemannian geometry or Yang--Mills, commutators are needed since $\nabla_M$ does not commute with $\mathcal{H}$ due to the presence of special conformal generators $K^A$. In Riemannian geometry or Yang--Mills, one finds that $[\nabla_M,U^{-1}]$ is equal to $\nabla_MU^{-1}$ and hence, the gauge transformation of the connection simplifies to the typical adjoint group action on $\omega$ plus a contribution from the Cartan--Maurer form. This is not the case anymore, since we cannot make this replacement. From here, the finite $\mathcal{H}$-gauge transformation of $\nabla_M$ is clear. Its infinitesimal counterpart can be found by setting $U=1+\Lambda$ and truncating to terms linear in $\Lambda$ --- while writing the action of $\Lambda$ on $\nabla_M$ as $\delta_{\mathcal{H}}\nabla_M$. We find
\begin{align}
\delta_{\mathcal{H}}\nabla_M = [\Lambda,\nabla_M]
\,.
\end{align}
Another argument that does not need to reference the finite result is given by treating infinitesimal gauge transformations by $\Lambda$ as obeying a product rule
\begin{equation}\label{GaugeTransformProductRule}
\Lambda(\nabla_M\phi) = (\Lambda\nabla_M)\phi + \nabla_M(\Lambda\phi) = (\delta_{\mathcal{H}}\nabla_M)\phi + \nabla_M(\Lambda\phi)
\,.
\end{equation}
In either case, we can expand to find $\delta_{\mathcal{H}}\nabla_M = \Lambda^{\ua} [X_{\ua},\nabla_M] - (\nabla_M\Lambda^{\ua})X_{\ua}$.
Treating $\Lambda=\Lambda^{\ua} X_{\ua}$ as a vector-valued superfield w.r.t.~the adjoint representation and enforcing that the gauge transformation is entirely in the connection, equivalent to $\delta_{\mathcal{H}}\nabla_M~=~(\partial_M~-~\delta_{\mathcal{H}}\omega_M)$, gives
\begin{equation}
\delta_{\mathcal{H}}\omega_M = (\partial_M \Lambda^{\uc} + \omega_M{}^{\ub} \Lambda^{\ua} f_{\ua\ub}{}^{\uc}) X_{\uc} - \Lambda^{\ua} [ X_{\ua}, \nabla_M]
\,,
\end{equation}
and implicitly enforces that $[ X_{\ua}, \nabla_M]$ is $X_{\ua}$-valued. We now enforce this explicitly by defining structure functions $F_{\ua B}{}^{\uc}$ encapsulating this dependence via
\begin{align}
[X_{\ua},\nabla_B] &= [X_{\ua}, E_B{}^M \nabla_M]
\nonumber\\
&=
(-1)^{\ua(B+M)}E_B{}^M [X_{\ua},\nabla_M] + (X_{\ua}E_B{}^M) \nabla_M
\nonumber\\
&=
(-1)^{\ua(B+M)}E_B{}^M [X_{\ua},\nabla_M] - f_{\ua B}{}^C \nabla_C
\nonumber\\
&=
- F_{\ua B}{}^{\uc} X_{\uc} - f_{\ua B}{}^{C} \nabla_C
\,.
\end{align}
At this point $F_{\ua B}{}^{\uc}$ are understood as some general superfield coefficients --- we have no reason to constrain them further yet. Using this, we can backtrack and prove that the gauge transformations take the form
\begin{subequations}\label{HGaugeTransformations}
\begin{align}
\delta_{\mathcal{H}} E_M{}^A &= E_M{}^C \Lambda^{\ub} f_{\ub C}{}^A
\,,\\
\delta_{\mathcal{H}}\omega_M{}^{\uc} &= 
\partial_M \Lambda^{\uc}
+ \omega_M{}^{\ub} \Lambda^{\ua} f_{\ua\ub}{}^{\uc} 
+ E_M{}^B \Lambda^{\ua} F_{\ua B}{}^{\uc} 
\,,
\end{align}
\end{subequations}
as typically seen in previous conformal superspace papers, e.g.\ \cite{Butter:2014xxa}.

The next step is to derive the curvature and torsion. We find it easiest to take a geometric approach using differential forms. The conventions we use for superforms are described in Appendix~\ref{appendix:superforms}. In particular, we mention that the exterior (covariant) derivative is taken as acting on (covariant) $n$-forms as
\begin{subequations}
\begin{alignat}{2}
d &= (-1)^n dz^M \partial_M
\,,&\qquad
\nabla &= (-1)^n dz^M \nabla_M
\,,\\
d &= (-1)^n E^A E_A
\,,&\qquad
\nabla &= (-1)^n E^A \nabla_A
\,.
\end{alignat}
\end{subequations}
Evaluating the exterior covariant derivative on some $\mathcal{H}$-covariant (0-form) superfield $\phi$ gives the curvature
\begin{align}\label{CurvatureDef}
\nabla\nabla\phi = - R^{\ua} X_{\ua} \phi 
&=
-d\omega\phi - \omega^{\ua} [\nabla,X_{\ua}]\phi - \omega^{\ua} \wedge \omega X_{\ua} \phi
\nonumber\\
&=-\left(d\omega^{\ua}
- E^C \wedge \omega^{\ub}  F_{\ub C}{}^{\ua} 
- \frac{1}{2}\omega^{\uc} \wedge \omega^{\ub} f_{\ub\uc}{}^{\ua} \right)X_{\ua} \phi 
\,.
\end{align}
Like the gauge transformation of the connection, the curvature has an extra term originating from $[X_{\ua},\nabla_M] \neq 0$.
Like in Riemannian geometry, the torsion is simply the exterior covariant derivative of the frame
\begin{equation}\label{TorsionDef}
T^A = \nabla E^A = dE^A - E^C \wedge \omega^{\ub} f_{\ub C}{}^A
\,.
\end{equation}
Due to the 2-form nature of $\nabla\nabla\phi$ we get $[\nabla_M,\nabla_N] = - R_{MN}$ where $R_{MN}$ are the superform components of $R$. Moreover, one can use $E_M{}^A$ to rewrite this to its in-frame version. This yields the gauged algebra
\begin{subequations}\label{GaugedAlgebra}
\begin{align}
[X_{\underline{a}} , X_{\underline{b}}] &= -f_{\underline{a} \underline{b}}{}^{\underline{c}} X_{\underline{c}} 
\,,\\
[X_{\underline{a}} , \nabla_{{B}} ] &=
- f_{\underline{a} {{B}}}{}^{{C}} \nabla_{{C}}
 -F_{\underline{a} {{B}}}{}^{\underline{c}} X_{\underline{c}}
\,,\\
[\nabla_A,\nabla_B] &= - T_{AB}{}^C \nabla_C 
- R_{AB}{}^{\uc} X_{\uc}
\,,
\end{align}
\end{subequations}
as typically seen in previous conformal superspace papers.
Now we consider the $\mathcal{H}$-gauge transformations of the curvature and torsion. They are calculated in the usual way using our gauge transformations of the connection 1-form and frame \eqref{HGaugeTransformations}. Whilst calculating $\delta_{\mathcal{H}} R^{\ua} = \Lambda^{\ub}(X_{\ub} R^{\ua})$, we must impose that the superfield-valued structure functions $F_{\ua B}{}^{\uc}$ are $\mathcal{H}$-covariant so that $X_{\ua}$ has a well-defined action on them. We find\footnote{We use the vertical lines in $\tilde{f}_{[\uc\ud}{}^{\tilde{e}} \tilde{f}_{|\tilde{e}|C]}{}^{\ua}$ to denote that the index $\tilde{e}$ is not being counted in the graded antisymmetrisation. In general, a dummy index being summed over will never be included in (anti)symmetrisations in this paper.}
\begin{subequations}\label{TorsionCurvatureGaugeTransforms1}
\begin{align}
\delta_{\mathcal{H}} R^{\ua}
&=
R^{\ub} \Lambda^{\uc} f_{\uc\ub}{}^{\ua}
+ T^B \Lambda^{\uc} F_{\uc B}{}^{\ua}
- E^C \wedge E^B \Lambda^{\ud} F_{\ud B}{}^{\ub} F_{\ub C}{}^{\ua}
+ E^B \wedge \Lambda^{\uc} \nabla F_{\uc B}{}^{\ua}
\nonumber\\&\quad
- E^C \wedge \omega^{\ud} \Lambda^{\uc} \left(
2 X_{[\uc} F_{\ud]C}{}^{\ua} + 3 \tilde{f}_{[\uc\ud}{}^{\tilde{e}} \tilde{f}_{|\tilde{e}|C]}{}^{\ua}
\right)
\,,\\
\delta_{\mathcal{H}} T^{A} &=
T^C\Lambda^{\ub} f_{\ub C}{}^A
- E^{C} \wedge E^D \Lambda^{\ua} F_{\ua D}{}^{\ub} f_{\ub C}{}^A
\,,
\end{align}
\end{subequations}
where the $\tilde{a}$ and $\tilde{f}_{\tilde{a}\tilde{b}}{}^{\tilde{c}}$ notation is shorthand for combining the covariant derivatives with the $\mathcal{H}$-generators as $X_{\tilde{a}} = (\nabla_{A},X_{\ua})$ with
\begin{equation}
[X_{\tilde{a}},X_{\tilde{b}}] = - \tilde{f}_{\tilde{a}\tilde{b}}{}^{\tilde{c}} X_{\tilde{c}}
\,.
\end{equation}
For completeness, the expansion is given by
\begin{align}
3\tilde{f}_{[\ua\ub}{}^{\tilde{e}} \tilde{f}_{|\tilde{e}|C]}{}^{\ud}
&=
f_{\ua\ub}{}^{\ue} F_{\ue C}{}^{\ud}
+(-1)^{\ua(C+\ub)}  F_{\ub C}{}^{\ue}  f_{\ue\ua}{}^{\ud}
+(-1)^{C (\ub +\ua) } F_{C \ua}{}^{\ue} f_{\ue\ub}{}^{\ud}
\nonumber\\&\quad
+(-1)^{\ua(C+\ub)}f_{\ub C}{}^E  F_{E \ua}{}^{\ud}
+(-1)^{C(\ub+\ua)}   f_{C \ua}{}^{E}  F_{E \ub}{}^{\ud}
\,.
\end{align}
There are some interesting points to be made. First, we again have extra terms in \eqref{TorsionCurvatureGaugeTransforms1} appearing due to $[X_{\ua},\nabla_M] \neq 0$. The next is that the gauged algebra Jacobi identities encode these gauge transformations. For example,
\begin{align}
0 &= -\left( [X_{\ua},[\nabla_B,\nabla_C]] + (\text{graded perms}) \right)
\nonumber\\
&=
-3 [X_{[\ua},[\nabla_B,\nabla_{C]}]]
\nonumber\\
&=  3\left( X_{[\ua} f_{B C]}{}^{\ud} 
+ \tilde{f}_{[\ua B}{}^{\tilde{e}} \tilde{f}_{|\tilde{e}|C]}{}^{\ud} 
\right) X_{\ud}
+ 3 \left(  X_{[\ua} f_{B C]}{}^{D} 
+\tilde{f}_{[\ua B}{}^{\tilde{e}} \tilde{f}_{|\tilde{e}|C]}{}^{D} 
\right) \nabla_D
\nonumber\\
&=  \left( X_{\ua} R_{B C}{}^{\ud} 
+ (-1)^{\ua(B+C)} 2\nabla_{[B} F_{C] \ua}{}^{\ud}
+ 3\tilde{f}_{[\ua B}{}^{\tilde{e}} \tilde{f}_{|\tilde{e}|C]}{}^{\ud} 
\right) X_{\ud}
\nonumber\\&\quad
+ \left(  X_{\ua} T_{B C}{}^{D} 
+3\tilde{f}_{[\ua B}{}^{\tilde{e}} \tilde{f}_{|\tilde{e}|C]}{}^{D} 
\right) \nabla_D
\,,
\end{align}
where we used $\nabla_B f_{C\ua}{}^{D} = 0$ since these are just structure constants of $\mathcal{H}$. Noting $\mathcal{H}$-covariance, so that we may freely add in $\Lambda^{\ua}$, and write $\d_{\mathcal{H}} = \Lambda^{\ua} X_{\ua}$, this Jacobi identity is equivalent to the 2-form equation
\begin{subequations}\label{TorsionCurvatureGaugeTransforms2}
\begin{align}
\delta_{\mathcal{H}} R^{\ua}
&=
R^{\ub} \Lambda^{\uc} f_{\uc\ub}{}^{\ua}
+ T^B \Lambda^{\uc} F_{\uc B}{}^{\ua}
- E^C \wedge E^B \Lambda^{\ud} F_{\ud B}{}^{\ub} F_{\ub C}{}^{\ua}
+ E^B \wedge \Lambda^{\uc} \nabla F_{\uc B}{}^{\ua}
\,,\\
\delta_{\mathcal{H}} T^{A} &=
T^C\Lambda^{\ub} f_{\ub C}{}^A
- E^{C} \wedge E^D \Lambda^{\ua} F_{\ua D}{}^{\ub} f_{\ub C}{}^A
\,.
\end{align}
\end{subequations}
This matches our original gauge transformations iff the following gauged algebra Jacobi identity holds
\begin{align}
0 &= -\left( [X_{\ua},[X_{\ub},\nabla_C]] + (\text{graded perms}) \right)
\nonumber\\
&=
-3 [X_{[\ua},[X_{\ub},X_{C]}]]
\nonumber\\
&=
3\left( X_{[\ua} f_{\ub C]}{}^{\ud} 
+ \tilde{f}_{[\ua\ub}{}^{\tilde{e}} \tilde{f}_{|\tilde{e}|C]}{}^{\ud} 
\right) X_{\ud}
+ 3\tilde{f}_{[\ua\ub}{}^{\tilde{e}} \tilde{f}_{|\tilde{e}|C]}{}^{D} \nabla_D
\nonumber\\
&=
\left( 2 X_{[\ua} F_{\ub] C}{}^{\ud} 
+ 3\tilde{f}_{[\ua\ub}{}^{\tilde{e}} \tilde{f}_{|\tilde{e}|C]}{}^{\ud} 
\right) X_{\ud}
\,,
\end{align}
where we used $\nabla_Cf_{\ua\ub}{}^{\ud} = 0$, $X_{\ua} f_{\ub C}{}^D = 0$ as they are structure constants. We also used $\tilde{f}_{[\ua\ub}{}^{\tilde{e}} \tilde{f}_{|\tilde{e}|C]}{}^{D} = 0$ which is just a graded Jacobi identity of the superconformal algebra.
We need to fix the gauge transformation of the $\mathcal{H}$-covariant structure functions $F_{\ua B}{}^{\uc}$. This can only be partially done at the first principles-level. In particular, since we originally imposed that $[K^A,\nabla_M] \neq 0$ and that $\nabla_M$ commutes with the other generators, then this implies $F_{\ua B}{}^{\uc} = 0$ if $\ua \equiv M_{ab},J_{ij},\bbD$. Thus, the above Jacobi identity immediately gives
\begin{equation}
X_{\ua} F_{\ub C}{}^{\ud} 
=
- 3\tilde{f}_{[\ua\ub}{}^{\tilde{e}} \tilde{f}_{|\tilde{e}|C]}{}^{\ud}
\,,\quad
\text{if $\ua \equiv M_{ab},J_{ij},\bbD$}
\,.
\end{equation}
This $\mathcal{H}$-action is extremely notationally condensed. One can check that the above is equivalent to the expected
\begin{subequations}
\begin{align}
\bbD F_{\ub C}{}^{\ud}
&=
( \bbD(\ub) + \bbD(C) - \bbD(\ud) ) F_{\ub C}{}^{\ud}
\,,\\
M_{ab} F_{\ub C}{}^{\ud}
&\equiv
M_{ab}(\ub)F_{\ub C}{}^{\ud} + M_{ab}(C)F_{\ub C}{}^{\ud} + M_{ab}(\ud)F_{\ub C}{}^{\ud}
\,,\\
J_{ij} F_{\ub C}{}^{\ud}
&\equiv
J_{ij}(\ub)F_{\ub C}{}^{\ud} + J_{ij}(C)F_{\ub C}{}^{\ud} + J_{ij}(\ud)F_{\ub C}{}^{\ud}
\,,
\end{align}
\end{subequations}
where $\bbD(\tilde{a})$ is the dilatation dimension of the operator, i.e\ $[\bbD,X_{\tilde{a}}] = \bbD(\tilde{a}) X_{\tilde{a}}$. This $\bbD$-action ensures that $[X_{\ua},\nabla_B]$ has a well-defined dilatation dimension.
The notation $M_{ab}(\ub)F_{\ub C}{}^{\ud}$ denotes the usual action of $M_{ab}$ on the vector or spinor indices in $\ub$, with the same applying for $J_{ij}(\ub)F_{\ub C}{}^{\ud}$ w.r.t.\ $\USp(4)$ indices. These $M,J$-actions are just the tensor product of representations applied to the defining representations $M_{ab} v_c = 2\eta_{c[a}v_{b]}$, $M_{ab}v_\g = - (1/2) (\g_{ab})_\g{}^\d v_\d$ and $J^{ij}v^k = \O^{k(i} v^{j)}$. The $S,K$-actions are only partially constrained by the Jacobi identity, with $X_{(\ua} F_{\ub)C}{}^{\ud}$ being an unknown.\footnote{As far as the authors can tell, for $S,K$-actions $X_{(\ua} F_{\ub)C}{}^{\ud}$ is dependent on the specific superspace at hand, and not known from this first principles analysis.} As we will see later, this information is sufficient for building our conformal superspace. The $S,K$-actions are then known in our case by the $S,K$-actions on the standard Weyl multiplet.
We now impose that both of these Jacobi identities hold true. One can consider this an extension of the Riemannian geometry case with the algebra generated by an in-frame (using $e_m{}^a$)  connection $\nabla_a$ and Euclidean rotation generators $M_{ab}$. These Jacobi identities hold trivially in this case as they amount to those of the Euclidean algebra.
Gauging just the conformal algebra will also result in some similar results to the above.

The Bianchi identity of the curvature differs from that of Riemannian geometry, but the torsion one remains the same. Given the $\mathcal{H}$-action on $R^{\ua}$ obtained from the Jacobi identities, they are
\begin{equation}
\nabla R^{\ua} = - E^C \wedge R^{\ub}\, F_{\ub C}{}^{\ua}
\,,\qquad
\nabla T^A = R \wedge E^A
\,,
\end{equation}
and they are exactly equivalent to the last Jacobi identity of the gauged algebra
\begin{equation}
[\nabla_A,[\nabla_B,\nabla_C]] + (\text{graded perms}) = 0
\,.
\end{equation}
If one makes the replacement 
\begin{equation}\label{structurefunctiondecomp}
    F_{\ua B}{}^{\uc} = f_{\ua B}{}^{\uc} + \boldsymbol{F}_{\ua B}{}^{\uc}
    \,,
\end{equation}
where $f_{\ua B}{}^{\uc}$ are the structure constants of the superconformal algebra and $\boldsymbol{F}_{\ua B}{}^{\uc}$ some $\mathcal{H}$-covariant superfields. Then, after using $X_{\ua} f_{\ub C}{}^{\ud} = 0$ and applying Jacobi identities of the superconformal algebra, we find that all the constraints above effectively remain the same with $F_{\ua B}{}^{\uc}$ replaced by $\boldsymbol{F}_{\ua B}{}^{\uc}$. So, without loss of generality, one can assume that the structure constants still appear in the gauged algebra with the superfield corrections $\boldsymbol{F}_{\ua B}{}^{\uc}$ obeying the constraints listed above.

Before we continue, it is worth mentioning that we can consider some basic consistency checks. One can consider discarding the $S,K$-generators which also sets $F_{\ua B}{}^{\uc}=0$, at which point we have a gauged Poincar\'e algebra with R-symmetry and all the formulas match Riemannian geometry (keeping in mind we carry also a connection valued in the R-symmetry algebra and its curvature). We could instead set all the connections to zero $\nabla_A \to E_A = E_A{}^M \partial_M$, then the curvature vanishes and the structure functions $F_{\ua B}{}^{\uc}$ along with the torsion become the structure constants of the superconformal algebra. The equations \eqref{TorsionCurvatureGaugeTransforms2} then collapse to zero by virtue of Jacobi identities of the superconformal algebra. The remaining algebra \eqref{GaugedAlgebra} is just the one generated by local superconformal Killing vectors on a supermanifold with an arbitrary supermetric.

The last step to completing the general gauging procedure is to define appropriate superdiffeomorphism actions. This is essential, because the fermionic part of a superdiffeomorphism should correspond roughly to a $Q$-supersymmetry transformation in components. In particular, we want actions on the frame $E^A$ and the connections $\omega^{\ua}$.
Geometrically, an infinitesimal superdiffeomorphism along a vector field $\xi = \xi^M \partial_M = \xi^A E_A$ acts on a superfield by a Lie derivative. For differential forms, we have the convenient Cartan formula
\begin{align}
\mathcal{L}_{\xi} = d\iota_\xi + \iota_\xi d
\,.
\end{align}
If what we wanted was a generic superdiffeomorphism action, then we could use this and be done. Instead, what we desire is a $\mathcal{H}$-covariant superdiffeomorphism action --- this is because we want $\nabla_\a^i$ to descend to $Q_\a^i$. On any $\mathcal{H}$-covariant $n$-form, this is given by the covariant Lie derivative
\begin{align}
\mathcal{L}_{\xi}^{\text{cov}} = \nabla\iota_\xi + \iota_\xi\nabla
\,.
\end{align}
For example, a $0$-form superfield $\phi$ and the frame $E^A$ give
\begin{subequations}
\begin{align}
\mathcal{L}_{\xi}^{\text{cov}}\phi &= \xi^A \nabla_A \phi
\,,\\
\mathcal{L}_{\xi}^{\text{cov}}E^A &= \nabla\xi^A + \iota_{\xi} T^A = d\xi^A + \omega^{\ub} \xi^C f_{C\ub}{}^A + E^C \xi^B T_{BC}{}^A
\,.
\end{align}
\end{subequations}
Though we mainly work with differential forms here, the covariant Lie derivative can be used on any tensor field in the usual way matching Riemannian geometry. One just needs to replace partial derivatives with covariant ones and be careful of the signs from superspace's grading.\footnote{For example, the covariant Lie derivative of a vector field $V = V^M \partial_M$ is given by 
the following rule $\mathcal{L}_{\xi}^{\text{cov}} V^M = \xi^N \nabla_N V^M - (-1)^{MN} (\nabla_N \xi^M) V^N$
 where one parity sign is introduced on the RHS due to the index $M$ moving through the index $N$ as compared to the LHS where no indices move through $M$.}
Finding the appropriate covariant superdiffeomorphism action on the non-covariant connection 1-form $\omega^{\ua}$ is not as intuitive. Similar to deriving the $\mathcal{H}$-gauge transformations \eqref{GaugeTransformProductRule},  one method is to impose that a \emph{combined} $\d_{\mathcal{G}} = \xi^A\nabla_A + \Lambda^{\ua} X_{\ua}$ transformation obeys a product rule on a superfield $\phi$
\begin{align}
\d_{\mathcal{G}}(\nabla_B\phi) &= (\d_{\mathcal{G}}\nabla_B)\phi + \nabla_B(\d_{\mathcal{G}}\phi)
\,,\\
\d_{\mathcal{G}}\nabla_M &= \partial_M - \d_{\mathcal{G}}\omega_M
\,,
\end{align}
where both the frame and connection implicitly transform due to
\begin{align}
\d_{\mathcal{G}}\nabla_B &= - (\d_{\mathcal{G}} E_M{}^B) \nabla_B + E_B{}^M \d_{\mathcal{G}} \nabla_M
\,.
\end{align}
Working through these equations leads us to impose, as mentioned in \cite{Butter:2014xxa}, that
\begin{subequations}
\begin{align}
\nabla_A \xi^B &= E_A\xi^B + \omega_A{}^{\uc} \xi^D f_{D\uc}{}^B
\,,\\
\nabla_A\Lambda^{\ub} &= E_A \Lambda^{\ub} + \omega_A{}^{\uc} \xi^D F_{D\uc}{}^{\ub} + \omega_A{}^{\uc} \Lambda^{\ud} f_{\ud\uc}{}^{\ub}
\,.
\end{align}
\end{subequations}
The first condition is just the in-frame covariant derivative of the vector field given by $\xi = \xi^M \partial_M$. However, the second is less transparent, we must allow the covariant derivative of the Lie algebra-valued parameter field to also produce a $\xi$ along with the usual adjoint action.
Leveraging these conditions gives the desired combined transformation rule for the connection and reproduces the transformation rule for the frame.
We find that $\d_{\mathcal{G}} = \xi^A\nabla_A + \Lambda^{\ua} X_{\ua}$ acts as
\begin{subequations}
\begin{align}
\d_{\mathcal{G}} E^A &= d\xi^A + E^B \Lambda^{\uc} f_{\uc B}{}^A + \omega^{\ub} \xi^C f_{C \ub}{}^A + E^B \xi^C T_{CB}{}^A
\,,\\
\d_{\mathcal{G}} \omega^{\ua} &= d\Lambda^{\ua} + \omega^{\ub} \Lambda^{\uc} f_{\uc \ub}{}^{\ua} + \omega^{\ub} \xi^{C} F_{C\ub}{}^{\ua} + E^B \Lambda^{\uc} F_{\uc B}{}^{\ua} + E^B \xi^C R_{CB}{}^{\ua}
\,.
\end{align}
\end{subequations}

We have finished the gauging. We now briefly list, summarise, and give some useful formulae in our conventions. The covariant derivatives expand as
\begin{align}
\nabla_A = E_A 
- \frac{1}{2} \Omega_A{}^{cd} M_{cd} 
- \Phi_A{}^{ij} J_{ij}
- B_A \bbD
- \mathfrak{F}_A{}_\g^k S_k^\g
- \mathfrak{F}_A{}^c K_c
\,.
\end{align}
Let $F_{\ua B}{}^{\uc} = f_{\ua B}{}^{\uc} + \boldsymbol{F}_{\ua B}{}^{\uc}$ be the general expansion with $f_{\ua B}{}^{\uc}$ structure constants of the superconformal algebra and $\boldsymbol{F}_{\ua B}{}^{\uc}$ some $\mathcal{H}$-covariant superfields. We describe this expansion and the constraints it imposes on $\boldsymbol{F}_{\ua B}{}^{\uc}$, in the paragraph below \eqref{structurefunctiondecomp}. The commutators of the gauged algebra are
\begin{subequations}
\begin{align}
[X_{\underline{a}} , X_{\underline{b}}] &= -f_{\underline{a} \underline{b}}{}^{\underline{c}} X_{\underline{c}} 
\,,\\
[X_{\underline{a}} , \nabla_{{B}} ] &=
- f_{\underline{a} {{B}}}{}^{{C}} \nabla_{{C}}
 -f_{\underline{a} {{B}}}{}^{\underline{c}} X_{\underline{c}}
 -\boldsymbol{F}_{\underline{a} {{B}}}{}^{\underline{c}} X_{\underline{c}}
\,,\\
\label{gaugedalgebra3}
[\nabla_A,\nabla_B] &= - T_{AB}{}^C \nabla_C 
- \frac{1}{2} R(M)_{AB}{}^{cd} M_{cd} 
- R(J)_{AB}{}^{ij} J_{ij}
\nonumber\\&\quad
- R(D)_{AB} \bbD
- R(S)_{AB}{}_\g^k S_k^\g
- R(K)_{AB}^c K_c
\,,
\end{align}
\end{subequations}
where $f_{\ua B}{}^{\uc}$, $f_{\ua B}{}^C$, $f_{\ua\ub}{}^{\uc}$ are structure constants inherited from the superconformal algebra.\footnote{In the 6D case, but also in 4D \cite{Butter:2019edc}, maximal conformal supersymmetry inevitably forces that $F_{\underline{a} {{B}}}{}^{\underline{c}} = f_{\underline{a} {{B}}}{}^{\underline{c}} + \boldsymbol{F}_{\ua B}{}^{\uc}$ where $\boldsymbol{F}_{\ua B}{}^{\uc} \neq 0$ represents non-zero superfield corrections. Whereas, in the $\mathrm{D}>3$ non-maximal supersymmetry cases, there seems to always be a change of vector connections $\nabla_a \to \hat{\nabla}_{a}$, such that $[X_{\ua},\hat{\nabla}_B]$ has structure constants of the superconformal algebra $F_{\underline{a} {{B}}}{}^{\underline{c}} = f_{\underline{a} {{B}}}{}^{\underline{c}}$.}
Following Cartan's structure equations \eqref{CurvatureDef} and \eqref{TorsionDef}, we insert the structure constants of the $6D$ $\cN=(2,0)$ superconformal algebra and find that they take the form
\begin{subequations}\label{CartanStructureEqns}
\begin{align}
T^a &= d E^a + E^b \wedge \Omega_b{}^a + E^a \wedge B 
\,, \\[1ex]
T^\a_i &= d E^\a_i 
+ E^\b_i \wedge \Omega_\b{}^\a 
+ \hf E^\a_i \wedge B - E^{\a j} \wedge \Phi_{ji} 
- \ri \, E^c \wedge \mathfrak{F}_{\b i} (\tilde{\g}_c)^{\a\b} 
\,,\\[1ex]
R(D) &= d B + 2 E^a \wedge \mathfrak{F}_a + 2 E^\a_i \wedge \mathfrak{F}_\a^i 
- E^C \wedge \omega^{\ub} \boldsymbol{F}_{\ub C}
\,,\\[1ex]
R(M)^{ab} &= d \Omega^{ab} + \Omega^{ac} \wedge \Omega_c{}^b 
- 4 E^{[a} \wedge \mathfrak{F}^{b]} 
+ 2 E^\a_j \wedge \mathfrak{F}_\b^j (\g^{ab})_\a{}^\b
\\&\quad
- E^C \wedge \omega^{\ub} \boldsymbol{F}_{\ub C}{}^{ab}
\,,\\[1ex]
R(J)^{ij} &= d \Phi^{ij} - \Phi^{k (i} \wedge \Phi^{j)}{}_k - 8 E^{\a (i} \wedge \mathfrak{F}_{\a}^{j)}
- E^C \wedge \omega^{\ub} \boldsymbol{F}_{\ub C}{}^{ij}
\,,\\[1ex]
R(K)^a &= d \mathfrak{F}^a 
+ \mathfrak{F}^b \wedge \Omega_b{}^a 
- \mathfrak{F}^a \wedge B 
- \ri \mathfrak{F}_\a^k \wedge \mathfrak{F}_{\b k} (\tilde{\g}^a)^{\a\b} 
- E^C \wedge \omega^{\ub} \boldsymbol{F}_{\ub C}{}^{a}
\,,\\[1ex]
R(S)_\a^i &= d \mathfrak{F}_\a^i 
- \mathfrak{F}_\b^i \wedge \Omega_\a{}^\b
- \hf \mathfrak{F}_\a^i \wedge B 
- \mathfrak{F}_\a^j \wedge \Phi_j{}^i 
- \ri E^{\b i} \wedge \mathfrak{F}^c (\g_c)_{\a\b}
\\&\quad
- E^C \wedge \omega^{\ub} \boldsymbol{F}_{\ub C}{}_\a^i
\,.
\end{align}
\end{subequations}
In general, any $M$-valued field comes with a factor of a half in our conventions, e.g.\
\begin{equation}
    T = \frac{1}{2} T^{ab} M_{ab}
    \,.
\end{equation}
It is good to recall that the structure function contributions from $\boldsymbol{F}_{\ub C}{}^{\ua}$ solely originate from the commutator $[K^B,\nabla_C]$ and not $[X_{\ub},\nabla_C]$ with $X_{\ub} = M_{ab},J^{ij},\bbD$. Hence, once we move on to solving our 6D $\cN=(2,0)$ gauged conformal algebra, it will be obvious which parts of $\boldsymbol{F}_{\ub C}{}^{\ua}$ are non-zero and contribute to the Cartan structure equations above. We find that contributions only arise for $R(K)^a$ and $R(S)_\a^i$ and that they come from $\{S_i^\a,\nabla_\b^j\}$ and $[S^\a_i,\nabla_b]$ as seen below in \eqref{StructureFunctionsSect2}.

We note that a \emph{superconformal primary} field $\phi$ of dimension $\Delta \in \bbR$ satisfies
\begin{align}
    S_i^\a\phi = 0
    \,,\qquad
    \bbD\phi = \Delta\phi
    \,,
\end{align}
where $\phi$ can be vector-valued, so that it may have any indices. Moreover, it is important to note that the structure functions can never affect $\{S_i^\a,S_j^\b\} = -2\ri \O_{ij} (\tilde{\gamma}^c)^{\a\b} K_{c}$. Thus, $S_i^\a \phi = 0$ implies $K_a\phi = 0$. 
A field $\phi$ annihilated by $K_{a}$, but not necessarily annihilated by the generator $S_i^\a$, along with $\bbD\phi = \Delta\phi$ is called a \emph{conformal primary} of dimension $\Delta$.

\subsection{\texorpdfstring{6D $\cN=(2,0)$ conformal superspace}{6D N=(2,0) conformal superspace}} \label{sect:6DN20conformalsuperspace}
In this section, we give the results, but postpone much of the logic and derivation until Section~\ref{sect:DerivationOf6DConformalSuperspace}.

The off-shell component description of 6D $\cN=(2,0)$ conformal supergravity, through the so-called standard Weyl multiplet, admits the independent fields \cite{Bergshoeff:1999}
\begin{equation}
e_m{}^a\,,\quad
\psi_m{}^\a_i\,,\quad
b_m\,,\quad
V_m{}^{ij}\,,\quad
T_{abc}{}^{ij}\,,\quad
\chi^{\a i,jk}\,,\quad
D^{ij,kl}
\,.
\end{equation}
The first four are gauge fields associated with pullbacks of the frame $E^A = (E^a,E_i^\a)$ and the $\bbD,J$-connections $B$, $\Phi^{ij}$ to the underlying manifold $\iota : \mathcal{M}^6 \to \mathcal{M}^{6|16}$. The remaining three are matter fields, of dimensions $1$, $3/2$, and $2$, respectively,
\begin{align}
\bbD T_{abc}{}^{ij} = T_{abc}{}^{ij}\,,\quad
\bbD \chi^{\a i,jk} = \frac{3}{2} \chi^{\a i,jk}\,,\quad
\bbD D^{ij,kl} = 2 D^{ij,kl}
\,.
\end{align}
The field $T_{abc}{}^{ij}$ is a superconformal primary $S_k^\g T_{abc}{}^{ij} = 0$ and the others are descendants. The key idea of building 6D $\cN=(2,0)$ conformal superspace comes from embedding a superfield $W_{abc}{}^{ij}$, who pulls back to $T_{abc}{}^{ij}$, into the curvatures and torsions of the gauged superconformal algebra. One then requires that \emph{all} curvatures and torsions are built from $W_{abc}{}^{ij}$ and its descendants. The superfield $W_{abc}{}^{ij}$ is called the \emph{super-Weyl tensor}. 
Similar procedures have been done in various dimensions \cite{Butter:2013goa,Butter:2009cp,Butter:2011sr,Kuzenko:2023qkg,Butter:2019edc,Butter:2014xxa,Butter:2016,Kuzenko:2022qnb}, but the 6D $\cN=(2,0)$ case is particularly hard due to its gauged algebra differing from super Yang--Mills. The representation theory of $\Spin(1,5) \times \USp(4)$ also makes things difficult. 
We will give more thorough details about our derivation in the next section. For now, we just state the useful results.

First, note that $T_{abc}{}^{ij}=T_{[abc]}{}^{[ij]}$ is an $SO(1,5)$ anti-self-dual 3-form and a $\USp(4)$-traceless 2-form. Anti-self-duality in our index notations is given by
\begin{align}
    -T_{abc}{}^{ij} = (*T)_{abc}{}^{ij} = \frac{1}{3!} \ve_{abcdef} T^{def\,ij}
    \,.
\end{align}
We require that the superfield $W_{abc}{}^{ij}$ is the same. Our curvatures and torsions for two spinor derivatives\footnote{All of these results are in-frame in the sense that $\nabla_A = (\nabla_a,\nabla_\a^i) = E_A{}^M \nabla_M$.} are
\begin{align}\label{QQAnsatzSect2} 
\{\nabla_\a^i,\nabla_\b^j\} &= -2\ri \O^{ij} \nabla_{\a\b} - W_{\a\b}{}^{ij} - G_{\a\b}{}^{ij}
\nonumber\\
&= -2\ri \O^{ij} (\g^{a})_{\a\b} \nabla_{a} - (\g^{a})_{\a\b} W_{a}{}^{ij} - \frac{1}{6} (\g^{abc})_{\a\b} G_{abc}{}^{ij}
\,,
\end{align}
where $W_{\a\b}{}^{ij}=W_{[\a\b]}{}^{[ij]}$ is antisymmetric in both pairs of indices and $\USp(4)$-traceless $W_{\a\b i}{}^i~=~0$, and $G_{\a\b}{}^{ij}=G_{(\a\b)}{}^{(ij)}$ is symmetric in both pairs of indices. Leveraging the Bianchi identities, one can show that the remaining commutators of covariant derivatives are
\begin{subequations}\label{QP&PPCurvatureDef}
\begin{align}
\label{QPCurvatureDef}
[\nabla_{\a}^i, \nabla_b] &=
\frac{1}{10}\ri (\g_b)^{\b\g} \left( 
[\nabla_{\b k}, W_{\a\g}{}^{ik}]
-\frac{1}{4}  [\nabla_{\a k}, W_{\b\g}{}^{ik}]
+ \frac{1}{2}  [\nabla_{\b k}, G_{\a\g}{}^{ik}]
\right)
\,,\\
\label{PPCurvatureDef}
[\nabla_a, \nabla_b] &=
-\frac{1}{512}(\g_{ab})_{\a}{}^{\b} \ve^{\a\d\e\g} \left(
[\nabla_{\b j},[\nabla_{\d l},W_{\e\g}{}^{jl}]]
+\frac{3}{5} 
[\nabla_{\d j},[\nabla_{\b l},W_{\e\g}{}^{jl}]]
\right.
\nonumber\\&\quad
\left.
-\frac{2}{5}  
[\nabla_{\d j},[\nabla_{\e l},W_{\b\g}{}^{jl}]]
+\frac{4}{5} 
[\nabla_{\d j},[\nabla_{\e l},G_{\b\g}{}^{jl}]]
\right)
\,.
\end{align}
\end{subequations}
The curvature operators expand as
\begin{subequations}
\begin{align}
W_{a}{}^{ij} &= 
\frac{1}{2} W_{abc}{}^{ij} M^{bc}
+ \WS_{a, \g}{}^{i j, k} S_{\g}^k
+ \WK_{a,d}{}^{ij} K^d
\,,\\
G_{abc}{}^{ij} &=
\GK_{abc,d}{}^{ij} K^d
\,,\\
%Spinor versions
W_{\a\b}{}^{ij} &= 
\frac{1}{2} \ve_{\a\b\g\e} W^{\g\d ij} M_{\d}{}^{\e}
+ \WS_{\a\b,\g}{}^{i j, k} S_{\g}^k
+ \frac{1}{4} \WK_{\a\b,\g\d}{}^{ij} K^{\g\d}
\,,\\
G_{\a\b}{}^{ij} &=
\frac{1}{4}\GK_{\a\b,\g\d}{}^{ij} K^{\g\d}
\,,
\end{align}
\end{subequations}
where $W^{\a\b ij} = 1/6\, (\tilde{\g}^{abc})^{\a\b} W_{abc}{}^{ij}$ is symmetric in its spinor indices. See Appendix~\ref{appendix:gammamatrixandsymplecticalgebra} for formulas for converting the vector $\bbR^6$ indices to the $\bbC^4$ spinor indices and back. The Bianchi identities enforce that
\begin{subequations}
\begin{align}
\WS_{\a\b,\g}{}^{ij,k} &= 
\frac{1}{40} \ve_{\a\b\g\q} X^{\q k,ij}
\,,\quad
\WK_{\a\b,\g\d}{}^{ij} =  
\frac{3}{20}\ve_{\a\b\q[\g} \nabla_{\d]\z} W^{\q\z ij}
\,,\\
\GK_{\a\b,\g\d}{}^{ij} &=
\frac{1}{240} \ri \ve_{\g\d\q(\a} Y_{\b)}{}^{\q ij}
\,,
\end{align}
\end{subequations}
where we have introduced some of the descendant superfields of the super-Weyl tensor $W^{\a\b ij}$. The full set of independent descendant superfields essentially forms the standard Weyl multiplet\footnote{This is very close to the standard Weyl multiplet, as each covariant field below is essentially the superspace version of the component curvatures or a covariant matter field. Hence, one just needs to add in the gauge fields. The title of Section~\ref{sect:constructionofstandardweylmultiplet} references things in this manner.} and are all annihilated by $K_{a}$. They are
\begin{subequations}\label{XYDefs}
\begin{align}
X_\a{}^{\b\g i} &= \nabla_\a^j W^{\b\g i}{}_j - \frac{2}{5} \d_\a^{(\b} \nabla^j_\q W^{\g)\q i}{}_j
\,,\\
X^{\a i,jk} &= \nabla_\b^i W^{\a \b jk}  - \frac{1}{5} \O^{jk} \nabla_\b^l W^{\a \b i}{}_l - \frac{4}{5} \O^{i[j} \nabla_\b^l W^{\a \b k]}{}_l
\,,\\
Y_{\a\b}{}^{\g\d} &= \nabla_{(\a}^{k} X_{\b)}{}^{\g\d}{}_k - \frac{1}{3} \d_{(\a}^{(\g} \nabla_\q^k X_{\b)}{}^{\d)\q}{}_k
\,,\\
Y_{\a}{}^{\b ij} &= \nabla_\a^k X^{\b (i,j)}{}_k - \frac{1}{4} \d_\a^\b \nabla_\q^k X^{\q (i,j)}{}_k  = - \nabla_\q^{(i} X_{\a}{}^{\q \b j)}
\,,\\
Y^{ij,kl} &= \frac{1}{2} \nabla_\q^{[i} X^{\q j], kl} + \frac{1}{2} \nabla_\q^{[k} X^{\q l], ij}
+ \frac{1}{12} \left(
\O^{ij} \nabla_\q^{i_1} X^\q{}_{i_1}{}^{kl}
+ \O^{kl}  \nabla_\q^{i_1} X^\q{}_{i_1}{}^{ij}
\right.\nonumber\\&\quad\left.
+ \O^{l[i}  \nabla_\q^{i_1} X^\q{}_{i_1}{}^{j]k}
- \O^{k[i}  \nabla_\q^{i_1} X^\q{}_{i_1}{}^{j]l}
\right)
\,.
\end{align}
\end{subequations}
Like $W^{\a\b ij}$, all of its descendant superfields live in $\Spin(1,5) \times \USp(4)$ irreps --- the combinations not present in the projections \eqref{XYDefs} are zero or expressible as $W^{\a\b ij} W^{\g \d kl}$ or $\nabla_a W^{\a\b ij}$ due to Bianchi identities.
The expansions above are further simplified by the Bianchi identities, they turn off all the traces that we minus away. Hence, we get
\begin{subequations}\label{XYDefsPostBI}
\begin{align}
X_\a{}^{\b\g i} &= \nabla_\a^j W^{\b\g i}{}_j
\,,\qquad
X^{\a i,jk} = \nabla_\b^i W^{\a \b jk} 
\,,\\
Y_{\a\b}{}^{\g\d} &= \nabla_{(\a}^{k} X_{\b)}{}^{\g\d}{}_k
\,,\\
Y_{\a}{}^{\b ij} &= \nabla_\a^k X^{\b (i,j)}{}_k = - \nabla_\q^{(i} X_{\a}{}^{\q \b j)}
\,,\\
Y^{ij,kl} &= \frac{1}{2} \nabla_\q^{[i} X^{\q j], kl} + \frac{1}{2} \nabla_\q^{[k} X^{\q l], ij}
\,.
\end{align}
\end{subequations}
It can be helpful to complexify and think in terms of $\SL(4,\bbC) \times \Sp(4,\bbC)$. In particular, the Weyl spinor indices $\a,\b$, etc.\ are then just $\bbC^4$ indices of $\SL(4,\bbC)$.
The symmetries of the indices are: $X_\a{}^{\b\g i}=X_\a{}^{(\b\g) i}$ is symmetric in $\b,\g$ and $\SL(4,\bbC)$-traceless; $X^{\a i,jk}=X^{\a i,[jk]}$ is antisymmetric in $j,k$, vanishes when antisymmetrised over $i,j,k$ and is $\Sp(4,\bbC)$-traceless; $Y^{ij,kl}=Y^{[ij],[kl]}$ is antisymmetric in $i,j$ and in $k,l$, vanishes when antisymmetrised over any 3 indices and is $\Sp(4,\bbC)$-traceless; $Y_{\a\b}{}^{\g\d}=Y_{(\a\b)}{}^{(\g\d)}$ is symmetric in $\a,\b$ and $\g,\d$ and is $\SL(4,\bbC)$-traceless; $Y_{\a}{}^{\b ij}$ is $\SL(4,\bbC)$-traceless and symmetric in $i,j$. If one restores reality conditions, these index symmetries remain the same and we find
\begin{subequations}
    \begin{alignat}{3}
    \overline{W^{\a\b ij}} &= - W^{\a\b}{}_{ij}
    \,,&\qquad
    \overline{X^{\a i,jk}} &= X^{\a}{}_{i,jk}
    \,,&\qquad
    \overline{X_{\a}{}^{\b\g i}} &= X_{\a}{}^{\b\g}{}_i
    \,,\\
    \overline{Y_{\a\b}{}^{\g\d}} &= Y_{\a\b}{}^{\g\d}
    \,,&\qquad
    \overline{Y_{\a}{}^{\b ij}} &= Y_{\a}{}^{\b}{}_{ij}
    \,,&\qquad
    \overline{Y^{ij,kl}} &= Y_{ij,kl}
    \,.
\end{alignat}
\end{subequations}
More detail on reality conditions is given in Appendix~\ref{appendix:gammamatrixandsymplecticalgebra}.

The Bianchi identities constrain the $Q$-actions on the multiplet to be
\begin{subequations}\label{QstandardWeylMultiplet-sect2}
\begin{align}
\nabla_\a^i W^{\b\g jk} &= 
\frac{1}{5} \O^{jk} X_\a{}^{\b\g i} 
+ \frac{4}{5} \O^{i[j} X_\a{}^{\b\g k]}
+ \frac{2}{5} \d_\a^{(\b} X^{\g) i,jk}
\,,\\[1ex]
\nabla_\a^i X^{\b j,kl}
&=
\frac{1}{4} \d_\a^\b Y^{ij,kl}
+ \frac{1}{6} \O^{kl} Y_\a{}^{\b ij}
+ \frac{5}{6} \O^{i[k} Y_\a{}^{\b l] j}
- \frac{1}{6} \O^{j[k} Y_\a{}^{\b l] i}
\nonumber\\
&\quad
-  4\ri\O^{i j} \nabla_{\a\q} W^{\q\b kl}
+ 2\ri\O^{i [k} \nabla_{\a\q} W^{\q\b l]j}
+ 2\ri\O^{j [k} \nabla_{\a\q} W^{\q\b l]i}
\,,\\[1ex]
\nabla_\a^i X_\b{}^{\g\d j}
&=
- \frac{1}{4} \O^{ij} Y_{\a\b}{}^{\g\d} 
- \frac{5}{12} \d_\a^{(\g} Y_\b{}^{\d) ij} 
+ \frac{1}{12} \d_\b^{(\g} Y_\a{}^{\d) ij}
+ \frac{5}{4} \ve_{\a\b\q\e} W^{\q(\g ki} W^{\d)\e j}{}_k
\nonumber\\
&\quad
+ 5\ri \nabla_{\a\b} W^{\g\d ij} 
+ 5\ri \d_\a^{(\g} \nabla_{\b\q} W^{\d)\q ij}
- 3\ri \d_\b^{(\g} \nabla_{\a\q} W^{\d)\q ij}
\,,\\[1ex]
\label{QYijkl}
\nabla_{\a}^{i} Y^{jk,li_1} &= 
-2\ri \O^{j k} \nabla_{\a \q}{X^{\q i, l {i_{1}}}}
-2\ri \O^{l {i_{1}}} \nabla_{\a \q}{X^{\q i, j k}}
-8\ri \O^{i [j} \nabla_{\a \q}{X^{\q k], l i_1}}
\nonumber\\
&\quad
-8\ri \O^{i [l} \nabla_{\a \q}{X^{\q {i_{1}}], j k}}
\,,\\[1ex]
\nabla_\a^i Y_{\b\g}{}^{\d\e} &=
-8\ri \nabla_{\a (\b}{X_{\g)}{}^{\d \e i}}
+16\ri \d_{\a}^{(\d} \nabla_{\q (\b}{X_{\g)}{}^{\e) \q i}}
+\frac{8}{3}\ri \d_{(\b}^{(\d} \nabla_{\g) \q}{X_{\a}{}^{\e) \q i}}
+\frac{16}{3}\ri \d_{(\b}^{(\d} \nabla_{|\a \q|}{X_{\g)}{}^{\e) \q i}}
\nonumber\\
&\quad
+4 \ve_{\a (\b |\q \z|} W^{\q (\d k i} X_{\g)}{}^{\e) \z}{}_k
+\frac{2}{3}\d_{(\b}^{(\d} \ve_{\g) \a \q \z} W^{|\q \eta| k i} X_{\eta}{}^{\e) \z }{}_k
\,,\\[1ex]
\nabla_\a^i Y_\b{}^{\g jk} &=
\frac{16}{5}\ri \O^{i (j} \nabla_{\b \d}{X_{\a}{}^{\g \d k)}}
+\frac{32}{5}\ri \O^{i (j} \nabla_{\a \d}{X_{\b}{}^{\g \d k)}}
-\frac{24}{5}\ri \nabla_{\a \b}{X^{\g (j,k) i}}
\nonumber\\
&\quad
-\frac{24}{5}\ri \d_{\a}^{\g} \nabla_{\b \d}{X^{\d (j,k) i}}
+ \frac{12}{5}\ri \d_{\b}^{\g} \nabla_{\a \d}{X^{\d (j,k) i}}  
+ \frac{4}{5} \ve_{\a \b \d \e} W^{\g \d l i} X^{\e (j,k)}{}_l
\nonumber\\
&\quad
+\frac{2}{5} \ve_{\a \b \d \e} W^{\g \d l (j} X^{\e}{}^{k), i}{}_l  
+\frac{1}{5} \ve_{\a \b \d \e} \O^{i (j} W^{\g \d}{}_{l i_1} X^{\e k),l i_1}
\nonumber\\
&\quad
+\frac{4}{5} \ve_{\a \b \d \e} \O^{i (j} W^{\d \z k) l} X_{\z}{}^{\g \e}{}_l
\,.
\end{align}
\end{subequations}
The $S$-actions are constrained to be
\begin{subequations}\label{SstandardWeylMultiplet-sect2}
\begin{align}
S^{l \d} X^{\a i, jk}
&=
16\O^{i l} W^{\a \d j k}
+8\O^{l[j} W^{\a \d k]i}
+8\O^{i [j}  W^{\a \d k] l}
\,,\\[1ex]
S^{l \d} X_{\a}{}^{\b \g i}
&=
20\d_{\a}^{\d} W^{\b \g i l}
-8\d_{\a}^{(\b} W^{\g)\d i l}
\,,\\[1ex]
S^{l \d} Y^{i j, k i_1}
&=
- 4 \O^{i j} X^{\d l k i_1}
- 4 \O^{k i_1} X^{\d l i j}
- 16 \O^{l [i} X^{\d j] k i_1} 
- 16 \O^{l [k} X^{\d i_1] i j} 
\,,\\[1ex]
S^{l \d} Y_{\a\b}{}^{\g\e}
&=
48\d_{(\a}^{\d} X_{\b)}{}^{\g \e l}
-16\d_{(\a}^{(\g} X_{\b)}{}^{\e) \d l}
\,,\\[1ex]
S^{l \d} Y_{\a}{}^{\b ij}
&=
\frac{96}{5} \O^{l (i} X_{\a}{}^{\b \d j)}
+\frac{96}{5}\d_{\a}^{\d} X^{\b (i,j) l} 
- \frac{24}{5}\d_{\a}^{\b} X^{\d (i,j) l}
\,.
\end{align}
\end{subequations}
There are also structure functions, but only in the following commutators
\begin{subequations}\label{StructureFunctionsSect2}
\begin{align}
\{S_i^\a,\nabla_\b^j\} &= 
2\d_\b^\a \d_i^j \bbD 
- 4 \d_i^j M_\b{}^\a 
+ 8 \d_\b^\a J_i{}^j 
+ \frac{1}{10} \ri (\g_{c})_{\b \g} W^{\a \g}{}_i{}^j K^{c}
\,,\\
 [S^{i\a}, \nabla_{b}] &= 
 -\ri (\tilde{\g}_{b})^{\a\b} \nabla_{\b}^{i} 
+ \frac{1}{40} \ri (\g_b)_{\b \g} W^{\a \b i j} S_{j}^{\g}
+ \frac{1}{160} (\g_{b c})_{\b}{}^{\g} X_{\g}{}^{\a \b i} K^c
\,.
\end{align}
\end{subequations}
The rest are structure constants matching with the superconformal algebra; aside from the commutators $[\nabla_A,\nabla_B]$ containing curvatures and torsions.

It is worth mentioning that the structures above are completely fixed by the self-consistency of the algebra and its Bianchi identities --- in the sense that there is no freedom left, so further constraints cannot be applied to keep the geometry representing the $(2,0)$ off-shell standard Weyl multiplet. It is also worth mentioning that this is true even without solving the entire collection of Bianchi identities. We give a more detailed analysis of this in Section~\ref{sect:partialsolutionofbianchiidentities}, but quickly mention some key ideas here.

Usually the analysis of the Bianchi identities is done by showing that, given the $\{\nabla_\a^i,\nabla_\b^j\}$ ansatz \eqref{QQAnsatzSect2}, one obtains the expansions $[\nabla_a,\nabla_\b^j]$, $[\nabla_a,\nabla_b]$ in \eqref{QP&PPCurvatureDef} and some constraints on the ansatzed curvatures by imposing the two Bianchi identities
\begin{subequations}\label{QQQ&PQQBISect2}
\begin{align}
[\nabla_\a^i,\{\nabla_\b^j,\nabla_\g^k\}]
+ [\nabla_\g^k,\{\nabla_\a^i,\nabla_\b^j\}]
+ [\nabla_\b^j,\{\nabla_\g^k,\nabla_\a^i\}]
&=0
\,,\\
[\nabla_{a},\{\nabla_\b^j,\nabla_\g^k\}]
- \{\nabla_\g^k,[\nabla_{a},\nabla_\b^j]\}
+ \{\nabla_\b^j,[\nabla_\g^k,\nabla_a]\}
&=0
\,.
\end{align}
\end{subequations}
Then, after imposing \eqref{QQQ&PQQBISect2} and the Jacobi identities, one aims to show that the remaining two Bianchi identities are satisfied as a consequence. Due to the complexity of the $(2,0)$ algebra, we have only checked that \eqref{QQQ&PQQBISect2} are completely satisfied using all the information above, but not that the dimension 5/2 and 3 are satisfied, as we expect.
Our analysis in Section~\ref{sect:DerivationOf6DConformalSuperspace} suggests that the Bianchi identities only impose constraints at dimension 3/2 and that all other constraints are implied by this. This constraint is
\begin{align}
    \nabla_\a^i W^{\b\g jk} &= 
\frac{1}{5} \O^{jk} X_\a{}^{\b\g i} 
+ \frac{4}{5} \O^{i[j} X_\a{}^{\b\g k]}
+ \frac{2}{5} \d_\a^{(\b} X^{\g) i,jk}
\,.
\end{align}
Equivalently, we could rewrite as $\nabla_\a^i W^{\a\b j}{}_i = 0$ and $\nabla_\a^{i} W^{\b\g jk} - (\text{all traces}) = 0$.

We now list the values of the curvatures and torsions obtained throughout our analysis. 
First, some $[\nabla_a,\nabla_b]$ curvatures and torsions correspond directly to the descendants of $W_{abc}{}^{ij}$ as follows
\begin{subequations}\label{CurvaturesAndTorsions0Sect2}
\begin{align}
T_{ab}{}_k^\g = R(Q)_{ab}{}^\g_k &= \frac{1}{80} (\gamma_{ab})_{\a}{}^{\b} X_{\b}{}^{\a \g}{}_k = \frac{1}{40} X_{ab}{}^\g{}_k
\,,\\
R(J)_{ab}{}^{ij} &= - \frac{1}{192} (\gamma_{ab})_{\a}{}^{\b} Y_{\b}{}^{\a ij} = - \frac{1}{96} Y_{ab}{}^{ij}
\,,\\
R(M)_{ab}{}^{cd} &= \frac{1}{160} Y_{ab}{}^{cd}
+ \frac{1}{16} \d_{[a}^{[c} W_{b]ef}{}^{ij} W^{d]ef}{}_{ij}
\,,
\end{align}
\end{subequations}
where $Y_{ab}{}^{cd} = 1/4 \, (\g_{ab})_\a{}^\g (\g^{cd})_\b{}^\g Y_{\g\d}{}^{\a\b}$.
The remaining non-zero curvatures and torsions are given by
\begin{subequations}\label{CurvaturesAndTorsions1Sect2}
    \begin{align}
\{\nabla_\a^i,\nabla_\b^j\}
&=
-2\ri \O^{i j} \nabla_{\a \b}
- \frac{1}{8} \ve_{\a \b \g \d} (\g^{ab})_\e{}^\g W^{\d \e i j} M_{ab}
+ \frac{1}{40} \ve_{\a \b \g \d} X^{\g k, i j} S_k^\d
\nonumber\\&\quad
+ \frac{3}{40} (\g_{a})_{\g [\a} \nabla_{\b]\d}  W^{\d \g i j} K^{a}
- \frac{1}{480} \ri (\g_{a})_{\g (\a} Y_{\b)}{}^{\g ij} K^{a}
\,,\\[1ex]
[\nabla_\a^i,\nabla_b]
&=
\frac{1}{8} \ri (\g_{b})_{\a \b} W^{\b \g i j} \nabla_{\g j}
+ \frac{1}{80} \ri (\g_{b})_{\a \b} (\g^{cd})_\d{}^\g X_\g{}^{\b\d i} M_{cd}
- \frac{1}{160} \ri (\g_{b}{}^{cd})_{\b\g} X_\a{}^{\b\g i} M_{cd}
\nonumber\\&\quad
- \frac{1}{20} \ri (\g_{b})_{\a \b} X^{\b k, l i} J_{kl}
- R(S)_{\a b}^i{}^k_\g S^\g_k 
- R(K)_{\a b c}^i K^c 
\,,\\[1ex]
[\nabla_a,\nabla_b]
&=
- R(Q)_{ab}{}_k^\g \nabla_\g^k
- \frac{1}{2} R(M)_{ab}{}^{cd} M_{cd}
- R(J)_{ab}{}^{ij} J_{ij}
\nonumber\\&\quad
- R(S)_{ab}{}_\g^k S_\g^k
- R(K)_{abc} K^c
\,.
\end{align}
\end{subequations}
The expansions of the rest of the curvatures are
\begin{subequations}\label{CurvaturesAndTorsions2Sect2}
\begin{align}
R(S)_{\a b}^i{}^k_\g 
&=
\frac{1}{32} (\g_{b})_{\a \b} \nabla_{\g \d} W^{\b \d i k}
+ \frac{1}{160} (\g_{b})_{\b \g} \nabla_{\a \d} W^{\b \d i k}
- \frac{1}{384} \ri (\g_{b})_{\a \b} Y_{\g}{}^{\b i k}
\nonumber\\&\quad
- \frac{1}{640} \ri (\g_{b})_{\b \g} Y_{\a}{}^{\b i k}
\,,\\[1ex]
R(K)_{\a b c}^i
&=  
 - \frac{1}{960}\mathrm{i} (\gamma_{b c}{}^d)_{\beta \gamma} \nabla_{d}{X_{\alpha}{}^{\beta \gamma i}}
 +\frac{1}{240}\mathrm{i} (\gamma_{b c})_{\beta}{}^{\delta} \nabla_{\alpha \gamma}{X_{\delta}{}^{\beta \gamma i}} 
\nonumber\\&\quad
- \frac{1}{240}\mathrm{i} (\gamma_{b}{}^d)_{\beta}{}^{\delta} (\gamma_{c})_{\alpha \gamma} \nabla_{d}{X_{\delta}{}^{\beta \gamma i}} 
 - \frac{1}{768}(\gamma_{b})_{\alpha \beta} (\gamma_{c})_{\delta \epsilon} W^{\delta \gamma i j} X_{\gamma}{}^{\beta \epsilon}{}_{j} 
  \nonumber\\&\quad
 - \frac{1}{1280}(\gamma_{b})_{\beta \delta} (\gamma_{c})_{\alpha \epsilon} W^{\beta \gamma i j} X_{\gamma}{}^{\delta \epsilon}{}_{j} 
  - \frac{1}{2560}(\gamma_{b})_{\alpha \beta} (\gamma_{c})_{\delta \gamma} W^{\beta \delta j k} X^{\gamma i}{}_{j k} 
  \nonumber\\&\quad
  - \frac{1}{2560}(\gamma_{b})_{\beta \delta} (\gamma_{c})_{\alpha \gamma} W^{\beta \gamma j k} X^{\delta i}{}_{j k}
 \,,\\[1ex]
R(S)_{ab}{}_\g{}^k
&=
 - \frac{1}{480}\mathrm{i} (\gamma_{a b}{}^{c})_{\alpha \beta} \nabla_{c}{X_{\gamma}{}^{\alpha \beta k}} 
 - \frac{1}{240}\mathrm{i} (\gamma_{a b})_{\alpha}{}^{\beta} \nabla_{\g \d}{X_{\beta}{}^{\alpha \delta k}}
   \nonumber\\&\quad
 - \frac{1}{960}(\gamma_{[a})_{\alpha \beta} (\gamma_{b]})_{\gamma \delta} W^{\alpha \epsilon k i} X_{\epsilon}{}^{\beta \delta}{}_{i}
 \,,\\[1ex]
R(K)_{a b c}
 &=
\frac{1}{3840}(\gamma_{a b})_{\alpha}{}^{\beta} (\gamma_{c}{}^d)_{\delta}{}^{\gamma} \nabla_{d}{Y_{\beta \gamma}{}^{\alpha \delta}}
+\frac{1}{19200}\mathrm{i} (\gamma_{a b c})_{\alpha \beta} X_{\delta}{}^{\alpha \gamma i} X_{\gamma}{}^{\beta \delta}{}_{i} 
\nonumber\\&\quad
- \frac{1}{512}(\gamma_{a})_{\alpha \beta} (\gamma_{c})_{\delta \gamma} W^{\alpha \delta i j} \nabla_{b}{W^{\beta \gamma}{}_{i j}}
+\frac{1}{512}(\gamma_{b})_{\alpha \beta} (\gamma_{c})_{\delta \gamma} W^{\alpha \delta i j} \nabla_{a}{W^{\beta \gamma}{}_{i j}}
\nonumber\\&\quad
+\frac{1}{6400}\mathrm{i} (\gamma_{a b})_{\alpha}{}^{\beta} (\gamma_{c})_{\delta \epsilon} X_{\beta}{}^{\delta \gamma i} X_{\gamma}{}^{\alpha \epsilon}{}_{i}
\,.
\end{align}
\end{subequations}

\section{\texorpdfstring{Derivation of 6D $\cN=(2,0)$ conformal superspace}{Derivation of 6D N=(2,0) conformal superspace}}\label{sect:DerivationOf6DConformalSuperspace}
As mentioned in the last section, the key to building 6D $\cN=(2,0)$ conformal superspace comes from embedding a primary superfield $W_{abc}{}^{ij}$, describing the super-Weyl tensor, into the curvatures and torsions of the gauged superconformal algebra. One then requires that \emph{all} curvatures and torsions are built from $W_{abc}{}^{ij}$ and its descendants, defined by multiple action of spinor covariant derivatives. Our initial embedding of this superfield comes from the component calculations in \cite{Bergshoeff:1999} equation (3.5). In our conventions it reads\footnote{The conversion of conventions is discussed in Section~\ref{sect:component-reduction} and Appendix~\ref{appendix:conversionofconventions}, with Table~\ref{table:conversionofconventions} giving a summary.}
\begin{align}\label{RM_arXiv:9904085v2}
R(M)_{mn}{}^{ab} &=
2 \partial_{[m} \omega_{n]}{}^{ab}
-2\omega_{[m}{}^{ae} \omega_{n]e}{}^b
+ 8 e_{[m}{}^{[a} \mathfrak{f}_{n]}{}^{b]}
- \psi_{[m j} \g^{ab} \phi_{n]}{}^j
\nonumber\\&\quad
+ 2\ri  \psi_{[m l}\g^{[a} R(Q)_{n]}{}^{b] l} 
+ \ri \psi_{[m l}\g_{n]} R(Q)^{a b l}  
- \frac{1}{2} \ri \psi_{[m k} \g_c \psi_{n] l} T^{abckl}
\,.
\end{align}
Reduction to components from our superspace\footnote{We refer the reader to \cite{Butter:2017} for an in-depth description of how to obtain the structure of the 6D $\mathcal{N}=(1,0)$ standard Weyl multiplet in components from the conformal superspace construction of \cite{Butter:2016}  ---  see also \cite{Kuzenko:2022skv,Kuzenko:2022ajd} for recent reviews and for more references. The derivations in the 6D $\mathcal{N}=(2,0)$ case are completely analogous. We will give details useful for our paper in Section~\ref{sect:component-reduction}.} gives the corresponding result
\begin{align}\label{RM_superspace}
R(M)_{mn}{}^{ab} &=
2 \partial_{[m} \omega_{n]}{}^{ab}
-2\omega_{[m}{}^{ae} \omega_{n]e}{}^b
+ 8 e_{[m}{}^{[a} \mathfrak{f}_{n]}{}^{b]}
- \psi_{[m j} \g^{ab} \phi_{n]}{}^j
\nonumber\\&\quad
+ e_{[m}{}^c \psi_{n] j}{}^\b R(M)_{\b c}^j{}^{ab}|
+ \frac{1}{4} \psi_{[m k}{}^{\g} \psi_{n] j}{}^\b R(M)_\b^j{}_\g^k{}^{ab} |
\,,
\end{align}
where the bar denotes the pullback through $\iota : \mathcal{M}^6 \to \mathcal{M}^{6|16}$, i.e.\ setting the fermionic coordinates to zero. We are led to conclude that the superspace curvature $\{\nabla_\a^i,\nabla_\b^j\}$ must be of the form
\begin{align}\label{QQinitialansatz}
\{\nabla_\a^i,\nabla_\b^j\} = - 2 \ri \O^{i j} (\g^{a})_{\a\b} \nabla_{a} - \frac{1}{2} (\g^{a})_{\a\b} W_{a b c}{}^{i j} M^{b c} + \dots 
\end{align}
and we will have $T_{abc}{}^{ij} = \frac{1}{2} \ri W_{a b c}{}^{i j} |$. We mention that reality condition-wise these two differ by a sign due to the factor of $\ri$
\begin{equation}
    \overline{W_{abc}{}^{ij}} = -W_{abc \, ij}
    \,,\qquad
    \overline{T_{abc}{}^{ij}} = T_{abc \, ij}
    \,.
\end{equation}

The constant torsion term is just inherited from the superconformal algebra. Recall that the curvatures, torsions and structure functions have well-defined dilatation dimensions by our first principles construction of conformal superspace.
This means no other torsion terms in $\{\nabla_\a^i,\nabla_\b^j\}$ are possible because of our assumption that the dimension 1 primary $W_{a b c}{}^{i j}$ ($K^D W_{a b c}{}^{i j}=0$) and its descendants $\nabla_{\a}^{i}W_{a b c}{}^{kl}, \nabla_\a^i\nabla_\b^j W_{abc}{}^{kl}$ (of dimension $3/2,2$) generate all non-constant curvatures and torsions. 
Such torsions would be of dimension 0 and 1/2. However, we should expect the possibility of structure functions because $[K^A,\nabla_B]$ has parts with dimension greater than or equal to 1.

\subsection{Ansatzes through representation theory}\label{sect:ansatzesthroughrepresentationtheory}
The next step is to determine which other curvatures can exist in $\{\nabla_\a^i,\nabla_\b^j\}$. We need to do the same for structure functions. This can be argued using representation theory to give rigorous general ansatzes in terms of $W_{a b c}{}^{i j}$ and its descendants.  We digress to give an explanation of how we implement this.

Since we are working with vector-valued superfields like $W_{abc}{}^{ij}$ and $\nabla_{\a}^{i} W_{abc}{}^{kl}$ with vector and spinor indices as well as $\USp(4)$ R-symmetry indices, we will heavily leverage the representation theory of $\SL(4,\bbC) \times \Sp(4,\bbC)$ which is the complexification of $\Spin(1,5) \times \USp(4)$.\footnote{Indeed, $\Spin(1,5) \cong \SL(2,\bbH) \cong \SU^*(4)$ is a real Lie group. It can be defined by $\SU^*(4) = \{ U \in {\rm Mat}_4(\bbC) \mid U^*J = J U \text{ and } \det(U)=1\}$ where $J$ is a matrix representing a conjugate-linear map such that $J J^*=-1$; that is, a quaternionic structure on $\bbC^4$. From this p.o.v.\ where we take a quaternionic matrix and map it to a complex one of double the size, multiplication by $J$ is viewed as multiplication by $j \in \bbH$ and a natural choice of matrix representation is given by
\begin{equation*}
    J = \begin{pmatrix}
    0&-I\\
    I&0
\end{pmatrix}
\,.
\end{equation*}
At the level of Lie algebras,  $X \in \mathfrak{su}^*(4)$ just obeys $X^*J = JX$ and $\tr(X)=0$ and complexification essentially removes this conjugation condition --- much like it does for $\mathfrak{su}(4)$. Hence, $\mathfrak{su}^*(4)\otimes \bbC$ is then just $\mathfrak{sl}(4,\bbC)$. We give some related discussion in Appendix~\ref{appendix:gammamatrixandsymplecticalgebra}.
}
It is beneficial to work with the complexification due to the following results\footnote{Algebraic closure of the field $k$ is necessary for the proposition. See the example with $G=H=U(1)$ and $k=\bbR$ given in \cite{Kowalski:2014} Example 2.7.33.}
\begin{theorem}[Weyl's theorem on complete reducibility]
Consider a semisimple Lie algebra over a field $k$ of characteristic zero. All its finite-dimensional representations over $k$ are direct sums of irreps.
\end{theorem}
\begin{proposition}\label{prop:IrrepsOfProductGroups}
Consider a group $G \times H$ and an algebraically closed field $k$. Then every finite-dimensional irrep $V$ over $k$ can be expressed uniquely as $V \cong U \otimes W$ with $U$ and $W$ irreps of $G$ and $H$ respectively.
\end{proposition}
There is also no obstruction to working with the complexification because the analysis involved in constructing our 6D $\cN=(2,0)$ conformal superspace never leverages reality conditions. We may essentially treat our fields as their complexifications, so that we are looking at complex reprensentations of complex groups --- reality conditions can be reapplied later through using our symplectic form $\O^{ij}$ to impose the real and quaternionic structures we have forgotten, as in Appendix~\ref{appendix:gammamatrixandsymplecticalgebra}. Another point to mention is that $\SL(4,\bbC)$ and $\Sp(4,\bbC)$ are simply-connected, so no distinctions need to be made between them and their Lie algebras. Hence, we can freely use both results.
The only other general result we need is Schur's lemma. The form useful for us is that there are no homomorphisms between irreps over $\bbC$ unless they are isomorphic, in which case there is a unique isomorphism up to scaling. Equivalently, for any group $G$ and $V$, $W$ complex irreps
\begin{equation}\label{SchursLemma1}
\dim \Hom_G(V,W) = \begin{cases}
1 \quad\text{if $V \cong W$}\,,\\
0 \quad\text{otherwise}\,.
\end{cases}
\end{equation}
Fix any group $G$, restriction and projection lead to the natural isomorphisms
\begin{subequations}
\begin{align}
\Hom_G(V\oplus W,U) &= \Hom_G(V,U) \oplus \Hom_G(W,U)
\,,\\
\Hom_G(U,V\oplus W) &= \Hom_G(U,V) \oplus \Hom_G(U,W)
\,.
\end{align}
\end{subequations}
Hence, we can take an arbitrary vector-valued superfield $T$ with some amount of $\SL(4,\bbC)$ and $\Sp(4,\bbC)$ indices, break into a sum of irreps and see if any of the irreps coincide with those of $W_{abc}{}^{ij}$, $\nabla_{\a}^{i} W_{abc}{}^{kl}$, $\nabla_{\a}^{i}\nabla_{\b}^{j} W_{abc}{}^{kl}$, $\dots$, etc. If they do, then we apply Schur's lemma and count dimensions \eqref{SchursLemma1} to determine how many unique ways $T$ can be built from $W_{abc}{}^{ij}$ and its descendants, then give a completely general ansatz for $T$. This method allows us to ensure the Ansatz is general, but also not over-constrained.

Practically, all that is needed now is an understanding of the irreps of $\SL(4,\bbC)$ and of $\Sp(4,\bbC)$ and their respective fusion rules. That is, a system to find the irreps of an arbitrary field $T$ with some $\SL(4,\bbC)$ and $\Sp(4,\bbC)$ indices. To do this for $\SL(4,\bbC)$, we use the well-known bijective correspondence between finite-dimensional complex irreps of $\SL(n+1,\bbC)$ and Young diagrams via Young symmetrisers. Each box in a Young diagram can be associated with the index of a tensor, and the symmetries of the tensor can be read off from the diagram. For example
\begin{align}\label{FirstYoungDiagramExample}
T_{[\a\b]} \equiv 
\begin{ytableau}
\a \\
\b
\end{ytableau}
\,,\quad
T_{(\a\b)} \equiv 
\begin{ytableau}
\a & \b
\end{ytableau}
\,,\quad
T_{\a,[\b\g]}
\equiv 
\begin{ytableau}
\b & \a \\
\g
\end{ytableau}
\,,\quad
T_{[\a\b],[\g\d]}
\equiv 
\begin{ytableau}
\a & \g \\
\b & \d
\end{ytableau}
\,,
\end{align}
where $T_{\a,\b\g}$ also obeys $T_{[\a,\b\g]} = 1/3 ( T_{\a,\b\g} + T_{\g,\a\b} + T_{\b,\g\a} )=0$ and $T_{\a\b,\g\d}$ has precisely the symmetries of a Riemann curvature tensor, i.e.\ $T_{[\a\b,\g]\d} = 0$. We are primarily using conventions where the Young symmetrisers or Schur functors associated with the diagrams are defined by symmetrising over the rows, then antisymmetrising over the columns of a diagram afterward; if the other convention is used we will state so explicitly and add a superscript ``sym'' on the diagram. This choice is arbitrary because transposing the Young diagram by reflecting it about its diagonal will generate the isomorphic irrep in the antisymmetrise, then symmetrise convention and vice-versa. 

In general, all of the multi-term symmetries or Garnir relations of a tensor are generated by antisymmetrising over a suitable collection of indices and obtaining zero. Such suitable collections are known from constructions using the symmetric group $S_k$. We discuss this relation in more detail in Appendix~\ref{appendix:reptheory-dimensions&highestweights}.
For practicality, we give a visual example which generalises to all Young diagrams. A tensor $T_{\a_1\a_2\a_3,\,\a_4\a_5\a_6,\,\a_7\a_8,\,\a_9}$ associated to the Young diagram below will vanish when antisymmetrised over all indices in the black boxes below. There are multiple choices
\begin{subequations}\label{GarnirDiagramsExample}
    \begin{alignat}{3}
&\begin{ytableau}
    \blackalpha{\a_1}&\blackalpha{\a_4}&\a_7&\a_9\\
    \blackalpha{\a_2}&\a_5&\a_8\\
    \blackalpha{\a_3}&\a_6
\end{ytableau}
\,,&\qquad
&\begin{ytableau}
    \a_1&\blackalpha{\a_4}&\a_7&\a_9\\
    \blackalpha{\a_2}&\blackalpha{\a_5}&\a_8\\
    \blackalpha{\a_3}&\a_6
\end{ytableau}
\,,&\qquad
&\begin{ytableau}
    \a_1&\blackalpha{\a_4}&\a_7&\a_9\\
    \a_2&\blackalpha{\a_5}&\a_8\\
    \blackalpha{\a_3}&\blackalpha{\a_6}
\end{ytableau}
\,,\\[3ex]
&\begin{ytableau}
    \a_1&\blackalpha{\a_4}&\blackalpha{\a_7}&\a_9\\
    \a_2&\blackalpha{\a_5}&\a_8\\
    \a_3&\blackalpha{\a_6}
\end{ytableau}
\,,&\qquad
&\begin{ytableau}
    \a_1&\a_4&\blackalpha{\a_7}&\a_9\\
    \a_2&\blackalpha{\a_5}&\blackalpha{\a_8}\\
    \a_3&\blackalpha{\a_6}
\end{ytableau}
\,,&\qquad
&\begin{ytableau}
    \a_1&\a_4&\blackalpha{\a_7}&\blackalpha{\a_9}\\
    \a_2&\a_5&\blackalpha{\a_8}\\
    \a_3&\a_6
\end{ytableau}
\,.
\end{alignat}
\end{subequations}
For example, the first and last diagrams indicate that we have Garnir relations
\begin{equation}\label{GarnirTensorsExample}
    T_{[\a_1\a_2\a_3,\,\a_4]\a_5\a_6,\,\a_7\a_8,\,\a_9} = 0
    \,,\qquad
    T_{\a_1\a_2\a_3,\,\a_4\a_5\a_6,\,[\a_7\a_8,\,\a_9]} = 0
    \,.
\end{equation}
Some other antisymmetrisations will vanish, e.g.\ $\a_1,\a_2,\a_3,\a_5$, or antisymmetrisations over five or more indices, but they are all immediate from the list above. The general construction is the same, one takes any Young diagram for $\SL(n+1,\bbC)$, and picks two adjacent columns. Then one takes the left column, starts from the bottom and selects some boxes. One then takes the right column, starts from the top and selects boxes until it meets the top selected box of the left column --- giving a sort of path from the bottom to the top of the diagram. The indices in the boxes along the path will cause the tensor to vanish upon antisymmetrisation.

All the symmetries of a tensor associated with a Young diagram are characterised by the inherent antisymmetry between indices in the same column, and the Garnir relations. The Garnir relations capture the consequences of symmetrising over the rows prior to antisymmetrising over the columns.

When the opposite convention is used i.e.\ where we antisymmetrise the rows, then symmetrise the columns, one must instead symmetrise over the black indices.

Fusion rules then come from the multiplication of Young diagrams via the Littlewood-Richardson rule. 
We give the Young symmetrisers (projectors) of the above diagrams in \eqref{FirstYoungDiagramExample} directly in index notation for reference
\begin{subequations}\label{YoungSymmetrisersExample}
    \begin{align}
    \begin{ytableau}
        \a\\
        \b
    \end{ytableau}
    \qquad&\longleftrightarrow\qquad
    \d_{[\a}^{\a_1} \d_{\b]}^{\b_1} = \frac{1}{2} \left( \d_{\a}^{\a_1} \d_{\b}^{\b_1} - \d_{\b}^{\a_1} \d_{\a}^{\b_1}\right)
    \,,\\
    \begin{ytableau}
        \a&\b
    \end{ytableau}
    \qquad&\longleftrightarrow\qquad
    \d_{(\a}^{\a_1} \d_{\b)}^{\b_1} = \frac{1}{2} \left( \d_{\a}^{\a_1} \d_{\b}^{\b_1} + \d_{\b}^{\a_1} \d_{\a}^{\b_1}\right)
    \,,\\
    \begin{ytableau}
        \a&\g\\
        \b
    \end{ytableau}
    \qquad&\longleftrightarrow\qquad
    \frac{4}{3} \d_{[\a}^{(\a_1} \d_{\b]}^{|\b_1|} \d_{\g}^{\g_1)} = \frac{4}{3} \left( \d_{[\a}^{\a_1} \d_{\b]}^{\b_1} \d_{\g}^{\g_1} \right) \left( \d_{(\a_1}^{\a_2} \d_{\g_1)}^{\g_2} \d_{\b_1}^{\b_2} \right)
    \,,\\
    \begin{ytableau}
        \a&\g\\
        \b&\d
    \end{ytableau}
    \qquad&\longleftrightarrow\qquad
     \frac{4}{3} \left( \d_{[\a}^{\a_1} \d_{\b]}^{\b_1} \d_{[\g}^{\g_1} \d_{\d]}^{\d_1} \right) \left( \d_{(\a_1}^{\a_2} \d_{\g_1)}^{\g_2} \d_{(\b_1}^{\b_2}  \d_{\d_1)}^{\d_2} \right)
     \,.
\end{align}
\end{subequations}
These projectors are to be interpreted by contraction with $T_{\a_1\b_1}$, $T_{\a_2\b_2\g_2}$ and $T_{\a_2\b_2\g_2\d_2}$ respectively. If these tensors are taken to be arbitrary, i.e.\ without any index symmetries, then they will inherit precisely those symmetries described above after projection. The scaling factor of $4/3$ in the last two projectors is a general feature, and as the Young diagrams get more complicated it will change. We give the formula for it in \eqref{scalingfactor} and an explanation of it in Appendix~\ref{appendix:reptheory-dimensions&highestweights}.

Let us give an explicit example when the opposite convention is used, i.e.\ where we antisymmetrise the rows, then symmetrise the columns. The projector for the `hook' diagram above becomes
\begin{align}
    \begin{ytableau}
        \a&\g\\
        \b
    \end{ytableau}^{\,{\rm sym}}
    \qquad&\longleftrightarrow\qquad
    \frac{4}{3} \d_{(\a}^{[\a_1} \d_{\b)}^{|\b_1|} \d_{\g}^{\g_1]} = \frac{4}{3} \left( \d_{(\a}^{\a_1} \d_{\b)}^{\b_1} \d_{\g}^{\g_1} \right) \left( \d_{[\a_1}^{\a_2} \d_{\g_1]}^{\g_2} \d_{\b_1}^{\b_2} \right)
    \,.
\end{align}

For $\Sp(4,\bbC)$, we first note the theorem\footnote{For more details, see Section 17.2, 17.3 of  Fulton and Harris \cite{Fulton:1991}.}
\begin{theorem}
All complex irreps of $\Sp(2n,\bbC)$ can be realised as restrictions of complex irreps of $\SL(2n,\bbC)$ quotiented by all traces w.r.t.\ the symplectic form. Moreover, all such restrictions quotiented by all traces are irreps.

\end{theorem}
For example, we have the following $\Sp(4,\bbC)$ irreps
\begin{align}
T_{[ij]} \equiv 
\begin{ytableau}
i \\
j
\end{ytableau}^{\,{\rm trcls}}
\,,\quad
T_{(ij)} \equiv 
\begin{ytableau}
i & j
\end{ytableau}
\,,\quad
T_{i,[jk]}
\equiv 
\begin{ytableau}
j & i \\
k
\end{ytableau}^{\,{\rm trcls}}
\,,\quad
T_{[ij],[kl]}
\equiv 
\begin{ytableau}
i & k \\
j & l
\end{ytableau}^{\,{\rm trcls}}
\,,
\end{align}
where the index symmetries remain the same as above, but all symplectic traces vanish, e.g.\ $T_{i,jk} \O^{i j} = T_{i,jk} \O^{j k} = 0$. The fusion rules for $\Sp(4,\bbC)$ can be found by using the fusion rules for $\SL(4,\bbC)$ along with \emph{virtual representations} in the sense of the Grothendieck group, followed by branching rules of $\Sp(4,\bbC) \to \SL(4,\bbC)$. Practically, virtual representations means that instead of just taking sums of representations, we may also take differences. This procedure indeed yields $\Sp(4,\bbC)$ fusion rules by enabling all traceless diagrams to be rewritten as traceful diagrams minus their traces, whence we can apply the Littlewood-Richardson rule and branching rules. We compute the branching rules case-by-case, but this analysis is relatively easy; one just needs to leverage the symmetries of a Young diagram to determine its unique $\Sp(4,\bbC)$ traces. We give a complete example of doing the fusion rules. Consider the product of fields $T_{i,jk} V_{l}$ which corresponds to a tensor product
\begin{align}
\ytableausetup{smalltableaux, centertableaux}
T_{i,jk} V_{l}
\equiv
\begin{ytableau}
~&~\\
~
\end{ytableau}^{\,{\rm trcls}}
\otimes
\begin{ytableau}
~
\end{ytableau}
&=
\left(
\begin{ytableau}
~&~\\
~
\end{ytableau}
-
\begin{ytableau}
~
\end{ytableau}
\right)
\otimes
\begin{ytableau}
~
\end{ytableau}
\nonumber\\
&=
\begin{ytableau}
~&~\\
~
\end{ytableau}
\otimes
\begin{ytableau}
~
\end{ytableau}
-
\begin{ytableau}
~
\end{ytableau}
\otimes
\begin{ytableau}
~
\end{ytableau}
\nonumber\\
&=
\left(
\begin{ytableau}
~&~\\
~&~
\end{ytableau}
+
\begin{ytableau}
~&~\\
~\\
~
\end{ytableau}
+
\begin{ytableau}
~&~&~\\
~
\end{ytableau}
\right)
-
\left(
\begin{ytableau}
~&~
\end{ytableau}
+
\begin{ytableau}
~\\
~
\end{ytableau}
\right)
\nonumber\\
&=
\begin{ytableau}
~&~\\
~&~
\end{ytableau}^{\,{\rm trcls}} 
+
\begin{ytableau}
~&~&~\\
~
\end{ytableau}^{\,{\rm trcls}}
+
\begin{ytableau}
~\\
~
\end{ytableau}^{\,{\rm trcls}}
+
\begin{ytableau}
~&~
\end{ytableau}
\,.
\end{align}
The calculation is finished by leveraging the branching rules
\begin{subequations}\label{BranchingRulesExample}
\begin{alignat}{2}
\begin{ytableau}
~&~\\
~&~
\end{ytableau}
&=
\begin{ytableau}
~&~\\
~&~
\end{ytableau}^{\,{\rm trcls}}
+
\begin{ytableau}
~\\
~
\end{ytableau}^{\,{\rm trcls}}
+
\bullet
\,,\qquad\qquad&
\begin{ytableau}
~&~\\
~\\
~
\end{ytableau}
&=
\begin{ytableau}
~&~
\end{ytableau}
+
\begin{ytableau}
~\\
~
\end{ytableau}^{\,{\rm trcls}}
\,,\\
\begin{ytableau}
~&~&~\\
~
\end{ytableau}
&=
\begin{ytableau}
~&~&~\\
~
\end{ytableau}^{\,{\rm trcls}}
+
\begin{ytableau}
~&~
\end{ytableau}
\,,\qquad\qquad&
\begin{ytableau}
~\\
~
\end{ytableau}
&=
\begin{ytableau}
~\\
~
\end{ytableau}^{\,{\rm trcls}}
+
\bullet
\,,
\end{alignat}
\end{subequations}
where the bullet stands for the scalar representation, e.g.\ in the last diagram antisymmetric $\hat{T}^{ij} = T^{ij} - \frac{1}{4} \O^{ij} T$ where $T^{ij}$ is traceless and $T = \O_{i j} T^{i j}$. Translating this into indices gives the irrep decomposition
\begin{align}
T_{k,ij} V_{l} = T^{(1)}_{ij,kl} + T^{(2)}_{ij,kl} - \frac{3}{10} ( \O_{k l} T^{(1)}_{i j} + \O_{l[k} T^{(1)}_{i j]} )  
+ (\O_{ij} T^{(2)}_{kl} - \O_{[ij} T^{(2)}_{k]l} )
\,,
\end{align}
where we use the Young symmetrisers of the diagram along with extraction of the $\Sp(4,\bbC)$-traces to get
\begin{subequations} \label{irrepexpansionexample}
\begin{align} 
T^{(1)}_{i j} &= \O^{kl} T_{k,ij} V_{l}
\,,\\
T^{(2)}_{k i} &= \O^{jl} T_{(k,i)j} V_{l}
\,,\\
T^{(1)}_{ij,kl} &=
- \frac{1}{2} \left( V_{[k} T_{l],ij} + V_{[i} T_{j],kl} \right)
+ \frac{1}{12} \left( \O_{i j} T^{(1)}_{k l} + \O_{k l} T^{(1)}_{i j} + \O_{l[i} T^{(1)}_{j]k} - \O_{k[i} T^{(1)}_{j]l} \right)
\,,\\
T^{(2)}_{ij,kl} &=
\frac{1}{4} \left( 3 V_{(k} T_{l),ij} -2 T_{(k,l)[i} V_{j]} \right)
- \frac{1}{12}  \left( 
\O_{ij} T^{(2)}_{kl} - \O_{k[i} T^{(2)}_{j]l} - \O_{l[i} T^{(2)}_{j]k}
\right)
\,.
\end{align}
\end{subequations}
In particular, we fix $T^{(1)}_{i j}$ and $T^{(2)}_{i j}$ as above. Then, to build $T^{(1)}_{ij,kl}$ we first take $T_{k,ij} V_{l}$ and symmetrise over $i,k$ and $j,l$, then antisymmetrise over $i,j$ and $k,l$ and then multiply by $4/3$ --- just as the Young symmetriser in \eqref{YoungSymmetrisersExample} does. Hence, one obtains
\begin{align}
\ytableausetup{nosmalltableaux}
    \begin{ytableau}
        i&k\\
        j&l
    \end{ytableau}
    \equiv - \frac{1}{2} \left( V_{[k} T_{l],ij} + V_{[i} T_{j],kl} \right)
    \,.
\ytableausetup{smalltableaux}
\end{align}
To make it traceless, we first use the branching rules \eqref{BranchingRulesExample} to determine that $\ydiagram{2,2}$ has a 2-form trace. So, we take $\O_{ij} T^{(1)}_{kl}$ and project it onto $\ydiagram{2,2}$ as we did with $T_{k,ij} V_{l}$ except now we leave an undetermined coefficient $\lambda$ so that\footnote{In general, the branching rule indicates that $\ydiagram{2,2}$ has a 2-form trace which itself has a scalar trace. Here, the scalar trace vanishes (there is no scalar of $V_i T_{j,kl}$). If the scalar did not vanish, one would also have to introduce terms $\O^{ij}\O^{kl} A$ if $A$ represents the scalar. This happens later in Section~\ref{sect:truncationto6DN=(1,0)} when we calculate $\cN = (2,0)$ to $\cN = (1,0)$ truncations for the Bach tensor.}
\begin{align}
    T^{(1)}_{ij,kl} &=
- \frac{1}{2} \left( V_{[k} T_{l],ij} + V_{[i} T_{j],kl} \right)
+ \lambda \left( \O_{i j} T^{(1)}_{k l} + \O_{k l} T^{(1)}_{i j} + \O_{l[i} T^{(1)}_{j]k} - \O_{k[i} T^{(1)}_{j]l} \right)
\,.
\end{align}
The coefficient $\lambda = 1/12$ is fixed uniquely by setting $\O^{i j} T^{(1)}_{ij,kl} = 0$, which also sets all other traces to zero due to the symmetries of $T^{(1)}_{ij,kl}$ since it lives inside $\ydiagram{2,2}$. The same process works for $T^{(2)}_{ij,kl}$.

Sometimes, the expansions \eqref{irrepexpansionexample} are not even needed, depending on exactly what computation one is performing. For example, such a situation occurs in our analysis when one is calculating $\nabla_\a^i Y^{jk,l p}$, $\nabla_\a^i Y_{\b \g}{}^{\d \e}$, $\nabla_\a^i Y_{\b}{}^{\g j k}$ by splitting them into irreps. 
We give details on how to calculate the highest weights and dimensions of the irreps from their Young diagrams in Appendix~\ref{appendix:reptheory-dimensions&highestweights}.

We give an example of how one implements fusion rules with Schur's lemma in order to make a rigorous general ansatz. We consider just $\SL(4,\bbC)$ for simplicity. Let $T_{\a,\b\g}$ be antisymmetric in $\b,\g$, but with no other symmetries. Then $T_{\a,\b\g} = T^{(1)}_{\a\b\g} + T^{(2)}_{\a,\b\g}$ where $T^{(1)}_{[\a\b\g]}$ is the 3-form part and $T^{(2)}_{\a,[\b\g]}$ is antisymmetric in $\b,\g$, but contains no 3-form part $T^{(2)}_{[\a,\b\g]} = 0$. This is simply the vector space decomposition
\begin{align}\label{YoungDiagramAnsatzExample0}
    V
    =
    \ydiagram{1} \otimes \ydiagram{1,1}
    =
    \ydiagram{1,1,1} + \ydiagram{2,1}
    \,.
\end{align}
Now consider how many independent ways the two tensors $H_{\a}$, $G_{\b\g}$ with $\b,\g$ symmetric can be used to construct $T_{\a,\b\g}$. Let us stop the analysis at quartic order. At linear order we only have the individual tensors $H_{\a}$, $G_{\b\g}$ on their own and they are both irreps corresponding to the vector spaces $W^{(1)} = \bbC^4$ and $W^{(2)} = {\rm Sym}^2(\bbC^4)$. In terms of Young diagrams
\begin{align}
    W^{(1)} = \ydiagram{1}
    \,,\qquad
    W^{(2)} = \ydiagram{2}
    \,.
\end{align}
Each Young diagram corresponds to a distinct irrep (modulo removing columns of length four or more). Hence, Schur's lemma with $G=\SL(4,\bbC)$ tells us that
\begin{subequations}
    \begin{alignat}{2}
    \Hom_G \left(\ydiagram{1},\ydiagram{1,1,1} \right) &= 0
    \,,&\qquad
    \Hom_G \left(\ydiagram{1},\ydiagram{2,1} \right) &= 0
    \,,\\
    \Hom_G \left(\ydiagram{2},\ydiagram{1,1,1} \right) &= 0
    \,,&\qquad
    \Hom_G \left(\ydiagram{2},\ydiagram{2,1} \right) &= 0
    \,.
\end{alignat}
\end{subequations}
Equivalently, $\Hom_G(W^{(1)},V) = 0 = \Hom_G(W^{(2)},V)$ states that there exists no way of embedding $H_{\a}$, $G_{\b\g}$ on their own into $T_{\a,\b\g}$.
At quadratic order, we have tensor products arising. They correspond to embedding $H_{\a} H_{\b}$, $H_{\a} G_{\b\g}$ and $G_{\a\b}G_{\g\d}$. Their diagrams are
\begin{subequations} \label{YoungDiagramAnsatzExample1}
    \begin{align} 
    W^{(3)} = \ydiagram{1} \otimes \ydiagram{1} &= \ydiagram{1,1} + \ydiagram{2} \label{YoungDiagramAnsatzExample1.1}
    \,,\\
   W^{(4)} = \ydiagram{1} \otimes \ydiagram{2} &= \ydiagram{3} + \ydiagram{2,1}
    \,,\\
    W^{(5)} = \ydiagram{2} \otimes \ydiagram{2} &= \ydiagram{4} + \ydiagram{3,1} + \ydiagram{2,2}
    \,.
\end{align}
\end{subequations}
Applying Schur's lemma amounts to comparing each of the seven RHS diagrams in \eqref{YoungDiagramAnsatzExample1} with the two RHS diagrams in \eqref{YoungDiagramAnsatzExample0}. For example, none of the diagrams in $W^{(3)}$ or $W^{(5)}$ match those of $V$ so $\Hom_G(W^{(3)},V) = 0 = \Hom_G(W^{(5)},V)$ do not contribute. One diagram of $W^{(4)}$ matches with one of $V$'s. Thus, we have
\begin{subequations}
    \begin{align}
     \Hom_G(W^{(4)},V) &= \Hom_G \left(\ydiagram{2,1},\ydiagram{2,1} \right)
     \,,\\
     \dim\Hom_G(W^{(4)},V) &= \dim\Hom_G \left(\ydiagram{2,1},\ydiagram{2,1} \right) = 1
     \,.
    \end{align}
\end{subequations}
Thus, up to quadratic order, the ansatz has \emph{one independent complex coefficient $\lambda_1$} so that
\begin{align}
    T_{\a,\b\g} = \lambda_1 H_{[\b} G_{\g]\a}
    \,.
\end{align}
It is important to note that the vector space p.o.v.\ provided by using Young diagrams does \emph{not} account for the inherent symmetry when a tensor is multiplied with itself, e.g.\ $H_\a H_\b = H_\b H_\a$ is symmetric, but \eqref{YoungDiagramAnsatzExample1.1} acts as if $H_{[\a} H_{\b]}$ is non-zero because it is treating the tensor product as if it was between distinct tensors. In practice, all this means is that some diagrams may vanish due to such symmetries.

At cubic order, we have $H_{\a} H_{\b} H_{\g}$, $H_{\a} H_{\b} G_{\g \d}$, $H_{\a} G_{\b \g} G_{\d \e}$ and $G_{\a\b}G_{\g\d}G_{\e\z}$ and diagrams given by
\begin{subequations}
    \begin{align}
    W^{(6)} = \ydiagram{1}^{\, 3} &= \ydiagram{1,1,1} + 2\, \ydiagram{2,1} + \ydiagram{3}
    \,,\\
    W^{(7)} = \ydiagram{1}^{\, 2} \otimes \ydiagram{2} &= \ydiagram{4} + 2\, \ydiagram{3,1} + \ydiagram{2,2} + \ydiagram{2,1,1}
    \,,\\
    W^{(8)} = \ydiagram{1} \otimes \ydiagram{2}^{\, 2} &= \ydiagram{5} + 2\,\ydiagram{4,1} + 2\,\ydiagram{3,2} + \ydiagram{3,1,1} + \ydiagram{2,2,1}
    \,,\\
    W^{(9)} = \ydiagram{2}^{\, 3} &= \ydiagram{6} + 2\,\ydiagram{5,1} + 3\,\ydiagram{4,2} + \ydiagram{4,1,1} 
    \nonumber\\&\quad
   + \ydiagram{3,3}  + 2\,\ydiagram{3,2,1} + \ydiagram{2,2,2}
    \,.
\end{align}
\end{subequations}
Hence, there seem to be three potential independent contributions that arise from $\dim \Hom_G(W^{(6)},V) = 3$, but all three vanish due to $H_{[\a} H_{\b]} = 0$. 
At quartic order, things become somewhat unwieldy. However, one can check that contributions only come from $H_{\a}G_{\b\g}G_{\d\e}G_{\z\q}$ in the sense that
\begin{align}
    W^{(10)} = \ydiagram{1} \otimes \ydiagram{2}^{\, 3} = \dots + 2\, \ydiagram{3,2,1,1} + \ydiagram{2,2,2,1} = \dots + 2\, \ydiagram{2,1} + \ydiagram{1,1,1}
    \,.
\end{align}
Hence, due to $\dim \Hom_G(W^{(10)},V) = 3$, there seems to be three potential independent contributions. One then needs to consider the inherent symmetry involved by using the same tensor $G_{\a\b}$ three times. In this case, the symmetry means that only the 3-form diagram is non-zero. Thus, up to quartic order, the ansatz has two independent complex coefficients so that
\begin{align}
    T_{\a,\b\g} = \lambda_1 H_{[\b} G_{\g]\a} + \lambda_2 \ve^{\d\e\z\q} H_{\d} G_{\e\a} G_{\z\b} G_{\q\g} 
    \,.
\end{align}
To illustrate a more general scenario, consider cubic order only and suppose we instead had three independent tensors $H^{(1)}_{\a}$, $H^{(2)}_{\a}$, $H^{(3)}_{\a}$ instead of just one $H_{\a}$. In this case, the `hook' diagram appears twice in $W^{(6)}$ and once in $V$. Thus, at \emph{exactly cubic order}, the ansatz has three independent complex coefficients so that
\begin{align}
    T_{\a,\b\g} &=
     \lambda_1 H^{(1)}_{[\a} H^{(2)}_{\b} H^{(3)}_{\g]}
    + \lambda_2 \left( H^{(1)}_{(\a} H^{(2)}_{\b)} H^{(3)}_{\g} - H^{(1)}_{(\a} H^{(2)}_{\g)} H^{(3)}_{\b} \right)
    \nonumber\\&\quad
    + \lambda_3 \left( H^{(1)}_{\a} H^{(2)}_{[\b} H^{(3)}_{\g]} + H^{(1)}_{[\b} H^{(2)}_{|\a|} H^{(3)}_{\g]} \right)
    \,.
\end{align}
A similar idea applies to the quartic case. If one considers independent $G^{(1)}_{\a\b}$, $G^{(2)}_{\a\b}$, $G^{(3)}_{\a\b}$, then, at exactly quartic order, the ansatz has three independent complex coefficents so that
\begin{align}
    T_{\a,\b\g} &=
    \lambda_1 \ve^{\d\e\z\q} H_{\d} G^{(1)}_{\e[\a} G^{(2)}_{|\z|\b} G^{(3)}_{|\q|\g]}
    \nonumber\\&\quad
    + \lambda_2 \ve^{\d\e\z\q} H_{\d}\left(  G^{(1)}_{\e(\a} G^{(2)}_{|\z|\b)} G^{(3)}_{\q\g} -  G^{(1)}_{\e(\a} G^{(2)}_{|\z|\g)} G^{(3)}_{\q\b} \right)
    \nonumber\\&\quad
    + \lambda_3 \ve^{\d\e\z\q} H_{\d} \left(  G^{(1)}_{\e\a} G^{(2)}_{\z[\b} G^{(3)}_{|\q|\g]} +  G^{(1)}_{\e[\b} G^{(2)}_{|\z\a} G^{(3)}_{\q|\g]} \right)
    \,.
\end{align}
As a final comment, since we are working with the superconformal algebra, we have more structure due to the generators $\bbD,K_a,S_i^\a$. Hence, we can further constrain ansatzes. In particular, the dilatation dimension $\bbD \phi = \Delta \phi$ gives a strong constraint. In the case above, if $T_{\a,\b\g}$ had dimension $\Delta_T$ and $H_{\a}$, $G_{\a\b}$ had dimensions $\Delta_H$ and $\Delta_G$, respectively, then for any ansatz to exist we would require $\Delta_T = n\Delta_H + m\Delta_G$ for some pairs $n,m \in \bbZ_{\geq 0}$. Indeed, the product rule of $\bbD$ associated to the tensor product of representations enforces this. This gives a strong constraint, e.g.\ $\Delta_T =2$ and $\Delta_H = 1 = \Delta_G$ means that only quadratic order contributions matter.

Returning to our analysis, note that $W^{\a\b ij} = 1/6\, (\tilde{\g}^{abc})^{\a\b} W_{abc}{}^{ij}$, the super-Weyl tensor, is a $\SL(4,\bbC)~\times~\Sp(4,\bbC)$ irrep. Using these tools, we find that the most general ansatz for the $\{\nabla_{\a}^i,\nabla_{\b}^j\}$ curvatures is
\begin{subequations}\label{QQansatz}
\begin{align}
\{\nabla_\a^i,\nabla_\b^j\} &= -2\ri \O^{ij} \nabla_{\a\b} - W_{\a\b}{}^{ij} - G_{\a\b}{}^{ij}
\,,\\
W_{\a\b}{}^{ij} &= 
\frac{1}{2} \ve_{\a\b\g\e} W^{\g\d ij} M_{\d}{}^{\e}
+ \WS_{\a\b,\g}{}^{i j, k} S_{\g}^k
+ \frac{1}{4} \WK_{\a\b,\g\d}{}^{ij} K^{\g\d}
\,,\\
G_{\a\b}{}^{ij} &=
\frac{1}{4}\GK_{\a\b,\g\d}{}^{ij} K^{\g\d}
\,.
\end{align}
\end{subequations}
All other curvatures \emph{cannot} be generated by $W^{\a\b ij}$ or its descendants and must vanish identically. We can also assume without loss of generality that $W_{\a\b}{}^{ij}$ is $\Sp(4,\bbC)$-traceless. If it was not, then its trace can be readily absorbed into the vector derivative $\nabla_a$, adjusting its $S$- and $K$-connections. We will make ansatzes for $W(S),W(K),G(K)$ in our later analysis --- it is overly cumbersome to bother now.
The most general structure function ansatzes are given by\footnote{Note that the need to deform the superconformal algebra's $\{S,Q\}$ and $[S,P]$ (anti-)commutation relations to obtain a consistent gauged algebra was noticed also in \cite{Butter:2019edc} for the case of 4D, $\cN=4$ conformal supergravity.} 
\begin{subequations}
\begin{align}
\{S^{\a i},\nabla_\b^j\} &= \dots 
+ \lambda_{1} W^{\a\q i j} K_{\q\b}
\,,\\
[S^{\a i}, \nabla_{\b \g}] &= \dots 
+ \lambda_{2} \ve_{\b \g \d \e} W^{\a \d i k} S_k^{\e}
+ \lambda_{3} \nabla_{[\b l} W^{\d \a l i} K_{\g]\d}
\nonumber\\&\quad
+ \lambda_{4} \nabla_{\d l} W^{\d \e l i} K_{\e [\b} \d_{\g]}^\a
+ \lambda_{5} \nabla_{\d l} W^{\d \a l i} K_{\b \g}
\,,
\end{align}
\end{subequations}
where the omitted terms are fully known, since they are the usual ones from the superconformal algebra \eqref{SuperconformalAlgebra} --- the discussion below \eqref{structurefunctiondecomp} explains this.

We note that almost all of the below calculations were performed using the open-source computer algebra software \emph{Cadabra} \cite{Peeters:2006,Peeters:2007,Peeters:2018}. We also used the codes of Gold and Khandelwal \cite{Gold:2024nbw,Gold:2024git}. Some calculations are feasible by hand, but are typically still cumbersome. Often one is taking large expressions and repeatedly substituting, performing gamma matrix manipulations, and then projecting onto various irreps via the Young diagram symmetries and kernels of symplectic traces. \emph{Cadabra} is well-equipped to handle these kinds of computations --- there are quite a few native algorithms which help, and another significant advantage is the flexibility for creating custom algorithms. The \texttt{Ex} and \texttt{ExNode} classes in which \emph{Cadabra} stores its symbolic expressions allow one to access and systematically change the indices and names of tensors, the indices and arguments of linear operators, and more.

\subsection{Construction of standard Weyl multiplet}\label{sect:constructionofstandardweylmultiplet}
Given our starting ansatz \eqref{QQansatz}, we would like to solve our conformal superspace by determining relations of the standard Weyl multiplet, here intended as the multiplet of field strengths and covariant fields generated by $W^{\a \b ij}$ and its descendants. Moreover, we need to find how the standard Weyl multiplet embeds into the structure functions, torsions and curvatures. We begin by analysing two Bianchi identities
\begin{subequations}
\begin{align}
\label{QQQBI}
[\nabla_\a^i,\{\nabla_\b^j,\nabla_\g^k\}]
+ [\nabla_\g^k,\{\nabla_\a^i,\nabla_\b^j\}]
+ [\nabla_\b^j,\{\nabla_\g^k,\nabla_\a^i\}]
&=0
\,,\\
\label{PQQBI}
[\nabla_{a},\{\nabla_\b^j,\nabla_\g^k\}]
- \{\nabla_\g^k,[\nabla_{a},\nabla_\b^j]\}
+ \{\nabla_\b^j,[\nabla_\g^k,\nabla_a]\}
&=0
\,.
\end{align}
\end{subequations}
Focusing on \eqref{QQQBI}, the first step is obtaining the expression \eqref{QPCurvatureDef} of $[\nabla_{\a}^i,\nabla_{b}]$ in terms of our ansatzed curvatures $W_{\a\b}{}^{ij}$, $G_{\a\b}{}^{ij}$. 
This is easily done by taking a $\Sp(4,\bbC)$ trace and using gamma matrix identities. The remaining part of \eqref{QQQBI} is then $\Sp(4,\bbC)$-traceless and gives genuine constraints on the form of the $\{\nabla_{\a}^i,\nabla_{\b}^j\}$ curvatures $W_{\a\b}{}^{ij}$, $G_{\a\b}{}^{ij}$. Leveraging tracelessness, we use the following decomposition into $\Sp(4,\bbC)$ irreps 
\begin{subequations}
    \begin{align}
[\nabla_\a^i,W_{\b\g}{}^{jk}] &= W_{\a,\b\g}{}^{i,jk} + (\text{$\Sp(4,\bbC)$-traces})
\,,\\
[\nabla_\a^i,G_{\b\g}{}^{jk}] &= G^{(1)}_{\a,\b\g}{}^{ijk} + G^{(2)}_{\a,\b\g}{}^{i,jk} + (\text{$\Sp(4,\bbC)$-traces})
\,.
\end{align}
\end{subequations}
Here the $\Sp(4,\bbC)$-traceless tensors correspond to the following Young diagrams
\begin{align}
\ytableausetup{nosmalltableaux}
W_{\a,\b\g}{}^{i,jk} \equiv
\begin{ytableau}
j&i\\
k
\end{ytableau}^{\,{\rm trcls}}
\,,\quad
G^{(1)}_{\a,\b\g}{}^{ijk} \equiv
\begin{ytableau}
i&j&k
\end{ytableau}
\,,\quad
G^{(2)}_{\a,\b\g}{}^{i,jk} \equiv
\begin{ytableau}
j&k\\
i
\end{ytableau}^{\,{\rm trcls},\,{\rm sym}}
\,,
\end{align}
where the last diagram follows the \emph{opposite} convention, i.e.\ it antisymmetrises $i,j$, then symmetries $j,k$ afterward. The remaining information in \eqref{QQQBI} is then given by
\begin{align}\label{QQQBI-constraints}
0&=
W_{\a,\b \g}{}^{i,j k}
+W_{\b,\a \g}{}^{j,i k}
+W_{\g,\a \b}{}^{k,i j}
+3G^{(1)}_{(\a,\b \g)}{}^{i j k}
\nonumber\\
&\quad
+G^{(2)}_{\a,\b \g}{}^{i,j k}
+G^{(2)}_{\b,\a \g}{}^{j,i k}
+G^{(2)}_{\g,\a \b}{}^{k,i j}
\,.
\end{align}
The symmetrisation over $i,j,k$ gives $G^{(1)}_{(\a,\b \g)}{}^{i j k} = 0$, and the remaining part of \eqref{QQQBI-constraints} is irreducible. We must now move down to the superfield expansions. For this, we need to know the expansions
\begin{subequations}
\begin{align}
W_{\g,\a \b}{}^{k,i j} 
&= 
[ \nabla_\g^k, W_{\a\b}{}^{ij}] + \frac{1}{5} \O^{i j} \left(  [ \nabla_{\g l}, W_{\a\b}{}^{kl}] + \frac{3}{2} [ \nabla_\g^k, W_{\a\b}{}^{l}{}_l]  \right)
\nonumber\\
&\quad
+ \frac{1}{5} \O^{k[i} \left(  4[ \nabla_{\g l}, W_{\a\b}{}^{j]l}] + [ \nabla_\g^{j]}, W_{\a\b}{}^{l}{}_l] \right)
\,, \\
G^{(1)}_{\g, \a \b}{}^{k i j} 
&= 
[ \nabla_\g^{(k}, G_{\a\b}{}^{ij)}]
\,, \\
G^{(2)}_{\g, \a \b}{}^{k, i j} 
&=
\frac{2}{3} [ \nabla_\g^{k}, G_{\a\b}{}^{ij}]
- \frac{2}{3} [ \nabla_\g^{(i}, G_{\a\b}{}^{j)k}]
- \frac{2}{5} \O^{k(i}  [ \nabla_{\g l}, G_{\a\b}{}^{j)l}]
\,.
\end{align}
\end{subequations}
Of course, one can also arrive at \eqref{QQQBI-constraints} by taking \eqref{QQQBI} and replacing both $\{\nabla_\a^i,\nabla_\b^j\}$ and $[\nabla_\a^i,\nabla_b]$ terms directly.
Something to note is that the structure functions will make an appearance in the expansion of \eqref{QQQBI-constraints} in the $S,K$-generator-valued parts. For now, we ignore those and focus on the constraints given by the other generators. Two useful constraints arise in the dimension-$3/2$, $M$-generator valued part
\begin{subequations}
    \begin{align}
    \nabla_{\a}^{i} W^{\b \g j k} - (\text{all traces}) &= 0
    \,,\\
    W(S)_{[\a\b,\g]}{}^{jk,i} - (\text{$\Sp(4,\bbC)$-traces}) &= \frac{1}{40} \ve_{\a \b \g \d} X^{\d i,jk}
    \,,
    \end{align}
\end{subequations}
where we have defined the $\Sp(4,\bbC)$-traceless descendant superfield
\begin{align}
X^{\a i,jk} = \nabla_{\b}^{i} W^{\a\b jk}  - (\text{$\Sp(4,\bbC)$-traces})
\,.
\end{align}
For reference, the $M$-valued part takes the form 
{\small
\begin{align}
    0 &=
    -4\delta_{\alpha}^{\epsilon} W(S)_{\beta \gamma \delta}{}^{j k i}
    -4\delta_{\beta}^{\epsilon} W(S)_{\alpha \gamma \delta}{}^{i k j}
    +\delta_{\delta}^{\epsilon} W(S)_{\alpha \beta \gamma}{}^{i j k}
    +\delta_{\delta}^{\epsilon} W(S)_{\alpha \gamma \beta}{}^{i k j}
    +\delta_{\delta}^{\epsilon} W(S)_{\beta \gamma \alpha}{}^{j k i}
    \nonumber\\&\quad
    -4\delta^{\epsilon}_{\gamma} W(S)_{\alpha \beta \delta}{}^{i j k}
    +\frac{1}{2}\varepsilon_{\alpha \beta \delta \zeta} \nabla_{\gamma}^{k}{W^{\epsilon \zeta i j}} 
    - \frac{1}{2}\varepsilon_{\alpha \delta \gamma \zeta} \nabla_{\beta}^{j}{W^{\epsilon \zeta i k}}
    - \frac{1}{2}\varepsilon_{\beta \delta \gamma \zeta} \nabla_{\alpha}^{i}{W^{\epsilon \zeta j k}}
    \nonumber\\&\quad
    - \frac{1}{10}\Omega^{i j} \varepsilon_{\alpha \beta \delta \zeta} \nabla_{\gamma}^{l}{W^{\epsilon \zeta k}{}_{l}} 
    - \frac{1}{5}\Omega^{i j} \varepsilon_{\alpha \delta \gamma \zeta} \nabla_{\beta}^{l}{W^{\epsilon \zeta k}{}_{l}}
    +\frac{1}{5}\Omega^{i j} \varepsilon_{\beta \delta \gamma \zeta} \nabla_{\alpha}^{l}{W^{\epsilon \zeta k}{}_{l}}
    \nonumber\\&\quad
    +\frac{1}{5}\Omega^{i k} \varepsilon_{\alpha \beta \delta \zeta} \nabla_{\gamma}^{l}{W^{\epsilon \zeta j}{}_{l}}
    +\frac{1}{10}\Omega^{i k} \varepsilon_{\alpha \delta \gamma \zeta} \nabla_{\beta}^{l}{W^{\epsilon \zeta j}{}_{l}} 
    - \frac{1}{5}\Omega^{i k} \varepsilon_{\beta \delta \gamma \zeta} \nabla_{\alpha}^{l}{W^{\epsilon \zeta j}{}_{l}}
    \nonumber\\&\quad
    - \frac{1}{5}\Omega^{j k} \varepsilon_{\alpha \beta \delta \zeta} \nabla_{\gamma}^{l}{W^{\epsilon \zeta i}{}_{l}}
    - \frac{1}{5}\Omega^{j k} \varepsilon_{\alpha \delta \gamma \zeta} \nabla_{\beta}^{l}{W^{\epsilon \zeta i}{}_{l}}
    +\frac{1}{10}\Omega^{j k} \varepsilon_{\beta \delta \gamma \zeta} \nabla_{\alpha}^{l}{W^{\epsilon \zeta i}{}_{l}} 
    \nonumber\\&\quad
    - \frac{8}{5}\delta_{\alpha}^{\epsilon} \Omega^{i j} W(S)_{\beta \gamma \delta}{}^{k l}{}_{l} 
+\frac{8}{5}\delta_{\alpha}^{\epsilon} \Omega^{i k} W(S)_{\beta \gamma \delta}{}^{j l}{}_{l} 
- \frac{4}{5}\delta_{\alpha}^{\epsilon} \Omega^{j k} W(S)_{\beta \gamma \delta}{}^{i l}{}_{l}
+\frac{8}{5}\delta_{\beta}^{\epsilon} \Omega^{i j} W(S)_{\alpha \gamma \delta}{}^{k l}{}_{l}
\nonumber\\&\quad
- \frac{4}{5}\delta_{\beta}^{\epsilon} \Omega^{i k} W(S)_{\alpha \gamma \delta}{}^{j l}{}_{l}+\frac{8}{5}\delta_{\beta}^{\epsilon} \Omega^{j k} W(S)_{\alpha \gamma \delta}{}^{i l}{}_{l}+\frac{1}{5}\delta_{\delta}^{\epsilon} \Omega^{i j} W(S)_{\alpha \beta \gamma}{}^{k l}{}_{l} 
- \frac{2}{5}\delta_{\delta}^{\epsilon} \Omega^{i j} W(S)_{\alpha \gamma \beta}{}^{k l}{}_{l}
\nonumber\\&\quad
+\frac{2}{5}\delta_{\delta}^{\epsilon} \Omega^{i j} W(S)_{\beta \gamma \alpha}{}^{k l}{}_{l} - \frac{2}{5}\delta_{\delta}^{\epsilon} \Omega^{i k} W(S)_{\alpha \beta \gamma}{}^{j l}{}_{l}+\frac{1}{5}\delta_{\delta}^{\epsilon} \Omega^{i k} W(S)_{\alpha \gamma \beta}{}^{j l}{}_{l} 
- \frac{2}{5}\delta_{\delta}^{\epsilon} \Omega^{i k} W(S)_{\beta \gamma \alpha}{}^{j l}{}_{l}
\nonumber\\&\quad
+\frac{2}{5}\delta_{\delta}^{\epsilon} \Omega^{j k} W(S)_{\alpha \beta \gamma}{}^{i l}{}_{l} - \frac{2}{5}\delta_{\delta}^{\epsilon} \Omega^{j k} W(S)_{\alpha \gamma \beta}{}^{i l}{}_{l}+\frac{1}{5}\delta_{\delta}^{\epsilon} \Omega^{j k} W(S)_{\beta \gamma \alpha}{}^{i l}{}_{l} 
- \frac{4}{5}\delta^{\epsilon}_{\gamma} \Omega^{i j} W(S)_{\alpha \beta \delta}{}^{k l}{}_{l}
\nonumber\\&\quad
+\frac{8}{5}\delta^{\epsilon}_{\gamma} \Omega^{i k} W(S)_{\alpha \beta \delta}{}^{j l}{}_{l} - \frac{8}{5}\delta^{\epsilon}_{\gamma} \Omega^{j k} W(S)_{\alpha \beta \delta}{}^{i l}{}_{l}
\,.
\end{align}
}

Next, we look at the second Bianchi identity \eqref{PQQBI}. The first step is obtaining the expression \eqref{PPCurvatureDef} of $[\nabla_a,\nabla_b]$ in terms of our ansatzed curvatures $W_{\a\b}{}^{ij}$, $G_{\a\b}{}^{ij}$ by taking a $\Sp(4,\bbC)$ trace. After this, the remaining part of the Bianchi identity is $\Sp(4,\bbC)$-traceless and gives constraints. Again, structure functions will make an appearance, but we need only look at the dimension-$3/2$, $\nabla_\a^i$-generator-valued parts, and they do not appear here. We actually reobtain both constraints above, but also get an important new one
\begin{align}
\nabla_{\b j} W^{\a\b i j} &= 0
\,.
\end{align}
For reference, the $\nabla_\a^i$-valued part takes the form 
{\small
\begin{align}
0 &=
    -\delta_{\alpha}^{\epsilon} W(S)_{\beta \delta \gamma}{}^{i j k}+\delta_{\alpha}^{\epsilon} W(S)_{\beta \delta \gamma}{}^{j k i}+\delta_{\alpha}^{\epsilon} W(S)_{\beta \gamma \delta}{}^{i j k}-\delta_{\alpha}^{\epsilon} W(S)_{\beta \gamma \delta}{}^{j k i} - \frac{1}{2}\delta_{\alpha}^{\epsilon} W(S)_{\delta \gamma \beta}{}^{i j k}
    \nonumber\\&\quad
    +\frac{1}{2}\delta_{\alpha}^{\epsilon} W(S)_{\delta \gamma \beta}{}^{j k i}
    +\delta_{\beta}^{\epsilon} W(S)_{\alpha \delta \gamma}{}^{i j k}+\delta_{\beta}^{\epsilon} W(S)_{\alpha \delta \gamma}{}^{i k j}-\delta_{\beta}^{\epsilon} W(S)_{\alpha \gamma \delta}{}^{i j k}
    -\delta_{\beta}^{\epsilon} W(S)_{\alpha \gamma \delta}{}^{i k j}
        \nonumber\\&\quad
    +\frac{1}{2}\delta_{\beta}^{\epsilon} W(S)_{\delta \gamma \alpha}{}^{i j k}+\frac{1}{2}\delta_{\beta}^{\epsilon} W(S)_{\delta \gamma \alpha}{}^{i k j} - \frac{5}{2}\delta_{\delta}^{\epsilon} W(S)_{\alpha \beta \gamma}{}^{i j k}+\frac{5}{2}\delta^{\epsilon}_{\gamma} W(S)_{\alpha \beta \delta}{}^{i j k} 
        \nonumber\\&\quad
    - \frac{5}{16}\varepsilon_{\alpha \delta \gamma \zeta} \nabla_{\beta}^{j}{W^{\epsilon \zeta i k}} - \frac{5}{16}\varepsilon_{\beta \delta \gamma \zeta} \nabla_{\alpha}^{i}{W^{\epsilon \zeta j k}}+\frac{1}{16}\Omega^{i j} \varepsilon_{\alpha \beta \delta \zeta} \nabla_{\gamma}^{l}{W^{\epsilon \zeta k}{}_{l}} - \frac{1}{16}\Omega^{i j} \varepsilon_{\alpha \beta \gamma \zeta} \nabla_{\delta}^{l}{W^{\epsilon \zeta k}{}_{l}} 
        \nonumber\\&\quad
    - \frac{1}{16}\Omega^{i j} \varepsilon_{\alpha \delta \gamma \zeta} \nabla_{\beta}^{l}{W^{\epsilon \zeta k}{}_{l}}
+\frac{1}{16}\Omega^{i j} \varepsilon_{\beta \delta \gamma \zeta} \nabla_{\alpha}^{l}{W^{\epsilon \zeta k}{}_{l}} - \frac{1}{8}\Omega^{i k} \varepsilon_{\alpha \beta \delta \zeta} \nabla_{\gamma}^{l}{W^{\epsilon \zeta j}{}_{l}}
\nonumber\\&\quad
+\frac{1}{8}\Omega^{i k} \varepsilon_{\alpha \beta \gamma \zeta} \nabla_{\delta}^{l}{W^{\epsilon \zeta j}{}_{l}} 
- \frac{1}{16}\Omega^{i k} \varepsilon_{\alpha \delta \gamma \zeta} \nabla_{\beta}^{l}{W^{\epsilon \zeta j}{}_{l}}
+\frac{1}{8}\Omega^{j k} \varepsilon_{\alpha \beta \delta \zeta} \nabla_{\gamma}^{l}{W^{\epsilon \zeta i}{}_{l}} 
\nonumber\\&\quad
- \frac{1}{8}\Omega^{j k} \varepsilon_{\alpha \beta \gamma \zeta} \nabla_{\delta}^{l}{W^{\epsilon \zeta i}{}_{l}} - \frac{1}{16}\Omega^{j k} \varepsilon_{\beta \delta \gamma \zeta} \nabla_{\alpha}^{l}{W^{\epsilon \zeta i}{}_{l}}
    \nonumber\\&\quad
+\frac{1}{4}\delta_{\alpha}^{\epsilon} \Omega^{i j} W(S)_{\beta \delta \gamma}{}^{k l}{}_{l} - \frac{1}{4}\delta_{\alpha}^{\epsilon} \Omega^{i j} W(S)_{\beta \gamma \delta}{}^{k l}{}_{l} - \frac{1}{2}\delta_{\alpha}^{\epsilon} \Omega^{j k} W(S)_{\delta \gamma \beta}{}^{i l}{}_{l} - \frac{1}{4}\delta_{\beta}^{\epsilon} \Omega^{i j} W(S)_{\alpha \delta \gamma}{}^{k l}{}_{l}
    \nonumber\\&\quad
+\frac{1}{4}\delta_{\beta}^{\epsilon} \Omega^{i j} W(S)_{\alpha \gamma \delta}{}^{k l}{}_{l} - \frac{1}{2}\delta_{\beta}^{\epsilon} \Omega^{i k} W(S)_{\delta \gamma \alpha}{}^{j l}{}_{l}+\frac{1}{4}\delta_{\delta}^{\epsilon} \Omega^{i j} W(S)_{\alpha \gamma \beta}{}^{k l}{}_{l} - \frac{1}{4}\delta_{\delta}^{\epsilon} \Omega^{i j} W(S)_{\beta \gamma \alpha}{}^{k l}{}_{l}
    \nonumber\\&\quad
+\delta_{\delta}^{\epsilon} \Omega^{i k} W(S)_{\beta \gamma \alpha}{}^{j l}{}_{l}+\delta_{\delta}^{\epsilon} \Omega^{j k} W(S)_{\alpha \gamma \beta}{}^{i l}{}_{l} - \frac{1}{4}\delta^{\epsilon}_{\gamma} \Omega^{i j} W(S)_{\alpha \delta \beta}{}^{k l}{}_{l}+\frac{1}{4}\delta^{\epsilon}_{\gamma} \Omega^{i j} W(S)_{\beta \delta \alpha}{}^{k l}{}_{l}
    \nonumber\\&\quad
-\delta^{\epsilon}_{\gamma} \Omega^{i k} W(S)_{\beta \delta \alpha}{}^{j l}{}_{l}-\delta^{\epsilon}_{\gamma} \Omega^{j k} W(S)_{\alpha \delta \beta}{}^{i l}{}_{l}
\,.
\end{align}
}

Representation theory tells us that $\nabla_\a^i W^{\b\g kl}$ splits into four $\SL(4,\bbC)~\times~\Sp(4,\bbC)$ irreps. With the above two conditions, we have now turned two of them off. Thus, we can conclude that there are at most two independent descendants of dimension 3/2 given by
\begin{subequations}\label{constr-W-0}
\begin{align}
X_\a{}^{\b\g i} &= \nabla_\a^j W^{\b\g i}{}_j
\,,\qquad
X^{\a i,jk} = \nabla_\b^i W^{\a \b jk}
\,,\\
\nabla_\a^i W^{\b\g jk} &= 
\frac{1}{5} \O^{jk} X_\a{}^{\b\g i} 
+ \frac{4}{5} \O^{i[j} X_\a{}^{\b\g k]}
+ \frac{2}{5} \d_\a^{(\b} X^{\g) i,jk}
\,.
\end{align}
\end{subequations}
The supersymmetry transformations of $W^{\a\b ij} | = -(1/3)\ri (\tilde{\g}^{abc})^{\a\b}T_{abc}{}^{ij}$ in components \cite{Bergshoeff:1999} confirm that this is right and that we should expect the descendants to correspond to $X_\a{}^{\b\g i} | \propto (\g^{ab})_{\a}{}^{\b} R(Q)_{ab}{}^{\g i}$ and $X^{\a i,jk} | \propto \chi^{\a i,jk}$.
Using that $W^{\a\b ij}$ is primary, together with the expansion of $\{S_i^\a,\nabla_\b^j\}$, we find the $S$-actions
\begin{subequations}
\begin{align}
S_{l}^{\d} X_{\a}{}^{\b \g i}
&=
20\d_{\a}^{\d} W^{\b \g i}{}_l
-8\d_{\a}^{(\b} W^{\g)\d i}{}_l
\,,\\[1ex]
S_{l}^{\d} X^{\a i, jk}
&=
-16\d^{i}_{l} W^{\a \d j k}
+8\d^{[j}_{l} W^{\a \d k]i}
+8\O^{i [j}  W^{\a \d k]}{}_l
\,.
\end{align}
\end{subequations}
Representation theory also tells us that the irrep $X_\a{}^{\b\g i}$ is not isomorphic to any of the irreps inside $W(S)_{\a\b,\g}{}^{ij,k}$. Hence, the previous result now strengthens to give
\begin{equation}
W(S)_{\a\b,\g}{}^{ij,k} = \frac{1}{40} \ve_{\a \b \g \d} X^{\d i,jk}
\,.
\end{equation}
Our ansatz for the structure functions also simplifies; we now have
\begin{subequations}
\begin{align}
\{S^{\a i},\nabla_\b^j\} &= \dots + \lambda_{1} W^{\a\q i j} K_{\q\b}
\,,\\
[S^{\a i}, \nabla_{\b \g}] &= \dots 
+ \lambda_{2} \ve_{\b \g \d \e} W^{\a \d i k} S_k^{\e}
+ \lambda_{3} X_{[\b}{}^{\d\a i} K_{\g]\d}
\,.
\end{align}
\end{subequations}
We now determine the structure functions. To do this, we analyse the Jacobi identity
\begin{equation}
[S_i^\a,\{\nabla_\b^j,\nabla_\g^k\}] + (\text{graded perms}) = 0
\,.
\end{equation}
Using the $S$-actions on $X_{\a}{}^{\b\g i}$ and $X^{\a i,jk}$ along with the expansion of $W(S)_{\a\b,\g}{}^{ij,k}$, we find that the solution of the $S$-generator-valued part of the Jacobi identity fixes $\lambda_1 = -2$, $\lambda_{2} = - \ri/10$. The $K$-generator-valued part gives us
\begin{align}
S^{k\d} \WK_{\a\b,\k\t}{}^{ij}
&=
\frac{3}{100} \ve_{\a\b\k\t} X^{\d k,ij}
-4 \left( \frac{1}{25} + \lambda_{3} \right) \ri \O^{ij} \ve_{\k\t\g[\a} X_{\b]}{}^{\g\d k}
\nonumber\\&\quad
- \frac{6}{25} \ri \O^{k[i} \ve_{\k\t\g[\a} X_{\b]}{}^{\g\d j]}
+ \frac{3}{25} \d_{[\a}^\d \ve_{\b]\k\t\g} X^{\g k,ij}
\,,\\
S^{k\d} \GK_{\a\b,\k\t}{}^{ij} 
&=
\frac{2}{25} \ri \O^{k(i} \ve_{\k\t\g(\a} X_{\b)}{}^{\d \g j)}
- \frac{2}{25} \ri  \d_{(\a}^\d \ve_{\b)\k\t\g} X^{\g (i,j)k}
\,.
\end{align}
Leveraging that $W_{\a\b}{}^{ij}$ is traceless implies $\O_{i j} \WK_{\a\b,\k\t}{}^{ij} = 0$ and we fix $\lambda_{3} = -1/40$. Now we have fixed the structure functions. The next task is to ansatz $\WK_{\a\b,\g\d}{}^{ij}$ and $\GK_{\a\b,\g\d}{}^{ij}$. A general ansatz is given by
\begin{subequations}
\begin{align}
W(K)_{\a\b,\g\d}{}^{ij}
&=
\lambda_{4}\, \ve_{\a\b\q[\g} \nabla_{\d]}^k X^{\q}{}_k{}^{ij}
+ \lambda_{5}\, \ve_{\a\b\g\d} \nabla_\q^k X^\q{}_k{}^{ij}
\nonumber\\&\quad
+ \lambda_{6}\, \ve_{\a\b\e[\g}
\left(
\nabla_{|\q|}^{[i} X_{\d]}{}^{\e\q j]} + \frac{1}{4} \nabla_{|\q|}^k X_{\d]}{}^{\ve\q}{}_k
\right)
\,,\\[1ex]
G(K)_{\a\b,\g\d}{}^{ij}
&=
\lambda_{7}\, \ve_{\g\d\q(\a} \nabla_{\b)}^k X^{\q (i,j)}{}_k
+ \lambda_{8}\, \ve_{\g\d\q(\a} \nabla_{|\z|}^{(i} X_{\b)}{}^{\q\z j)}\,.
\end{align}
\end{subequations}
Note that we know the $S$-actions on  $\WK_{\a\b,\g\d}{}^{ij}$ and $\GK_{\a\b,\g\d}{}^{ij}$ from the Jacobi identity. Hence, we can use our knowledge of the $S$-actions on $X_{\a}{}^{\b\g i}$ and $X^{\a i,jk}$ and the commutator $\{S_i^\a,\nabla_\b^j\}$ to act $S$ on the ansatz. Doing this and comparing the results sets
\begin{align}
\lambda_{4} &= - \frac{\ri}{104}
\,,\quad
\lambda_{5} = - \frac{9\ri}{1040}
\,,\quad
\lambda_{6} = - \frac{\ri}{1560}
\,,\quad
\lambda_{7} - \lambda_{8} = \frac{\ri}{240}
\,.
\end{align}
As we will soon see, our ansatz for $G(K)_{\a\b,\g\d}{}^{ij}$ is not uniquely fixed because the two terms are proportional to one another through a Bianchi identity. 
We now begin the final steps of constructing the superspace by finishing off the relations of the standard Weyl multiplet. The idea is simple, we have $W^{\a\b ij}$ and its first two descendants at dimension $3/2$. We would like to find the next set of descendants at dimension $2$. This information is encoded in the Bianchi identities, but it is also encoded in the simpler equation
\begin{equation}
\{\nabla_\a^i,\nabla_\b^j\} W^{\g\d kl} =
\nabla_\a^i\nabla_\b^j W^{\g\d kl}
+ \nabla_\b^j\nabla_\a^i W^{\g\d kl}
\,.
\end{equation}
This equation actually does not depend on the structure functions, nor any curvatures aside from $W^{\a\b ij}$ itself --- because it is $\{\nabla_\a^i,\nabla_\b^j\}$'s $M$-valued curvature, the other curvatures are $S,K$-valued and $K^AW^{\a\b ij} = 0$. Moreover, the LHS is entirely known and schematically gives contributions of the form $\nabla_a W^{\a\b ij}$ and $W^{\a\b ij} W^{\g\d kl}$. On the RHS, we input \eqref{constr-W-0}, which is the expansion of $\nabla_\a^i W^{\b\g jk}$ in terms of the dimension $3/2$ descendants $X_{\a}{}^{\b\g i}$ and $X^{\a i,jk}$. After inputting this, $\nabla_\a^i\nabla_\b^j W^{\g\d jk}$ is replaced and the equation gives constraints on $\nabla_\a^i X_{\b}{}^{\g\d j}$ and $\nabla_\a^iX^{\b j,kl}$ in terms of $\nabla_a W^{\a\b ij}$ and $W^{\a\b ij} W^{\g\d kl}$. Any unconstrained irreps inside $\nabla_\a^i X_{\b}{}^{\g\d j}$ and $\nabla_\a^iX^{\b j,kl}$ (i.e.\ those not determined by $\nabla_a W^{\a\b ij}$ and $W^{\a\b ij} W^{\g\d kl}$) will correspond to the dimension $2$ descendants. The irrep decompositions of $\nabla_\a^i X_{\b}{}^{\g\d j}$ and $\nabla_\a^iX^{\b j,kl}$ in terms of Young diagrams is given in Appendix~\ref{appendix:reptheory-irrepQX}.
We make a list of all the resulting identities from the constraints. The traces satisfy
\begin{subequations}\label{QXconstraints1}
\begin{align}
\nabla_{\a}^{(i} X^{\a j),kl} = 0
\,,\quad
\nabla_{\a}^{k} X^{\a i,j}{}_k &= 0
\,,\quad
\nabla_{[\a}^k X_{\b]}{}^{\g\d}{}_k = 0
\,,\quad
\nabla_{\q}^k X_{\a}{}^{\b\q}{}_k = 0
\,,\\
\nabla_\q^{(i} X_\a{}^{\q\b j)} + \nabla_\a^k X^{\b (i,j)}{}_k &= 0
\,,\quad
\nabla_\q^{[i} X_\a{}^{\b\q j]} - \frac{3}{8} \nabla_\a^k X^{\b}{}_k{}^{ij} = 0
\,.
\end{align}
\end{subequations}
The traceless parts satisfy
\begin{subequations}
\begin{align}
&\nabla_\q^{[i} X_\a{}^{\b\q j]} - \nabla_\a^k X^{\b}{}_k{}^{ij} = -10 \ri \nabla_{\a\q} W^{\q\b ij}
\,,\\
&\nabla_{[\a}^{(i} X_{\b]}{}^{\g \d j)} - (\text{all traces})
= \frac{5}{4} \ve_{\a\b\q\e} W^{\g\q k(i} W^{\d\e j)}{}_k
\,,\\
&\nabla_{[\a}^{[i} X_{\b]}{}^{\g \d j]} - (\text{all traces})
=  5\ri \left(
\nabla_{\a\b} W^{\g\d ij} + \d_{[\a}^{(\g} \nabla_{\b]\q} W^{\d)\q ij}
\right)
\,,\\
&\nabla_\a^{[i} X^{\b j],kl} + \nabla_\a^{[k} X^{\b l],ij} - (\text{all traces})
= 0
\,,\quad
\nabla_\a^{(i} X^{\b j,k)l} - (\text{all traces})
= 0
\,,\\
&\nabla_{(\a}^{(i} X_{\b)}{}^{\g \d j)} - (\text{all traces})
= 0
\,,\quad
\nabla_{(\a}^{[i} X_{\b)}{}^{\g \d j]} - (\text{all traces})
= 0
\,.
\end{align}
\end{subequations}
With some of these irreps turned off, the expansions become much simpler. 
We find that 3 of the 8 $\SL(4,\bbC) \times \Sp(4,\bbC)$ irreps of $\nabla_\a^i X^{\b j,kl}$ survive and they are
\begin{align}\label{QX_YExpansion1}
\nabla_\a^i X^{\b j,kl}
&=
\frac{1}{4} \d_\a^\b Y^{ij,kl}
+ \frac{1}{6} \O^{kl} Y_\a{}^{\b ij}
+ \frac{5}{6} \O^{i[k} Y_\a{}^{\b l] j}
- \frac{1}{6} \O^{j[k} Y_\a{}^{\b l] i}
\nonumber\\
&\quad
- \frac{1}{4} \O^{i j} \tilde{Y}_\a{}^{\b kl} 
+ \frac{1}{8} \O^{i[k} \tilde{Y}_\a{}^{\b l]j}
+ \frac{1}{8} \O^{j[k} \tilde{Y}_\a{}^{\b l]i}
\,.
\end{align}
Similarly, we find 5 of the 9 irreps of $\nabla_\a^i X_\b{}^{\g\d j}$ survive and they are
\begin{align}\label{QX_YExpansion2}
\nabla_\a^i X_\b{}^{\g\d j}
&=
- \frac{1}{4} \O^{ij} Y_{\a\b}{}^{\g\d} 
+ \frac{5}{12} \d_\a^{(\g} \hat{Y}_\b{}^{\d) ij} 
- \frac{1}{12} \d_\b^{(\g} \hat{Y}_\a{}^{\d) ij}
+ \frac{5}{12} \d_\a^{(\g} \hat{\tilde{Y}}_\b{}^{\d) ij} 
- \frac{1}{12} \d_\b^{(\g} \hat{\tilde{Y}}_\a{}^{\d) ij}
\nonumber\\
&\quad
+ Y_{\a\b}{}^{\g\d ij}
+ \tilde{Y}_{\a\b}{}^{\g\d ij}
\,.
\end{align}
The relations between the surviving irreps are
\begin{subequations}
\begin{align}
Y_\a{}^{\b ij} &= - \hat{Y}_\a{}^{\b ij}
\,,\quad
\hat{\tilde{Y}}_\a{}^{\b ij} = 6\ri \nabla_{\a\q} W^{\q\b ij}
\,,\quad
\tilde{Y}_\a{}^{\b ij} = 16\ri \nabla_{\a\q} W^{\q\b ij}
\,,\\
Y_{\a\b}{}^{\g\d ij} &= \frac{5}{4} \ve_{\a\b\q\e} W^{\g\q k(i} W^{\d\e j)}{}_k
\,,\quad
\tilde{Y}_{\a\b}{}^{\g\d ij} = 5\ri \left(
\nabla_{\a\b} W^{\g\d ij} + \d_{[\a}^{(\g} \nabla_{\b]\q} W^{\d)\q ij}
\right)
\,.
\end{align}
\end{subequations}
Given that many of the irreps are set to zero, we can give some simple, explicit definitions for the remaining non-zero irreps
\begin{subequations}
\begin{align}
Y_{\a}{}^{\b ij} &= \nabla_\a^k X^{\b (i,j)}{}_k
\,,\quad
\tilde{Y}_{\a}{}^{\b ij} = \nabla_\a^k X^\b{}_k{}^{ij}
\,,\quad
\hat{Y}_\a{}^{\b ij} = \nabla_\q^{(i} X_\a{}^{\b\q j)}
\,,\\
\hat{\tilde{Y}}_\a{}^{\b ij} &= \nabla_\q^{[i} X_\a{}^{\b\q j]}
\,,\quad
Y_{\a\b}{}^{\g\d} = \nabla_{(\a}^{k} X_{\b)}{}^{\g\d}{}_k 
\,,\\
Y^{ij,kl} &= \frac{1}{2} \nabla_\q^{[i} X^{\q j], kl} + \frac{1}{2} \nabla_\q^{[k} X^{\q l], ij}
\,,\quad
Y_{\a\b}{}^{\g\d ij} = \nabla_{[\a}^{(i} X_{\b]}{}^{\g\d j)} - \frac{1}{2} \d_{[\a}^{(\g} \hat{Y}_{\b]}{}^{\d) ij}
\,,\\
\tilde{Y}_{\a\b}{}^{\g\d ij} &= \nabla_{[\a}^{[i} X_{\b]}{}^{\g\d j]}   - \frac{1}{2} \d_{[\a}^{(\g} \hat{\tilde{Y}}_{\b]}{}^{\d) ij}
\,.
\end{align}
\end{subequations}
We now always write $\hat{Y}_\a{}^{\b ij}$ as $-Y_\a{}^{\b ij}$. Finally, we can implement all of this information at once by writing the $\nabla_\a^i$-actions as
\begin{subequations}\label{QX-0}
\begin{align}
\nabla_\a^i X^{\b j,kl}
&=
\frac{1}{4} \d_\a^\b Y^{ij,kl}
+ \frac{1}{6} \O^{kl} Y_\a{}^{\b ij}
+ \frac{5}{6} \O^{i[k} Y_\a{}^{\b l] j}
- \frac{1}{6} \O^{j[k} Y_\a{}^{\b l] i}
\nonumber\\
&\quad
-  4\ri\O^{i j} \nabla_{\a\q} W^{\q\b kl}
+ 2\ri\O^{i [k} \nabla_{\a\q} W^{\q\b l]j}
+ 2\ri\O^{j [k} \nabla_{\a\q} W^{\q\b l]i}
\,,\\[1ex]
\nabla_\a^i X_\b{}^{\g\d j}
&=
- \frac{1}{4} \O^{ij} Y_{\a\b}{}^{\g\d} 
- \frac{5}{12} \d_\a^{(\g} Y_\b{}^{\d) ij} 
+ \frac{1}{12} \d_\b^{(\g} Y_\a{}^{\d) ij}
+ \frac{5}{4} \ve_{\a\b\q\e} W^{\q(\g ki} W^{\d)\e j}{}_k
\nonumber\\
&\quad
+ 5\ri \nabla_{\a\b} W^{\g\d ij} 
+ 5\ri \d_\a^{(\g} \nabla_{\b\q} W^{\d)\q ij}
- 3\ri \d_\b^{(\g} \nabla_{\a\q} W^{\d)\q ij}
\,.
\end{align}
\end{subequations}
The above analysis shows that there are three unconstrained dimension 2 irreps $Y^{ij,kl}$, $Y_\a{}^{\b ij}$, $Y_{\a\b}{}^{\g\d}$. Practically, these three irreps correspond to the Lorentz scalar matter field of the Weyl multiplet, the $[\nabla_a,\nabla_b]$ $\USp(4)$ curvature and the $[\nabla_a,\nabla_b]$ Lorentz curvature, respectively.
Moreover, our analysis has also fixed $G(K)_{\a\b,\g\d}{}^{ij}$ since we now know the ansatz was linearly dependent due to \eqref{QXconstraints1}. 
Next, we need to compute the next set of supersymmetry transformations to close the multiplet. The $S$-actions can already be easily obtained by leveraging the gauged algebra commutators $[S_i^\a,\nabla_B]$ and the expansions of $Y^{ij,kl}$, $Y_\a{}^{\b ij}$, $Y_{\a\b}{}^{\g\d}$ in terms of $X_\a{}^{\b\g i}$ and $X^{\a i,jk}$.
To this end, we implement a similar strategy to the above and solve the equations
\begin{subequations}\label{QQY}
\begin{align}
\{\nabla_\a^i,\nabla_\b^j\} X^{\g k l i_1} &= \nabla_\a^i \nabla_\b^j X^{\g k l i_1} + \nabla_\b^j \nabla_\a^i X^{\g k l i_1}
\,,\\
\{\nabla_\a^i,\nabla_\b^j\} X_{\g}{}^{\d \e k} &= \nabla_\a^i \nabla_\b^j X_{\g}{}^{\d \e k} + \nabla_\b^j \nabla_\a^i X_{\g}{}^{\d \e k}
\,,
\end{align}
\end{subequations}
for constraints on $\nabla_\a^i Y_\b{}^{\g kl}$, $\nabla_\a^i Y^{jk,li_1}$, $\nabla_\a^i Y_{\b\g}{}^{\d\e}$. 
The irrep decompositions of $\nabla_\a^i Y_\b{}^{\g kl}$, $\nabla_\a^i Y^{jk,li_1}$, $\nabla_\a^i Y_{\b\g}{}^{\d\e}$ in terms of Young diagrams is given in Appendix~\ref{appendix:reptheory-irrepQY}.
Recall that to get the constraints, we need to use our expansions $\nabla_\a^i X_{\b}{}^{\g\d j}$ and $\nabla_\a^iX^{\b j,kl}$, eq.\,\eqref{QX-0}, which contain terms of the form $\nabla_{a}W^{\a\b ij}$. These will turn into $\nabla_\g^k \nabla_a W^{\a\b ij}$ when inserted into \eqref{QQY} and we need to rewrite this by commuting $\nabla_\g^k$ through $\nabla_{a}$. To this end, we need to use our expansion \eqref{QPCurvatureDef} of $[\nabla_\a^i,\nabla_b]$ in terms of the curvatures of $\{\nabla_\a^i,\nabla_\b^j\}$ that we have already fixed. In particular, at the $S$-curvature level one encounters more $\nabla_\g^k \nabla_a W^{\a\b ij}$ terms, but this is fine since we need only know the torsions and $M$, $J$, and $\bbD$-curvatures of  $[\nabla_\a^i,\nabla_b]$ to calculate $\nabla_\g^k \nabla_a W^{\a\b ij}$ since $W^{\a\b ij}$ is primary, i.e.\ $K^{A} W^{\a\b ij} = 0$. 
Using this information, we find that the following irreps are set to zero
\begin{subequations}\label{QYConstraints1}
\begin{align}
&\nabla_{(\a}^i Y_{\b\g)}{}^{\d\e} - (\text{all traces}) = 0
\,,\quad
\nabla_\a^i Y^{jk, li_1} - (\text{all traces}) =0
\,,\\
&\nabla_\q^{(i} Y_\a{}^{\q jk)} = 0
\,,\quad
\nabla_{(\a}^{(i} Y_{\b)}{}^{\g jk)} = 0
\,,\quad
\nabla_{(\a}^{[i} Y_{\b)}{}^{\q j]k} - (\text{all traces}) = 0
\,,\\
&\nabla_\q^k Y_{\a}{}^{\q i}{}_k = 0
\,.
\end{align}
\end{subequations}
Given that these 6 of the 14 irreps vanish, we find that\footnote{Here the $\bar{Z}$ are not complex conjugates, just names for the tensors.}
\begin{subequations}\label{QY_ZExpansions}
\begin{align}
\nabla_{\a}^{i} Y^{jk,li_1} &= 
\frac{1}{7} \O^{j k} Z_{\a}{}^{i, l i_1} 
+ \frac{1}{7} \O^{l i_1} Z_{\a}{}^{i, j k} 
+ \frac{4}{7} \O^{i [j} Z_{\a}{}^{k], l i_1} 
+ \frac{4}{7} \O^{i [l} Z_{\a}{}^{i_1], j k}
\,,\\
\nabla_\a^i Y_{\b\g}{}^{\d\e} &=
Z_{\a,\b\g}{}^{\d\e i}
+ \frac{3}{7} \d_\a^{(\d} Z_{\b\g}{}^{\e) i}
- \frac{1}{7} \d_{(\b}^{(\d} Z_{\g)\a}{}^{\e) i}
\,,\\
\nabla_\a^i Y_\b{}^{\g jk} &=
-\frac{2}{5} \O^{i(j} \bar{Z}_{\a\b}{}^{\g k)}
+ Z_{\a \b}{}^{\g ijk}
+ \hat{Z}_{\a \b}{}^{\g i, j k}
+ \frac{4}{15} \d_\a^\g \hat{Z}_\b{}^{i,jk}
- \frac{1}{15} \d_\b^\g  \hat{Z}_\a{}^{i,jk}
\,,
\end{align}
\end{subequations}
where all the non-zero irreps above are given by some $W^{\a\b ij} X$ or $\nabla_{a} X$ expression. Since some irreps vanish, the expansions of the remaining ones are mostly simple
\begin{subequations}
\begin{align}
Z_{\a}{}^{i,jk} &= \nabla_{\a}^{l} Y^{jk,i}{}_l
\,,\quad
Z_{\a\b}{}^{\g i} = \nabla_\q^i Y_{\a\b}{}^{\q\g}
\,,\quad
\bar{Z}_{\a\b}{}^{\g i} = \nabla_\a^k Y_\b{}^{\g i}{}_k
\,,\\
Z_{\a\b}{}^{\g ijk} &= \nabla_{[\a}^{(i} Y_{\b]}{}^{\g jk)}
\,,\quad
\hat{Z}_\a{}^{i,jk} = \nabla_\q^i Y_\a{}^{\q jk} 
\,,\\
\hat{Z}_{\a\b}{}^{\g i,jk} &= \nabla_{[\a}^i Y_{\b]}{}^{\q jk} 
+ \frac{2}{5} \O^{i(j} \nabla_{[\a}^{|l|} Y_{\b]}{}^{\g k)}{}_l
-\frac{1}{3} \d_{[\a}^\g \hat{Z}_{\b]}{}^{i,jk}
\,.
\end{align}
\end{subequations}
Note that $\bar{Z}_{\a\b}{}^{\g i}$ is not actually an irrep. Indeed, $\bar{Z}_{\a\b}{}^{\g i}$ is spinor traceless and readily decomposes into 2 irreps through (anti)symmetrisation of $\a,\b$. The rest are genuine irreps.
Finally, after putting the $Z$-expansions above back into \eqref{QQY}, each $Z$ is determined and we get 
\begin{subequations}\label{QY-0}
\begin{align}
\nabla_{\a}^{i} Y^{jk,li_1} &= 
-2\ri \O^{j k} \nabla_{\a \q}{X^{\q i, l {i_{1}}}}
-2\ri \O^{l {i_{1}}} \nabla_{\a \q}{X^{\q i, j k}}
-8\ri \O^{i [j} \nabla_{\a \q}{X^{\q k], l i_1}}
\nonumber\\
&\quad
-8\ri \O^{i [l} \nabla_{\a \q}{X^{\q {i_{1}}], j k}}
\,,\\[1ex]
\nabla_\a^i Y_{\b\g}{}^{\d\e} &=
-8\ri \nabla_{\a (\b}{X_{\g)}{}^{\d \e i}}
+16\ri \d_{\a}^{(\d} \nabla_{\q (\b}{X_{\g)}{}^{\e) \q i}}
+\frac{8}{3}\ri \d_{(\b}^{(\d} \nabla_{\g) \q}{X_{\a}{}^{\e) \q i}}
+\frac{16}{3}\ri \d_{(\b}^{(\d} \nabla_{|\a \q|}{X_{\g)}{}^{\e) \q i}}
\nonumber\\
&\quad
+4 \ve_{\a (\b |\q \z|} W^{\q (\d k i} X_{\g)}{}^{\e) \z}{}_k
+\frac{2}{3}\d_{(\b}^{(\d} \ve_{\g) \a \q \z} W^{|\q \eta| k i} X_{\eta}{}^{\e) \z }{}_k
\,,\\[1ex]
\nabla_\a^i Y_\b{}^{\g jk} &=
\frac{16}{5}\ri \O^{i (j} \nabla_{\b \d}{X_{\a}{}^{\g \d k)}}
+\frac{32}{5}\ri \O^{i (j} \nabla_{\a \d}{X_{\b}{}^{\g \d k)}}
-\frac{24}{5}\ri \nabla_{\a \b}{X^{\g (j,k) i}}
\nonumber\\
&\quad
-\frac{24}{5}\ri \d_{\a}^{\g} \nabla_{\b \d}{X^{\d (j,k) i}}
+ \frac{12}{5}\ri \d_{\b}^{\g} \nabla_{\a \d}{X^{\d (j,k) i}}  
+ \frac{4}{5} \ve_{\a \b \d \e} W^{\g \d l i} X^{\e (j,k)}{}_l
\nonumber\\
&\quad
+\frac{2}{5} \ve_{\a \b \d \e} W^{\g \d l (j} X^{\e}{}^{k), i}{}_l  
+\frac{1}{5} \ve_{\a \b \d \e} \O^{i (j} W^{\g \d}{}_{l i_1} X^{\e k),l i_1}
\nonumber\\
&\quad
+\frac{4}{5} \ve_{\a \b \d \e} \O^{i (j} W^{\d \z k) l} X_{\z}{}^{\g \e}{}_l
\,.
\end{align}
\end{subequations}

The analysis thus far has fixed the structure of the superspace in the sense that anything can be computed, and no more constraints can exist. As we gave explicitly in Section~\ref{sect:6DN20conformalsuperspace}, all of the curvatures $[\nabla_A,\nabla_B]$ are known, and all of the structure functions $[K^A,\nabla_B]$ are known. The remaining analysis is to check that the above structure satisfies the Bianchi identities. We have not fully checked this, but give some of the analysis we have performed in Section~\ref{sect:partialsolutionofbianchiidentities} below.

\subsection{Partial solution of Bianchi identities}\label{sect:partialsolutionofbianchiidentities}
In Section~\ref{sect:6DN20conformalsuperspace}, we commented that, even though our analysis completely fixes the algebra and the action of all generators, including $\nabla_\a^i$ and $S^\a_i$, on the standard Weyl multiplet, we have not explicitly checked all Bianchi identities. This alone is a good sign that all Bianchi identities are solved, because there is no freedom left to further constrain the structure.\footnote{Reality conditions are already baked in. Up to a sign, they are determined by the $\USp(4)$ indices held by a given vector-valued superfield, e.g.\ $\overline{B_{ij}} = \pm B^{ij}$ in general. More detail is given in Appendix~\ref{appendix:gammamatrixandsymplecticalgebra}.} In this sense, all Bianchi identities must be true, or there is an error in the equations we have given thus far.

We now explain exactly what we have checked, and also give some identities that arose. In 6D $\cN=(2,0)$, apriori, there are four independent Bianchi identities stemming from $\nabla_A = (\nabla_a,\nabla_\a^i)$
\begin{align}
[\nabla_A,[\nabla_B,\nabla_C]] + (\text{graded perms})
= 0
\,.
\end{align}
Based on other conformal superspaces \cite{Butter:2014xxa,Butter:2016}, the authors speculate that all four Bianchi identities are solved if the following two are solved
\begin{subequations}\label{bigeq1}
\begin{align}
\label{eq2}
[\nabla_\a^i,\{\nabla_\b^j,\nabla_\g^k\}]
+ [\nabla_\g^k,\{\nabla_\a^i,\nabla_\b^j\}]
+ [\nabla_\b^j,\{\nabla_\g^k,\nabla_\a^i\}]
&=0
\,,\\
\label{eq3}
[\nabla_{a},\{\nabla_\b^j,\nabla_\g^k\}]
- \{\nabla_\g^k,[\nabla_{a},\nabla_\b^j]\}
+ \{\nabla_\b^j,[\nabla_\g^k,\nabla_a]\}
&=0
\,.
\end{align}
\end{subequations}
However, we have no rigorous proof of this yet for the 6D $\mathcal{N}=(2,0)$ case.
One at minimum needs both of these because they are both needed to determine $[\nabla_\a^i,\nabla_b]$ and $[\nabla_a,\nabla_b]$ from our $\{\nabla_\a^i,\nabla_\b^j\}$ ansatz. 
Moreover, as we saw in Section~\ref{sect:constructionofstandardweylmultiplet}, these two Bianchi identities, the Jacobi identity generated by $[S_i^\a,[\nabla_\b^j,\nabla_\g^k]]$, and the ansatz for $\{\nabla_\a^i,\nabla_\b^j\}$ are sufficient to constrain the $\nabla_\a^i$ and $S^\a_i$ actions on the standard Weyl multiplet. In particular, this analysis shows that the key independent equation is the dimension 3/2 one
\begin{align}
    \nabla_\a^i W^{\b\g jk} &= 
\frac{1}{5} \O^{jk} X_\a{}^{\b\g i} 
+ \frac{4}{5} \O^{i[j} X_\a{}^{\b\g k]}
+ \frac{2}{5} \d_\a^{(\b} X^{\g) i,jk}
\,.
\end{align}
All the rest of the tower of $\nabla_\a^i$-actions, see the equations \eqref{QstandardWeylMultiplet-sect2},  can be proven as consequences of this by iteratively using the expansion $\{\nabla_\a^i,\nabla_\b^j\} = \nabla_\a^i\nabla_\b^j + \nabla_\b^j\nabla_\a^i$ and the fact that $\{\nabla_\a^i,\nabla_\b^j\}$ also expands into our ansatzed curvatures.

In proofs, one typically argues that \eqref{bigeq1} implies the remaining two Bianchi identities at the operator level by using the ansatzed curvatures $W_{\a\b}{}^{ij}$ and $G_{\a\b}{}^{ij}$ substituted into equation \eqref{bigeq1}. Then one uses the Jacobi identities generated by $[\nabla_A,[\nabla_B,X_{\uc}]]$ and $[\nabla_A,[X_{\ub},X_{\uc}]]$ to help shuffle terms, but the complexity of the $\SL(4,\bbC) \times \Sp(4,\bbC)$ representations of $W_{\a\b}{}^{ij}$, $G_{\a\b}{}^{ij}$ and $[\nabla_A,[\nabla_B,\nabla_C]]$ makes this difficult for the 6D $\mathcal{N}=(2,0)$ case.

We now check \eqref{bigeq1} is satisfied by expanding down to the superfield level and use our knowledge of $[\nabla_A,\nabla_B]$ and $\nabla_\a^i,S$-actions on the standard Weyl multiplet. This needs the use of identities, some of which we give below. The first Bianchi identity \eqref{eq2} is relatively easy to check, and most terms cancel immediately. However, at the $K$-generator level one must leverage a $\Sp(4,\bbC)$ identity\footnote{What we mean by ``$\Sp(4,\bbC)$ identity'' is that this identity can be generated by taking $W^{\a\b ij} X^{\g k,li_1}$ and projecting it suitably using (anti)symmetrisation of $\Sp(4,\bbC)$ indices and contractions with the symplectic form $\O^{ij}$. It is just some algebraic identity generated by symmetries of the $\Sp(4,\bbC)$ indices.} given by
\begin{align}\label{WXidentity}
 W^{\a\b l[i} X^{\g |k|, j]}{}_l 
=
\frac{1}{4}\O^{i j} W^{\a\b l i_1} X^{\g k}{}_{l i_1}
\,.
\end{align}
A related $\Sp(4,\bbC)$ identity is
\begin{align}
 W^{\g\q l[i} W^{\d\eta j]}{}_l = \frac{1}{4} \O^{ij} W^{\g\q kl} W^{\d\eta}{}_{kl}
\,.
\end{align}
This can be derived just using $\Sp(4,\bbC)$ properties on $W^{\a\b ij} W^{\g\d kl}$, but can also be derived by applying the $S$-generator to \eqref{WXidentity}. Applying the $S$-generator to \eqref{WXidentity} also gives one more $\Sp(4,\bbC)$ identity
\begin{align}\label{WWidentity2}
W^{\a \d k (i} W^{\b \g j) l} - W^{\a \d l (i} W^{\b \g j) k} &= 
-\Omega^{k (i} W^{\a \d j) i_1} W^{\b \g l}{}_{i_1}
+\Omega^{l (i} W^{\a \d j) i_1} W^{\b \g k}{}_{i_1}
\nonumber\\&\quad
-\Omega^{k l} W^{\a \d i_1 (i} W^{\b \g j)}{}_{i_1}
\,.
\end{align}

The second Bianchi identity \eqref{eq3} requires some algebraic identities, a direct application of $[\nabla_a,\nabla_b]$ and also a genuine 3-form Bianchi identity, i.e.\ one involving three vector indices $a,b,c$. Some of these identities are\footnote{The $W^3$ identity is just a $\Sp(4,\bbC)$ identity and can be generated by multiplying \eqref{WWidentity2} by $W^{\k \t}{}_{k l}$.}
\begin{subequations}
    \begin{align}
W^{\a \d k(i} W^{\b \g j) l} W^{\k \t}{}_{k l} 
&= - W^{\a \d k(i} W^{\b \g}{}_{kl} W^{\k \t j) l}
\,,\\
\nabla_{\q(\a} Y_{\b)}{}^{\q ij} &= 0
\,,\\
 \nabla^{\q(\a} Y_{\q}{}^{\b) ij}  &= \frac{6}{25} \ri X_{\q}{}^{\a\b k} X^{\q (i,j)}{}_k
\,,\\
(\g^{abc})_{\a \b} \nabla_b \nabla_c W^{\a \b i j}  &= - \frac{1}{96} (\g^{a})_{\a \b} W^{\a \q k [i} Y_{\q}^{\b j]}{}_k
\,.
    \end{align}
\end{subequations}
Recalling that $R(J)_{ab}{}^{ij} = -1/96\, Y_{ab}{}^{ij} = -1/192\, (\gamma_{ab})_\a{}^\b Y_\b{}^{\a ij}$, the second and third equations combine to give the following 3-form Bianchi identity.
\begin{align} \label{3-formBIexample}
    \nabla_{[a} Y_{bc]}{}^{ij} &= \frac{1}{12}(\tilde{\g}_{a b c})^{\a \b} \nabla_{\q(\a} Y_{\b)}{}^{\q ij} - \frac{1}{12} (\gamma_{a b c})_{\a \b} \nabla^{\q(\a} Y_{\q}{}^{\b) ij}
    \nonumber\\
    &= - \frac{1}{50} \ri (\g_{abc})_{\a\b} X_{\g}{}^{\a\b k} X^{\g (i,j)}{}_k
    \,.
\end{align}

The algebraic equations just depend on the symmetries of the tensors. The differential equations are different. One can resolve $\nabla_{[b} \nabla_{c]} W^{\a \b i j}$ quickly by using the expansion of $[\nabla_a,\nabla_b]$. In general, a vector derivative acting on $W^{\a\b ij}$, $X_\a{}^{\b ij}$, $X^{\a ijk}$, $Y_\a{}^{\b ij}$, $Y_{\a\b}{}^{\g\d}$, $Y^{ij,kl}$ is equivalent to
\begin{align}
    \nabla_a \phi = - \frac{1}{32} \ri \O_{i j} (\tilde{\gamma}_a)^{\a\b} \left( \nabla_\a^i\nabla_\b^j \phi + \nabla_\b^j\nabla_\a^i \phi \right) 
    \,.
\end{align}
Hence, we can use the $\nabla_\a^i$-actions on each of these fields \eqref{QstandardWeylMultiplet-sect2} derived in Section~\ref{sect:constructionofstandardweylmultiplet} to expand $\nabla_a\phi$ for a general $\phi$. In this case we can recover all 3-form Bianchi identities like \eqref{3-formBIexample} as consequences of the supersymmetry transformations.

The above gives some evidence that the Bianchi identities are all solved. Aside from the computational arguments, a key point is that the structure cannot be constrained further --- so consistency conditions, like Bianchi identities, must be solved, or there is an error in our equations. Another consistency check we make is by comparing to the component results of \cite{Bergshoeff:1999}. We do this in Section~\ref{sect:component-reduction} and find that everything matches, up to changing conventions. More evidence is provided by our truncation to 6D $\cN=(1,0)$ supersymmetry in Section~\ref{sect:truncationto6DN=(1,0)2} which matches with the results of \cite{Butter:2016,Butter:2017}. Lastly, more evidence comes from our analysis of the Bach tensor in Section~\ref{sect:BachTensor} --- which also leverages some of these truncation results.

\section{Component reduction}
\label{sect:component-reduction}

As a consistency check of our results, we reduce to components and compare with the results of \cite{Bergshoeff:1999}. The conversion of our conventions to those of \cite{Bergshoeff:1999} is given in Table~\ref{table:conversionofconventions} in Appendix~\ref{appendix:conversionofconventions} and is identical to Table~2 of \cite{Butter:2017} (6D $\cN=(1,0)$ case), up to replacing our $\USp(4)$ symplectic form $\O^{ij}$ with the $\SU(2)$ symplectic form $\ve^{ij}$. 

The frame and connection 1-forms of conformal superspace reduce to 1-forms on the underlying manifold $\iota : \mathcal{M}^6 \to \mathcal{M}^{6|16}$ via the pullback. The frame 1-forms $E^A = (E^a,E_i^\a)$ pullback to the vielbein and gravitino as
\begin{equation}
    e^a = dx^m e_m{}^a = E^a||
    \,,\qquad
    \psi_i^\a = dx^m \psi_m{}_i^\a = 2 E_i^\a||
    \,,
\end{equation}
that is, one turns off the fermionic coordinates $\theta^\mu_i$ and their 1-forms $d\theta^\mu_i$. Hence, we are just taking the lowest components of these 1-forms. The connection 1-forms are exactly the same\footnote{In both cases, it is just a conventional choice to rescale by a factor of $2$ for the gravitino and $S$-connection. This matches the notation of the 6D $\cN=(1,0)$ paper \cite{Butter:2017}.}
\begin{equation}
    V^{ij} = \Phi^{ij} ||
    \,,\qquad
    b = B||
    \,,\qquad
    \omega^{ab} = \Omega^{cd} ||
    \,,\qquad
    \phi_\a^i = 2 \mathfrak{F}_\a^i ||
    \,,\qquad
    \mathfrak{f}_a = \mathfrak{F}_a ||
    \,.
\end{equation}
The covariant matter fields are 0-forms, so we set $\theta^\mu_i$ to zero and they reduce to fields on $\mathcal{M}^6$ defined as 
\begin{align}\label{CovariantMatterFieldScaling}
    T_{abc}{}^{ij} = \frac{1}{2} \ri W_{abc}{}^{ij} |
    \,,\qquad
    \chi^{\a i,jk} = - \frac{3}{16} X^{\a i,jk} |
    \,,\quad
    D^{ij,kl} = \frac{3}{32} Y^{ij,kl} |
    \,.
\end{align}
For the other descendants of $W_{abc}{}^{ij}$ we abuse notation and denote $X_{\a}{}^{\b\g i} |$ as $X_{\a}{}^{\b\g i}$, $Y_\a{}^{\b ij} |$ as $Y_\a{}^{\b ij}$ and $Y_{\a\b}{}^{\g\d}|$ as $Y_{\a\b}{}^{\g\d}$. Recall that the exterior covariant derivative in superspace is
\begin{align}
    \nabla = E^A \nabla_A = d - \frac{1}{2} \Omega^{ab} M_{ab} - B\bbD - \Phi^{ij} J_{ij} - \mathfrak{F}_\a^i S_i^\a - \mathfrak{F}_a K^a
    \,.
\end{align}
The component covariant derivative $\nabla_a|$ is defined by taking the pullback of the 1-form operator
\begin{align}
    E^a \nabla_a = \nabla - E_i^\a \nabla_\a^i
    \,.
\end{align}
We abuse notation and write $\nabla_a$ for $\nabla_a|$ and the same for $M_{ab},\bbD,J_{ij},S_i^\a,K^a$. We then set $Q_\a^i = \nabla_\a^i |$ and find the expansion
\begin{align}
    e_m{}^a \nabla_a = \partial_m - \frac{1}{2} \psi_m{}_i^\a Q_\a^i - \frac{1}{2} \omega_m{}^{ab} M_{ab} - b_m \bbD - V_m{}^{ij} J_{ij} - \frac{1}{2} \phi_m{}_\a^i S_i^\a - \mathfrak{f}_{ma} K^a
    \,.
\end{align}
It is understood that any 0-form operator $\mathcal{O}$ in conformal superspace acts on component fields as follows. If $\phi$ is some superfield with lowest component $\phi|$, then $\mathcal{O}|(\phi|) := (\mathcal{O}\phi)|$.
In a similar fashion, the component curvatures $R_{ab}{}^{\uc}|$ are defined by taking the pullback of the 2-form
\begin{align}\label{2formComponents}
    \frac{1}{2} E^b \wedge E^a R_{ab}{}^{\uc} = R^{\uc} -  E_j^\b \wedge E^a R_{a}{}_\b^j{}^{\uc} - \frac{1}{2} E_j^\b \wedge E_i^\a R_\a^i{}_\b^j{}^{\uc}
    \,,
\end{align}
where one also replaces $R^{\uc}$ by its Cartan structure equation (\ref{CurvatureDef}) and (\ref{CartanStructureEqns}). We again abuse notation and denote $R_{ab}{}^{\uc}|$ by $R_{ab}{}^{\uc}$. For example, $R(M)_{ab}{}^{cd} |$ is denoted as just $R(M)_{ab}{}^{cd}$. 
Alternatively, the component curvatures $R_{ab}{}^{\uc}|$ can also be read off directly from the superspace results of $[\nabla_A,\nabla_B]$ in (\ref{CurvaturesAndTorsions0Sect2},~\ref{CurvaturesAndTorsions1Sect2},~\ref{CurvaturesAndTorsions2Sect2}). In this format, the equations are $0$-forms and so they descend to components directly. One needs only take into account the rescalings in \eqref{CovariantMatterFieldScaling}.  The same is true for the structure functions $[S_i^\a,\nabla_B]$ given in \eqref{StructureFunctionsSect2}. For example, we take $\nabla_\a^i| = Q_\a^i$ and find the component commutation relation
\begin{align}\label{QQcomponents}
    \{Q_\a^i,Q_\b^j\}
&=
-2\ri \O^{i j} (\g_{a})_{\a\b} \nabla_a
- \frac{1}{24} \ri \ve_{\a \b \g \d} (\tilde{\g}^{abc}\g^{de})^{\g\d} T_{abc}{}^{i j} M_{de}
- \frac{2}{15} \ve_{\a \b \g \d} \chi^{\g k, i j} S_k^\d
\nonumber\\&\quad
- \frac{1}{40} \ri  (\g_{a}\tilde{\g}^{cde}\g^{b})_{[\a\b]} \nabla_{b} T_{cde}{}^{i j} K^{a}
- \frac{1}{20} \ri   (\g^{bc}\g_{a})_{(\a \b)} R(J)_{bc}{}^{ij} K^{a}
\,.
\end{align}
Continuing this, we find \eqref{gaugedalgebra3} gives the component commutation relation
\begin{align}\label{PPcomponents}
    [\nabla_a,\nabla_b] &= - R(P)_{ab}{}^c \nabla_c - R(Q)_{ab}{}_k^\g Q_\g^k - \frac{1}{2} R(M)_{ab}{}^{cd} M_{cd} - R(J)_{ab}{}^{kl} J_{kl} 
    \nonumber\\&\quad
    - R(\bbD)_{ab} \bbD - R(S)_{ab}{}_\g^k S_k^\g - R(K)_{abc} K^c
    \,,
\end{align}
where we have used $R(P)_{ab}{}^c = T_{ab}{}^c|$ and $R(Q)_{ab}{}_k^\g = T_{ab}{}_k^\g|$. 
The structure functions give component commutation relations
\begin{subequations}\label{structurefunctionscomponents}
\begin{align}
\{S_i^\a,Q_\b^j\} &= 
2\d_\b^\a \d_i^j \bbD 
- 4 \d_i^j M_\b{}^\a 
+ 8 \d_\b^\a J_i{}^j 
+ \frac{1}{30} (\g_{d}\tilde{\g}^{a b c})_\b{}^\a  T_{abc}{}_i{}^j K^{d}
\,,\\
 [S^{i\a}, \nabla_{b}] &= 
 -\ri (\tilde{\g}_{b})^{\a\b} Q_{\b}^{i} 
+ \frac{1}{120} (\tilde{\g}^{cde}\g_b)^\a{}_\g T_{cde}{}^{i j} S_{j}^{\g}
+ \frac{1}{2} R(Q)_{bc}{}^{\a i} K^c
\,.
\end{align}
\end{subequations}
In the case of $R(P)_{ab}{}^c$, $R(Q)_{ab}{}_i^\a$, $R(M)_{ab}{}^{cd}$, comparing the two methods of descending to components gives the conventional constraints on the component curvatures \eqref{componentcurvatureconventionalconstraints}. We talk more on this below in Section~\ref{sect:ConventionalConstraintsAndDeformations}. For the other composite curvatures, it is sufficient to take their natural\footnote{These expansions are natural in the sense that they are enforced by solving the Bianchi identities. Rather than repeat this in components, one can take these directly from superspace.} covariant superspace expansions in (\ref{CurvaturesAndTorsions1Sect2},~\ref{CurvaturesAndTorsions2Sect2}) and project these $0$-form equations --- just like is 
 done in (\ref{QQcomponents},~\ref{PPcomponents},~\ref{structurefunctionscomponents}). One can then solve the conventional constraints given by $R(P)_{ab}{}^c$, $R(Q)_{ab}{}_i^\a$, $R(M)_{ab}{}^{cd}$ to determine the independent fields and how they build all the others. Then one can use these results to expand the covariant curvatures in components. We do not solve the conventional constraints for the independent fields in this paper, but it was done in the component paper we compare to \cite{Bergshoeff:1999}. We give some more information on this in Section~\ref{sect:CurvaturesAndTorsions}.

Our paper extends the results of \cite{Bergshoeff:1999} by giving the full expansions of the curvatures, torsions and structure functions in terms of covariant fields.

\subsection{Conventional constraints and deformations}\label{sect:ConventionalConstraintsAndDeformations}
The conventional constraints on the component curvatures arise already at the superspace level. Recall from Section \ref{sect:ConformalSuperspace} that we have the equations
\begin{subequations}
\begin{align}
R(P)_{ab}{}^c &= 0
\,,\\
T_{ab}{}_k^\g = R(Q)_{ab}{}^\g_k &= \frac{1}{80} (\gamma_{ab})_{\a}{}^{\b} X_{\b}{}^{\a \g}{}_k = \frac{1}{40} X_{ab}{}^\g{}_k
\,,\\
R(J)_{ab}{}^{ij} &= - \frac{1}{192} (\gamma_{ab})_{\a}{}^{\b} Y_{\b}{}^{\a ij} = - \frac{1}{96} Y_{ab}{}^{ij}
\,,\\
R(M)_{ab}{}^{cd} &= \frac{1}{160} Y_{ab}{}^{cd}
+ \frac{1}{16} \d_{[a}^{[c} W_{b]ef}{}^{ij} W^{d]ef}{}_{ij}
\,.
\end{align}
\end{subequations}
The symmetries of $X_{\a}{}^{\b\g i}$ and $Y_{\a\b}{}^{\g\d}$ immediately imply that
\begin{subequations}
\begin{align}\label{conventionalconstraintseq}
R(P)_{ab}{}^c &= 0
\,,\\
 (\g^a)_{\d\g} R(Q)_{ab}{}^\g_k &= 0
\,,\\
R(M)_{ad}{}^{cd} &= \frac{1}{16} W_{aef}{}^{ij} W^{cef}{}_{ij}
\,.
\end{align}
\end{subequations}
Hence, the conventional constraints on the component curvatures are given by
\begin{subequations}\label{componentcurvatureconventionalconstraints}
\begin{align}
R(P)_{ab}{}^c &= 0
\,,\\
 (\g^a)_{\d\g} R(Q)_{ab}{}^\g_k &= 0
\,,\\
R(M)_{ad}{}^{cd} &= -\frac{1}{4} T_{aef}{}^{ij} T^{cef}{}_{ij}
\,.
\end{align}
\end{subequations}
These conventional constraints match those of the component paper \cite{Bergshoeff:1999} exactly.

We can leverage our superspace formalism to get the most general form of conventional constraints. A change of conventional constraints at the component level is equivalent to a deformation of the gauged algebra in superspace. In theory, we can use covariant fields ($W^{\a\b ij}$ and its descendants) to deform any generator of the algebra $\nabla_a,\nabla_\a^i,M_{ab},\bbD,J_{ij},S_i^\a,K_a$ --- one only requires that the Jacobi and Bianchi identities are satisfied for the deformed generators. However, in practice, one never deforms $M_{ab},\bbD,J_{ij}$ as it is conveninent to have a well-defined dilatation dimension and it is convenient to have $M,J$-actions determined by the indices of the object they act on. Similarly, in general cases, deforming $S_i^\a,K_a$ may change which fields are (super)conformal primary, so this is often not desirable and even possible given the conformal weights of the covariant fields in a standard Weyl multiplet.

Keeping $M_{ab},\bbD,J_{ij}$ fixed means we can actively enforce the Jacobi identities generated by $[\bbD,[-,-]]$, $[M_{ab},[-,-]]$ and $[J_{ij},[-,-]]$ when making the deformations to $\nabla_a,\nabla_\a^i,S_i^\a,K_a$. That is, the deformations must have the appropriate dilatation dimension and the appropriate $\SL(4,\bbC)$ and $\Sp(4,\bbC)$ representation (index structure). 

In our 6D $\cN=(2,0)$ case, under these constraints, $S_i^\a,K_a$ cannot change as our covariant fields have dilatation dimension greater than one. Moreover, $\SL(4,\bbC)$ and $\Sp(4,\bbC)$ representations fix the possibilities for $\nabla_a,\nabla_\a^i$. In particular, $\nabla_\a^i$ cannot be deformed and $\nabla_a$ has a 1-parameter\footnote{A real parameter once reality conditions are enforced.} deformation. Hence, the most general deformation of the gauged algebra is by a single real constant $\lambda$ via
\begin{eqnarray}
    \hat{\nabla}_a = \nabla_a - \lambda\,(\gamma_{a})_{\a\b} (\gamma_{b})_{\g\d} W^{\a\g ij} W^{\b\d}{}_{ij} K^{b}
    \,.
\end{eqnarray}
Thus, we can see that the 6D $\cN=(2,0)$ gauged algebra is very rigid. Comparatively, in the 6D $\cN=(1,0)$ case there is a real 5-parameter deformation of $\nabla_a$ seen in \cite{Butter:2017} equation (A.17b).

Deforming $\nabla_a$ adjusts the superspace curvatures accordingly, but does \emph{not} affect the structure functions. We find that the corresponding conventional constraints are
\begin{subequations}\label{componentconventionalconstraints1param}
\begin{align}
\hat{R}(P)_{ab}{}^c &= 0
\,,\\
\hat{R}(M)_{ad}{}^{cd} &= -\left( \frac{1}{4} + 256 \lambda \right) T_{aef}{}^{ij} T^{cef}{}_{ij}
\,,\\
\g_{a \a \b} \hat{R}(Q)_{ab}{}^{\b i} &= 0
\,. 
\end{align}
\end{subequations}
Since we are comparing to the component paper \cite{Bergshoeff:1999} we will match their conventional constraints and set $\lambda = 0$. Another point which is worth to note is that the deformation given by  $\lambda = -1/1024$, which sets $\hat{R}(M)_{ad}{}^{cd} = 0$, corresponds to the choice made in the 6D $\cN=(1,0)$ component paper \cite{Butter:2017}. There it goes by the name \emph{traceless frame}.

\subsection{Supersymmetry transformations}
We now compare to the supersymmetry transformations of \cite{Bergshoeff:1999}. Recall the full superspace transformations of the frame $E^A$ (supervielbein) and connections $\omega^{\ua}$ w.r.t.\ both superdiffeomorphisms and $\mathcal{H}$-generators $\d_{\mathcal{G}} = \xi^A \nabla_A + \Lambda^{\ua} X_{\ua}$ is given by
\begin{subequations}
\begin{align}
\d_{\mathcal{G}} E^A &= d\xi^A + E^B \Lambda^{\uc} f_{\uc B}{}^A + \omega^{\ub} \xi^C f_{C \ub}{}^A + E^B \xi^C T_{CB}{}^A
\,,\\
\d_{\mathcal{G}} \omega^{\ua} &= d\Lambda^{\ua} + \omega^{\ub} \Lambda^{\uc} f_{\uc \ub}{}^{\ua} + \omega^{\ub} \xi^{C} F_{C\ub}{}^{\ua} + E^B \Lambda^{\uc} F_{\uc B}{}^{\ua} + E^B \xi^C R_{CB}{}^{\ua}
\,.
\end{align}
\end{subequations}
We will restrict ourselves to the class of transformations $\d = \xi^\a_i \nabla_\a^i + \Lambda_\a^i S_i^\a + \Lambda_a K^a$. We follow the conventions from 6D $\cN=(1,0)$ conformal superspace \cite{Butter:2017} and set
\begin{align}
(\xi^\a_i \nabla_\a^i) | &= \xi^\a_i Q_\a^i
\,,\quad
(\Lambda_\a^i  S^\a_i) | = \eta_\a^i S^\a_i
\,,\quad 
(\Lambda_a K^a) | = \lambda_a K^a
\,.
\end{align}
Finally, it is convenient to introduce the Lorentz, dilatation, and R-symmetry covariant derivative $\cD_{a}$ defined by
\begin{align}\label{LorentzDilatationRsymmetryCovDeriv}
    \cD_{m} =  \partial_m - \frac{1}{2} \omega_m{}^{cd} M_{cd} - b_m \bbD - V_m{}^{ij} J_{ij}
    \,.
\end{align}
Putting this together, the component gauge fields transform as
\begin{subequations}\label{componentfieldtransforms1}
\begin{align}
\d e_m{}^a &= - \ri \xi_k \g^a \psi_m{}^k
\,,\\
\d\psi_{m}{}^i &= 2 \mathcal{D}_m \xi^i 
+ \frac{1}{12} T_{abc}{}^{ik} \tilde{\gamma}^{abc} \gamma_m \xi_k
+ 2\ri \tilde{\gamma}_m \eta^i  
\,,\\
\d V_m{}^{kl} &= -4 \xi^{(k} \phi_m{}^{l)} + 4 \psi_m{}^{(k} \eta^{l)} - \frac{4}{15} \ri \xi_i \gamma_m \chi^{(k,l)i}
\,,\\
\d b_m &= \xi_i \phi_m{}^i + \psi_m{}^i \eta_i - 2e_m{}^a \lambda_a
\,.
\end{align}
\end{subequations}
The component matter fields are covariant, and their $Q,S$-transformations are given in the superspace information from earlier (\ref{QstandardWeylMultiplet-sect2},~\ref{SstandardWeylMultiplet-sect2}).
We find the component matter fields transform as
\begin{subequations}
\begin{align}
 \d T_{abc}{}^{ij} &=
-\frac{2}{15} \ri \xi_k \g_{abc} \chi^{k,ij}
- \frac{1}{8} \ri \xi_k \g^{de} \g_{abc}  \left( \O^{ij} R(Q)_{de}{}^{k} + 4\O^{k[i} R(Q)_{de}{}^{j]}
\right)
\,,\\
\d \chi^{i,jk} &= 
- \frac{1}{2}  D^{li,jk} \xi_l
+ \frac{3}{4}  \left( 
\O^{jk} R(J)_{ab}{}^{il} 
- \O^{i[j} R(J)_{ab}{}^{k]l} 
+ 5 \O^{l[j} R(J)_{ab}{}^{k]i} \right) \tilde{\g}^{ab} \xi_l
\nonumber\\&\quad
+ \frac{1}{16} \left(
5\O^{il} \nabla_{d} T_{abc}{}^{jk}
+4\O^{i[j} \nabla_{d} T_{abc}{}^{k]l}
+\O^{jk} \nabla_{d}T_{abc}{}^{il}
\right) \g^{abc} \g^{d} \xi_l
\nonumber\\&\quad
- \frac{1}{4} \ri  \left( 
5\O^{il}  T_{abc}{}^{jk}
+4\O^{i[j} T_{abc}{}^{k]l}
+\O^{jk} T_{abc}{}^{il}
\right) \g^{abc}\eta_l
\,,\\
\d D^{ij,kl} &=
\ri  \xi_{i_1} \g^{a} \left(
\O^{ij} \nabla_{a} \chi^{i_1,kl} 
+ 4 \O^{i_1[i} \nabla_{a} \chi^{j],kl}
+ (ij  \leftrightarrow kl ) 
\right) 
\\&\quad
- 2 \eta_{i_1}\left(\O^{ij} \chi^{i_1,kl} 
+ 4 \O^{i_1 [i} \chi^{j],kl}
+ (ij  \leftrightarrow kl ) 
\right) 
\,.
\end{align}
\end{subequations}
These calculations are independent of conventional constraints in the sense that $\hat{R}(Q)_{ab}{}_\g^k = R(Q)_{ab}{}_\g^k$ and $\hat{R}(J)_{ab}{}^{ij} = R(J)_{ab}{}^{ij}$ and one can freely replace $\nabla_a$ with $\hat{\nabla}_a$ in the above expressions --- remember the matter fields are annihilated by $K_{a}$. 

All of these supersymmetry transformations match exactly with \cite{Bergshoeff:1999} when one rescales the gamma matrices, curvatures and parameters according to Table~\ref{table:conversionofconventions}.

\subsection{Curvatures and torsions}\label{sect:CurvaturesAndTorsions}
We use equation \eqref{2formComponents} along with some of the superspace curvatures $R_\a^i{}_\b^j{}^{\uc}$, $R_\a^i{}_b{}^{\uc}$ of $\{\nabla_\a^i,\nabla_\b^j\}$ and $[\nabla_\a^i,\nabla_b]$. Note that equation \eqref{2formComponents} needs us to also use the superspace Cartan structure equations (\ref{CurvatureDef}) and (\ref{CartanStructureEqns}). Combining this together gives us the component curvatures prior to applying conventional constraints. They are
\begin{subequations}\label{unconstrainedcomponentcurvatures}
\begin{align}
R(P)_{mn}{}^a &= 2 \cD_{[m} e_{n]}{}^a
+ \frac{1}{2} \ri \psi_{[m i} \gamma^a \psi_{n]}{}^i
\,,\\
R(M)_{mn}{}^{ab} &=
2 \partial_{[m} \omega_{n]}{}^{ab}
-2\omega_{[m}{}^{ae} \omega_{n]e}{}^b
+ 8 e_{[m}{}^{[a} \mathfrak{f}_{n]}{}^{b]}
- \psi_{[m j} \g^{ab} \phi_{n]}{}^j
\nonumber\\&\quad
+ 2\ri  \psi_{[m l}\g^{[a} R(Q)_{n]}{}^{b] l} 
+ \ri \psi_{[m l}\g_{n]} R(Q)^{a b l}  
- \frac{1}{2} \ri \psi_{[m k} \g^c \psi_{n] l} T^{ab}{}_c{}^{kl}
\,,\\
R(Q)_{mn i} &=
\cD_{[m} \psi_{n] i}
+ \ri \tilde{\g}_{[m} \phi_{n] i}
+ \frac{1}{24}  T_{a b c i}{}^k   \g^{a b c} \g_{[m} \psi_{n] k}
\,.
\end{align}
\end{subequations}
All of these curvatures match exactly with \cite{Bergshoeff:1999} when one rescales the gamma matrices, curvatures and parameters according to Table~\ref{table:conversionofconventions}. 

Application of conventional constraints \eqref{componentconventionalconstraints1param} fixes $\omega_m{}^{ab}$, $\mathfrak{f}_n{}^a$ and $\phi_m{}_\a^i$ to be composite in terms of the independent fields
\begin{equation}
e_m{}^a\,,\quad
\psi_m{}^\a_i\,,\quad
b_m\,,\quad
V_m{}^{ij}\,,\quad
T_{abc}{}^{ij}\,,\quad
\chi^{\a i,jk}\,,\quad
D^{ij,kl}
\,.
\end{equation}
Their expressions can be derived by leveraging the conventional constraints, the above eq.~\eqref{unconstrainedcomponentcurvatures} and the invertibility of $e_m{}^a$. We do not give these expansions since the main purpose of the section is to compare with the results of \cite{Bergshoeff:1999} as a consistency check of our results.

\section{\texorpdfstring{Truncation to 6D $\cN = (1,0)$}{Truncation to 6D N=(1,0)}}\label{sect:truncationto6DN=(1,0)2}

In Section~\ref{sect:BachTensor}, we will construct the multiplet of equations of motion for 6D $\cN=(2,0)$ conformal supergravity associated to the 6D $(2,0)$ Bach tensor. Within the analysis in Section~\ref{sect:BachTensor} we perform a truncation of the  6D $\cN = (2,0)$ Bach tensor to 6D $\cN=(1,0)$ and compare with the results of \cite{Butter:2016,Butter:2017}. The reason for doing this is to take advantage of the fact that in \cite{Butter:2016} the $\cN=(1,0)$ Bach tensors were completely fixed, and, as we shall see later, this, together with results of \cite{Butter:2017}, can be used to uniquely determine the $\cN=(2,0)$ Bach tensor. In Section~\ref{sect:BachTensor} we also fix, up to one undetermined coefficient, the $\cN = (2,0)$ Bach tensor through constraints directly in our 6D $\cN=(2,0)$ conformal superspace. The agreement of this comparison of Bach tensors will serve as a further consistency check of our 6D $\cN=(2,0)$ conformal superspace results.

In this section, we will develop the truncation, in components, of the 6D $\cN=(2,0)$ standard Weyl multiplet to the 6D $\cN=(1,0)$ standard Weyl multiplet --- whilst setting to zero any other multiplets. In doing this, we compare directly to the 6D $\cN=(1,0)$ results in \cite{Butter:2016,Butter:2017}. 
In Section~\ref{sect:component-reduction}, we have already compared our results to the component paper \cite{Bergshoeff:1999}. See also an equivalent truncation analysis in Appendix~B of \cite{Bergshoeff:1999} where they truncate to match results in \cite{Bergshoeff:1985mz}. 
The reason why we re-perform this analysis is to match with the conventions and the gauged algebra that was used to construct the $\cN=(1,0)$ Bach tensor in \cite{Butter:2016}. In particular, one needs to deform the gauged algebra of \cite{Butter:2016} in order get that of \cite{Bergshoeff:1985mz,Bergshoeff:1999}. As we discussed in Section~\ref{sect:ConventionalConstraintsAndDeformations}, deforming the gauged algebra corresponds to deforming the conventional constraints on component curvatures. More generally, it can also deform the $Q$-supersymmetry transformations in components. This is the key point of this section and all of our analysis will hinge on going to components and matching $Q$-supersymmetry transformations and conventional constraints.

Working in components avoids technicalities of embedding a supermanifold $\mathcal{M}^{6|8}$ inside our supermanifold $\mathcal{M}^{6|16}$ in such a way that the gauged 6D $\cN=(2,0)$ superconformal algebra can be gauge-fixed and restricted appropriately. 
As the results of \cite{Butter:2016,Butter:2017} are based on their own 6D $\cN=(1,0)$ superconformal algebra on $\mathcal{M}^{6|8}$, we will still need to work in that superspace as well so we can descend to components from it.

We now discuss the truncation of the 6D $\cN=(2,0)$ standard Weyl multiplet in components. We follow \cite{Bergshoeff:1999,Butter:2017} for this part of the truncation procedure. However, since we now want to compare with \cite{Butter:2016,Butter:2017}, we will not rescale the matter fields $W_{abc}{}^{ij}|$, $X^{\a i,jk}|$ and $Y^{ij,kl}|$ like in Section~\ref{sect:component-reduction} equation \eqref{CovariantMatterFieldScaling}. We will instead abuse notation and denote them as $W_{abc}{}^{ij}$, $X^{\a i,jk}$ and $Y^{ij,kl}$.

The $(2,0)$ standard Weyl multiplet decomposes into the $(1,0)$ standard Weyl multiplet, two $(1,0)$ gravitini multiplets and a $(1,0)$ $\SU(2)$ Yang--Mills multiplet. We turn off the gravitini and SU(2) Yang--Mills multiplets. 
We split the USp(4) indices $i = 1, \cdots, 4$ to $(i = 1,2, ~i' = 1,2)$ and switch off the third and fourth gravitini $\psi_m{}_{i'}^{\a} = 0$. To preserve this last condition, we must restrict to $\SU(2) \times \SU(2) \subseteq \USp(4)$, i.e.\ we fix the block-diagonal form
\begin{align}\label{omegatruncation}
\Omega^{ij} &=
\begin{pmatrix}
\eps^{i j} & 0 \\
0 & \eps^{i' j'}
\end{pmatrix} \,,\qquad
\Omega_{ij} =
\begin{pmatrix}
\eps_{i j} & 0 \\
0 & \eps_{i' j'}
\end{pmatrix} \,.
\end{align}
Then, we must only consider $\SU(2) \times \SU(2)$  block-diagonal transformations
\begin{align} 
\L^{i}{}_{j} = \begin{pmatrix}
\Lambda^i{}_j & 0 \\
0 & \Lambda^{i'}{}_{j'}
\end{pmatrix}
\,.
\end{align}
The supersymmetry transformation of the gravitini $\psi_m{}_{i'}^{\a}= 0$ that we have turned off, requires us to set to zero part of the $\USp(4)$ connection $V_m{}^{ij'} = 0$. 
We also set $V_m{}^{i'j'} = 0$, turning off the $\SU(2)$ Yang--Mills multiplet. Hence, only the $\SU(2)$ R-symmetry connection survives.
Since we have turned off the extra gravitini, it is necessary to constrain some of the covariant fields so that the $Q$- and $S$-supersymmetry transformations are consistent. As in \cite{Bergshoeff:1999} and \cite{Butter:2017}, we make the following choice for the truncated supersymmetry parameters $\e^{i} \rightarrow (\e^{i} , 0)$ and $\eta^{i} \rightarrow (\eta^i, 0 )$.

Noting the truncation formulae in equation (B.2) of \cite{Bergshoeff:1999}, the scaling factors in our equation \eqref{CovariantMatterFieldScaling}, and the scaling factors in equation (2.4) of \cite{Butter:2017}, i.e.\
\begin{subequations}
\begin{alignat}{3}
T^-_{abc} &= -2W_{abc} |
\,,&\quad
\chi^i &= \frac{15}{2} X^i |
\,,&\quad
D &= \frac{15}{2} Y |
\,,\\
T_{abc}{}^{ij} &= \frac{1}{2}\ri\, W_{abc}{}^{ij} |
\,,&\quad
\chi^{i,jk} &= -\frac{3}{16} X^{i,jk} |
\,,&\quad
D^{ij,kl} &= \frac{3}{32} Y^{ij,kl} |
\,,
\end{alignat}
\end{subequations}
we find that the truncations of covariant matter fields of our $(2,0)$ gauged algebra in components are
\begin{subequations}\label{truncationrules2}
\begin{align}
W_{abc}{}^{i j} &= 4\ri \eps^{i j} W_{abc}\,, \qquad W_{abc}{}^{i'j'} = -4\ri \eps^{i'j'} W_{abc} \,, \\
X_i{}^{jk} &= - 40 \eps^{jk} X_i \,, \quad X_i{}^{j'k'} = 40 \eps^{j'k'} X_i \,, \quad
X_{i'}{}^{j'k} =  - 20 \d_{i'}^{j'} X^k \,, \\
Y^{ij}{}_{kl} &= - 80 \eps^{ij} \eps_{kl} Y \,, \quad
Y^{ij}{}_{k'l'} = 80\eps^{ij} \eps_{k'l'} Y \,, \quad
Y^{i'j'}{}_{k'l'} = - 80 \eps^{i'j'} \eps_{k'l'} Y \,, \\
Y^{ij'}{}_{kl'} &= - 40 \d^i_k \d^{j'}_{l'} Y  
\,.
\end{align}
\end{subequations}
We again abuse notation by denoting $W_{abc}|$, $X^{\a i}|$, $Y|$ by $W_{abc}$, $X^{\a i}$, $Y$.
The only other non-zero parts of the truncated fields are related by symmetry, for example $Y^{i'j}{}_{k'l} = -40 \d_l^j \d_{k'}^{i'} Y$. In fact, one can just take the first truncation equation for each field and use the $\USp(4)$ symmetries of said field to determine the others, for example, $W_{abc}{}^{i j} = 4\ri \eps^{i j} W_{abc}$ and $\O_{i j} W^{\a \b i j} = 0$ implies $W_{abc}{}^{i'j'} = -4\ri \eps^{i'j'} W_{abc}$ due to the block-diagonal decomposition of $\O^{ij}$ in \eqref{omegatruncation}.

To complete the truncation, we must further match the supersymmetry transformations\footnote{This is already partially done by turning off $V_m{}^{ij'} = 0$.} and conventional constraints. We develop this now.

\subsection{Supersymmetry transformations}

We initially look at the covariant fields $W^{\a\b ij}$, $X^{\a i,jk}$, $X_{\a}{}^{\b\g i}$, $Y_{\a\b}{}^{\g\d}$, $Y^{ij,kl}$, and $Y_{\a}{}^{\b ij}$. Then we look at the independent gauge fields $\psi_m{}_i^\a$, $V_m{}^{ij}$, and $b_m$. We postulate the following $(2,0)\to(1,0)$ component field truncations
\begin{subequations}\label{truncationrules3}
    \begin{align}
        W^{\a \b i j} &\longrightarrow 4 \ri \eps^{i j} W^{\a \b}
        \,,\\
        X^{\a i,jk} &\longrightarrow \lambda_{X_1} \ve^{jk} X^{\a i}
        \,,\\
        X_{\a}{}^{\b \g i} &\longrightarrow \lambda_{X_2} X_{\a}{}^{\b \g i} + \lambda_{X_3} \d_\a^{(\b} X^{\g) i}
        \,,\\
        Y_{\a\b}{}^{\g\d} &\longrightarrow \lambda_{Y_1} Y_{\a\b}{}^{\g\d} + \lambda_{Y_2} \d_{(\a}^{\g} \d_{\b)}^{\d} Y 
        \,,\\
        Y^{ij,kl} &\longrightarrow \lambda_{Y_3} \ve^{ij} \ve^{kl} Y
        \,,\\
        Y_{\a}{}^{\b ij} &\longrightarrow \lambda_{Y_4} Y_{\a}{}^{\b ij}
        \,,\\
        \psi_m{}_i^\a &\longrightarrow \lambda_{\psi} \psi_m{}_i^\a
        \,,\qquad
        V_m{}^{ij} \longrightarrow \lambda_{V} V_m{}^{ij}
        \,,\qquad
        b_m \longrightarrow \lambda_{b} b_m
        \,,
    \end{align}
\end{subequations}
where the LHS is our $(2,0)$ component fields restricted to unprimed $\SU(2)$ indices and the RHS is the $(1,0)$ component fields of \cite{Butter:2016,Butter:2017}. The other non-zero parts can of course be determined by $\USp(4)$ symmetries as described above. Anything unobtainable this way is zero, e.g.\ $X_{\a}{}^{\b \g i'}=0$,\footnote{Given our results in Section~\ref{sect:6DN20conformalsuperspace} this is equivalent to $R(Q)_{ab}{}^{\a i'} = 0$, which makes sense since we turn off its gauge field $\psi_m{}^{\a i'} = 0$.} $\psi_m{}_{i'}^\a =0$, $V_m{}^{ij'}=0$, $V_m{}^{i'j'}= 0$, $W^{\a \b i'j}=0$, $X^{\a i',jk} = 0$, $Y^{i'j,kl} = 0$. 

We now take our supersymmetry transformations from \eqref{QstandardWeylMultiplet-sect2}, which immediately descend to components as per our discussion in Section~\ref{sect:component-reduction}, and truncate them using the above rules \eqref{truncationrules2}. The only subtlety one needs to keep in mind is that the truncation of our $Q_\a^i=\nabla_\a^i|$ and $\nabla_a |$ need not match with the $(1,0)$ counterparts from the paper \cite{Butter:2016}. One only expects that $Q_\a^i=\nabla_\a^i|$ and $\nabla_a |$ should, in general, correspond to a deformation of the $(1,0)$ counterparts used in \cite{Butter:2016,Butter:2017}. At the \emph{superspace level}, this corresponds to deforming the $(1,0)$ gauged algebra of \cite{Butter:2016} by 
$\tensor[^{(1,0)}]{\nabla}{_A}  \rightarrow  \tensor[^{(1,0)}]{\hnabla}{_A} $, where $\tensor[^{(1,0)}]{\nabla}{_A}$ denotes the superspace covariant derivatives used in \cite{Butter:2016} while $\tensor[^{(1,0)}]{\hnabla}{_A}$ are deformed $(1,0)$ superspace covariant derivatives that our $(2,0)$ covariant derivatives truncate to. 
The most general deformation of the $(1,0)$ derivatives of \cite{Butter:2016} is\footnote{We match our coefficients $\lambda_1,\dots,\lambda_5$ exactly with \cite{Butter:2017} equation (A.17b). Moreover, it is useful to note the identity $\tensor[^{(1,0)}]{\widetilde{\nabla}}{_a} W^{\a \b} = \tensor[^{(1,0)}]{\nabla}{_a} W^{\a \b}$.}
\begin{subequations}\label{(1,0)generalframedeformation}
\begin{align}
  \tensor[^{(1,0)}]{\widetilde{\nabla}}{_a}  &= 
  \tensor[^{(1,0)}]{\nabla}{_a}
   -\frac{1}{16} \l_{1}(\g_{a}{}^{b c})_{\a \b} W^{\a \b}  M_{b c}
    -\ri \l_{2} (\g_{a})_{\a \b} X^{\a i}  S_{i}{}^{\b}
    -\l_{3} Y  K_{a}
   \nonumber\\&\quad
   -\frac{1}{8} \l_{4}(\g_{a b c})_{\a \b}  \tensor[^{(1,0)}]{\nabla}{^b}{W^{\a \b}} K^{c}
   -\frac{1}{8} \l_{5} (\g_{a})_{\a \b} (\g_{b})_{\d \g} W^{\a \d} W^{\b \g} K^{b}
   \,,\\
\tensor*[^{(1,0)}]{\widetilde{\nabla}}{*_\a^i} &= \tensor*[^{(1,0)}]{\nabla}{*_\a^i} - \ri \l_{6} (\g_{a})_{\a \b} X^{\b i} K^{a}
\,,
\end{align}
\end{subequations}
where here we used $\tensor[^{(1,0)}]{\widetilde{\nabla}}{_A} $ to denote a general deformation. 
At the component level, we have that $\nabla_a | = \tensor[^{(1,0)}]{\hnabla}{_a} |$ and $Q_\a^i = \tensor*[^{(1,0)}]{\hat{Q}}{*_\a^i}$ for a particular choice of $\lambda_1,\dots,\lambda_6$ which we need to fix. Matching nomenclature with \cite{Butter:2017}'s \emph{traceless frame}, we call the \emph{truncated frame} the choice of $\lambda_1,\dots,\lambda_6$ that sends $\tensor[^{(1,0)}]{\nabla}{_A}  \rightarrow  \tensor[^{(1,0)}]{\hnabla}{_A} $. We also call the choice of derivatives $\tensor[^{(1,0)}]{\nabla}{_A}$  in \cite{Butter:2016} the \emph{Yang--Mills frame}, as the gauged algebra takes the convenient form of the algebra of a 6D $\cN=(1,0)$ vector multiplet.

As we will prove, the truncated frame is very similar to the traceless frame, they deform the $(1,0)$ gauged algebra of \cite{Butter:2016} via the following choices of parameters in equation \eqref{(1,0)generalframedeformation}
\begin{subequations}
    \begin{align}
        \text{(Yang--Mills frame)}\,:& \quad (\lambda_1,\lambda_2,\lambda_3,\lambda_4,\lambda_5,\lambda_6) = \left(0,0,0,0,0,0\right)
        \,,\\
        \text{(Traceless frame)}\,:& \quad (\lambda_1,\lambda_2,\lambda_3,\lambda_4,\lambda_5,\lambda_6) = \left(2,-\frac{3}{8},\frac{1}{8},-\frac{1}{2},\frac{1}{2},0\right)
        \,,\\
        \text{(Truncated frame)}\,:& \quad
        (\lambda_1,\lambda_2,\lambda_3,\lambda_4,\lambda_5,\lambda_6) = \left(2,-\frac{3}{8},\frac{1}{8},-\frac{1}{2},0,-\frac{1}{4}\right)
        \,.
    \end{align}
\end{subequations}
In fact, the only practical difference for the purpose of our analysis in terms of $(1,0)$ component fields between the traceless and truncated frames is the supersymmetry transformation of the dilatation connection $b_m$ and the constraint on the Lorentz curvature $R(M)_{ab}{}^{cd}$.  Everything else is the same.

In the traceless frame, $\hat{R}(M)_{ad}{}^{cd} = 0$, but in the truncated frame the Lorentz curvature satisfies the conventional constraint $\hat{R}(M)_{ad}{}^{cd} = -4 W_{aef} W^{cef}$. In the traceless frame, $\d b_m$ has a term proportional to $X^{\a i}$ whereas in the truncated frame it is set to zero.\footnote{In Section~\ref{sect:ConventionalConstraintsAndDeformations} we showed that the $(2,0)$ gauged algebra admits only a 1-parameter deformation, and that this can be chosen so that the trace of the $(2,0)$ Lorentz curvature is zero. Hence, it is possible that we deform to this $(2,0)$ traceless frame. Correspondingly, its truncated frame would have $\lambda_5=1/2$.
However, these observations also showed that there is no deformation of the $(2,0)$ $Q$-supersymmetry generator $Q_\a^i = \nabla_\a^i |$. Hence, the choice $\lambda_6 = -1/4$ is necessary when comparing truncated results in components.}

We return to the analysis. To ease notation, let us always denote $\tensor[^{(1,0)}]{\hnabla}{_A}$ as $\hat{\nabla}_A$. Moreover, we follow the usual abuse of notation from Section~\ref{sect:component-reduction} and write $\hnabla_a |$ as $\hnabla_a$ and $\hnabla_\a^i|$ as $\hat{Q}_\a^i$.
We find the truncated component supersymmetry transformations by taking \eqref{QstandardWeylMultiplet-sect2} and truncating using \eqref{truncationrules3}, whilst understanding that the truncated $(2,0)$ $\nabla_a$ and $Q_\a^i$ are replaced with the $(1,0)$ $\hnabla_a$ and $\hat{Q}_\a^i$. To match against the corresponding $(1,0)$ component results of \cite{Butter:2016,Butter:2017}, we find
\begin{subequations}
    \begin{alignat}{3}
        \lambda_{X_1} &= -40\,,&\qquad
        \lambda_{X_2} &= 80\,,&\qquad
        \lambda_{X_3} &= 0
        \,,\\
        \lambda_{Y_1} &= 160\,,&\qquad
        \lambda_{Y_2} &= 0\,,&\qquad
        \lambda_{Y_3} &= 80
        \,,\\
        \lambda_{Y_4} &= -96\,,&\qquad
        \lambda_{1} &= 2
        \,.
    \end{alignat}
\end{subequations}
This is in agreement with \eqref{truncationrules2}, which we obtained for the matter fields from following the analysis of \cite{Bergshoeff:1999,Butter:2017}.
The truncated component $Q$-actions are then precisely those of the traceless frame, equation (A.32a) of \cite{Butter:2017}.\footnote{These formulae only depend on the choice of $\lambda_1$ and not any others.} They are given by
\begin{subequations}\label{Q-weyl-10-truncated}
\begin{align}
\hat{Q}_\a^i W^{\b\g}
&=
4\ri X_{\a}{}^{\b\g i} 
+ 4\ri \d_{\a}^{(\b} X^{\g) i}
\,,\\
%%%%
\hat{Q}_\a^i X^{\b j} 
&=
- \frac{2}{5} Y_\a{}^\b{}^{ij} 
- \frac{2}{5} \ve^{ij} \hat{\nabla}_{\a \g} W^{\g\b}
- \frac{1}{2} \ve^{ij} \d_\a^\b Y 
\,,\\
%%%
\hat{Q}_\a^i X_\b{}^{\g\d j}
&=
\frac{1}{2} \d^{(\g}_\a Y_\b{}^{\d) ij} 
- \frac{1}{10} \d_\b^{(\g} Y_\a{}^{\d)ij}
- \frac{1}{2} \ve^{ij} Y_{\a\b}{}^{\g\d}
- \frac{1}{4}\ve^{ij} \hat{\nabla}_{\a\b} W^{\g\d} 
\nonumber\\&\quad
+ \frac{3}{20} \ve^{ij} \d_\b^{(\g} \hat{\nabla}_{\a \r} W^{\d) \r}
-\frac{1}{4} \ve^{ij} \d^{(\g}_\a \hat{\nabla}_{\b \r} W^{\d) \r} 
\,,\\
\hat{Q}_\a^i Y 
&=
- 2 \ri \hat{\nabla}_{\a \b} X^{\b i} 
\,,\\
%%%
\hat{Q}_\g^k Y_{\a}{}^{\b ij} 
&=
\frac{2}{3} \ve^{k (i} \Big( 
- 8 \ri\hnabla_{\g\d}{X_{\alpha}{}^{\beta \delta j)}} 
- 4 \ri\hnabla_{\a\d}{X_{\gamma}{}^{\beta \delta j)}}
+ 3 \ri\hnabla_{\g\a}{X^{\beta j)}} 
+ 3 \ri\delta^{\beta}_{\gamma} \hnabla_{\a\d}{X^{\delta j)}} 
\nonumber\\&\quad
- \frac{3\ri}{2} \delta_{\alpha}^{\beta} \hnabla_{\g\d}{X^{\delta j)}} 
- 3 \ri \ve_{\alpha \gamma \delta \e} W^{\beta \delta} X^{\e j)} 
+ 4 \ri \ve_{\alpha \gamma \e \rho} W^{\delta \e} X_{\delta}{}^{\beta \rho j)} 
\Big)\,,\\
%%%
\hat{Q}_\e^l Y_{\a\b}{}^{\g\d} 
&=
- 4 \ri \,\hnabla_{\e(\a}{X_{\beta)}{}^{\gamma\delta l}} 
+ \frac{4\ri}{3} \delta_{(\alpha}^{(\gamma} \hnabla_{\b)\rho}{X_{\e}{}^{\delta) \rho l}} 
+ \frac{8\ri}{3} \delta_{(\alpha}^{(\gamma} \hnabla_{|\e\rho|}{X_{\beta)}{}^{\delta) \rho l}} 
+ 8 \ri \delta_{\e}^{(\gamma} \hnabla_{\rho(\alpha}{X_{\beta)}{}^{\delta) \rho l}} 
\nonumber\\&\quad
- \frac{4\ri}{3} \, W^{\rho \sigma} 
	\delta_{(\alpha}^{(\gamma} \ve_{\beta) \e \sigma \tau} X_{\rho}{}^{\delta) \tau l} 
- 8 \ri \ve_{\e \rho \sigma (\alpha} W^{\rho (\gamma} X_{\beta)}{}^{\delta) \sigma l}
\,.
\end{align}
\end{subequations}

Now we consider the gauge fields $\psi_m{}_i^\a$, $V_m{}^{ij}$, and $b_m$. Like before, we find the component truncated supersymmetry transformations by taking \eqref{componentfieldtransforms1} and truncating using \eqref{truncationrules3}, whilst understanding that the truncated $(2,0)$ $\nabla_a$ and $Q_\a^i$ are replaced with the $(1,0)$ $\hnabla_a$ and $\hat{Q}_\a^i$. One also has to be careful of the $S$-generator, but, as we discussed earlier, it is paired with a parameter $\eta^i = (\eta^i,0)$ with $i=3,4$ turned off. To match against the corresponding $(1,0)$ component results of \cite{Butter:2016,Butter:2017}, we find
\begin{subequations}
    \begin{alignat}{3}
    \lambda_{\psi} &= 1\,,&\qquad
    \lambda_{V} &= 1\,,&\qquad
    \lambda_{b} &=1
    \,,\\
    \lambda_{2} &= -\frac{3}{8}
    \,,&\qquad
    \lambda_{6} &= - \frac{1}{4}
    \,.
\end{alignat}
\end{subequations}
The truncated component results almost match the traceless frame ones in equation (2.45) of \cite{Butter:2017}, but not quite because $\lambda_6 \neq 0$ makes a contribution to $\d b_m$. We find them to be
\begin{subequations}
    \begin{align}
\delta \psi_m{}^i
&=
2\hat{\cD}_m \xi^i
-\frac{1}{6}W^{abc}\tilde{\g}_{abc}\g_m\xi^i
+2\ri\tilde{\g}_m\eta^i
\,,\\
\delta V_m{}^{kl} 
&=
-4 \xi^{(k}\hat{\phi}_m{}^{l)}
+4 \psi_m{}^{(k}\eta^{l)}
-2 \ri \xi^{(k}\g_mX^{l)} 
\,,\\
\delta b_m &=
\xi_i \hat{\phi}_m{}^i 
+\psi_m{}^i \eta_i
-2e_m{}^a  \l_a
\,.
    \end{align}
\end{subequations}
Note that $\hat{\phi}_m{}_\a^i$ is the $(1,0)$ $S$-connection in components in the truncated frame (it coincides with the traceless frame). That is, it is the $S$-connection of $\hat{\nabla}_a$. Moreover, $\hat{\cD}_m$ is a covariant derivative built from component Lorentz, dilatation, and $\SU(2)$ R-symmetry connections of $\hat{\nabla}_a$ --- much like \eqref{LorentzDilatationRsymmetryCovDeriv} which we used in the $(2,0)$ case. That is
\begin{align}
    \hat{\cD}_{m} =  \partial_m - \frac{1}{2} \hat{\omega}_m{}^{cd} M_{cd} - b_m \bbD - V_m{}^{ij} J_{ij}
    \,.
\end{align}

This concludes the information that we can gain from analysing the supersymmetry transformations of $W^{\a\b ij}$, $X^{\a i,jk}$, $X_{\a}{}^{\b\g i}$, $Y_{\a\b}{}^{\g\d}$, $Y^{ij,kl}$, $Y_{\a}{}^{\b ij}$ and $\psi_m{}_i^\a$, $V_m{}^{ij}$, $b_m$. We must now analyse the conventional constraints on the curvatures.

\subsection{Conventional constraints}

To compare conventional constraints, we must first truncate our own $(2,0)$ ones in \eqref{conventionalconstraintseq}. 
First, note that the truncation of the $(2,0)$ component curvatures $R(Q)_{ab}{}_k^\g$ and $R(M)_{ab}{}^{cd}$ are immediate from their expressions in terms of $W^{\a\b ij}$, $X_{\a}{}^{\b\g i}$, and $Y_{\a\b}{}^{\g\d}$ given in \eqref{CurvaturesAndTorsions0Sect2}. The $(2,0)$ component torsion $R(P)_{ab}{}^c = 0$ vanishes, so its truncation must too. 
With this in mind, we find that the truncations are
\begin{subequations}
\begin{align}\label{conventionalconstraintseq-sect5}
\hat{R}(P)_{ab}{}^c &= 0
\,,\\
 (\g^a)_{\d\g} \hat{R}(Q)_{ab}{}^\g_k &= 0
\,,\\
\hat{R}(M)_{ad}{}^{cd} &= -4 W_{aef} W^{cef}
\,.
\end{align}
\end{subequations}
To enforce this matches with the truncated frame results, we leverage the $(1,0)$ conformal superspace of \cite{Butter:2016,Butter:2017}. In particular, we apply our general deformation \eqref{(1,0)generalframedeformation} with $\lambda_1,\lambda_2,\lambda_6$ fixed from earlier and compute the superspace curvature $[\hnabla_a,\hnabla_b]$ to find $\hat{R}(P)_{ab}{}^c$, $\hat{R}(Q)_{ab}{}^\g_k$ and $\hat{R}(M)_{ab}{}^{cd}$. Matching against this gives
\begin{align}
    \lambda_3 = \frac{1}{8}\,,\qquad
    \lambda_4 = -\frac{1}{2}\,,\qquad
    \lambda_5 = 0
    \,.
\end{align}

This completes the analysis. We now know how to truncate the component fields and how to deform the $(1,0)$ covariant derivative and $Q$-supersymmetry of \cite{Butter:2016} to obtain the one that is present in our truncation. Hence, we may truncate our $(2,0)$ Bach tensor to compare against \cite{Butter:2016}'s $(1,0)$ Bach tensor. Moreover, all these calculations show that our $(2,0)$ theory reduces to the $(1,0)$ theory in \cite{Butter:2016}, giving us a further consistency check.

\section{\texorpdfstring{The 6D $\cN=(2,0)$ Bach tensor}{The 6D N=(2,0) Bach tensor}}\label{sect:BachTensor}

In this section, we first use our 6D $\cN=(2,0)$ conformal superspace to give a general ansatz for the $\cN=(2,0)$ Bach tensor $B^{ij,kl}$, or superconformal current for conformal supergravity. We partially solve the superconformal primary condition $S_p^\a B^{ij,kl} = 0$ and fix all coefficients of the ansatz, aside from one. More precisely, we solve its unique $\Sp(4,\bbC)$-trace $S_i^\a B^{ij,kl} = 0$ and the simpler condition $K_{a}B^{ij,kl}=0$. 
By using results in the previous section, we finish by truncating to 6D $\cN=(1,0)$ to compare to the associated $\cN=(1,0)$ Bach tensor. We find that the coefficients we fixed match the results of the truncation, giving us a consistency check. Moreover, the truncation fixes the sole undetermined coefficient to be zero.

\subsection{Bach tensor ansatz}

The Bach tensor $B^{ij,kl}$ corresponds to the variation of the $\cN=(2,0)$ action w.r.t.\ the highest dimension field of the Weyl multiplet $Y^{ij,kl}$. Due to this, it inherits the index symmetries of $Y^{ij,kl}$, is of dimension $4$, $\mathbb{D} B^{ij,kl} = 4B^{ij,kl}$, and is also superconformal primary, $S_p^\a B^{ij,kl} = 0$. Following \cite{Kuzenko:2017zsw}, it is expected to obey the following conservation equation
\begin{align}\label{ConservationEquation}
\nabla_\a^{p} B^{ij,kl} 
- \O^{p[i} \Psi_\a^{j],kl}
- \frac{1}{4} \O^{ij}  \Psi_\a^{p,kl}
+ (ij \leftrightarrow kl )
=0
\,,
\end{align}
where $\Psi_\a{}^{i,jk} = \frac{4}{7} \nabla_{\a l} B^{li,jk}$. Equivalently, the $\Sp(4,\bbC)$ irrep corresponding to $\ytableausetup{smalltableaux}\ydiagram{3,2}^{\,{\rm trcls}}$ is set to zero, whilst $\Psi_\a{}^{i,jk} \cong \ydiagram{2,1}^{\,{\rm trcls}}$ is unconstrained.

If we assume that $B^{ij,kl}$ is constructed by using only the super-Weyl tensor and all its descendants, the most general ansatz for the Bach tensor can be fixed by dilatation dimension and $\SL(4,\bbC) \times \Sp(4,\bbC)$ representation theory. Moreover, applying reality conditions leaves seven real parameters $b_i$ in the general ansatz. We find
\begin{align}\label{(2,0)BachTensor}
B^{ij,kl} &= 
b_1 \nabla^2 Y^{ij,kl} 
+ b_2 ( Y^{pq,ij} Y_{pq}{}^{kl} 
- Y^{pq,k[i}Y_{pq}{}^{j]l} - (\text{all traces}))
\nonumber\\&\quad
+ \ri\, b_3 ( X^{p,ij} \slashed{\nabla} X_p{}^{kl} 
+ X^{p,kl} \slashed{\nabla} X_p{}^{ij}
+ X^{p,l[i} \slashed{\nabla} X_p{}^{j]k} 
- X^{p,k[i} \slashed{\nabla} X_p{}^{j]l} 
 - (\text{all traces}) )
 \nonumber\\&\quad
 + b_4 ( Y_{\a}{}^{\b k [i} Y_{\b}{}^{\a j] l} - (\text{all traces}) )
  \nonumber\\&\quad
 + b_5 (
 \nabla_{\a \b} W^{\a \g i j}  \nabla_{\g \d} W^{\b \d k l}
 -  \nabla_{\a \b} W^{\a \g k [i}  \nabla_{\g \d} W^{\b \d j] l}
 - (\text{all traces}) )
  \nonumber\\&\quad
  + b_6 (
  W^{\a \g i j}  \nabla_{\a \b}\nabla_{\g \d} W^{\b \d k l}
 +  W^{\a \g k l}  \nabla_{\a \b}\nabla_{\g \d} W^{\b \d i j}
 +  W^{\a \g l [i}  \nabla_{\a \b}\nabla_{\g \d} W^{\b \d j] k}
 \nonumber\\&\quad
~~~~~~~ -  W^{\a \g k [i}  \nabla_{\a \b}\nabla_{\g \d} W^{\b \d j] l}
 - (\text{all traces}) )
 \nonumber\\&\quad
 + b_7 \ve_{\a_1 \a_2 \a_3 \a_4} \ve_{\b_1 \b_2 \b_3 \b_4} \Big( W^{\a_1 \b_1 i j} W^{\a_2 \b_2 k l} 
-  W^{\a_1 \b_1 k [i} W^{\a_2 \b_2 j] l}
 \nonumber\\&\quad
 ~~~~~~~~~~~~~~~~~~~~~~~~~~~~~~
 - (\text{all traces}) 
 \Big)W^{\a_3 \b_3 p q} W^{\a_4 \b_4}{}_{pq}
 \,,
\end{align}
with $\nabla^2:=\nabla^a\nabla_a$. We can fix several of the $b_i$ parameters by imposing the the superconformal primary condition $S_p^\a B^{ij,kl} = 0$.
Due to the substantial computational involvement in the 6D $\cN=(2,0)$ case, we checked only the $\Sp(4,\bbC)$-trace, $S_i^\a B^{ij,kl} = 0$, and also that $K_{a} B^{ij,kl} = 0$. We did not verify that the $\ytableausetup{smalltableaux}\ydiagram{3,2}^{\,{\rm trcls}}$ irrep of $S_p^\a B^{ij,kl}$ is zero, nor did we verify the conservation equation \eqref{ConservationEquation} holds true. In any case, $S_i^\a B^{ij,kl} = 0$ and $K_{a}B^{ij,kl} = 0$ alone enforce that the coefficients $b_i$ have to satisfy
\begin{alignat}{3}\label{Bach-constraints}
b_2 &=  - \frac{1}{80} b_1 \,, &\qquad
b_3 &= \frac{4}{15} b_1 \,, &\qquad
b_4 &= \frac{5}{144} b_1 \,, 
\nonumber\\
b_5 &=  \frac{5}{3} b_1 \,, &\qquad
b_6 &= - \frac{5}{3} b_1
\,.
\end{alignat}
As we will show in more detail later in this section, the truncation to 6D $\cN=(1,0)$ results independently gives the same and also fixes the last coefficient to be
\begin{equation}\label{Bach-constraints2}
    b_7 = 0
    \,.
\end{equation}
The fact that truncation to 6D $\cN=(1,0)$ results independently fixes all coefficients is not too surprising. Indeed, a key result of \cite{Butter:2016} is that only two independent $\cN=(1,0)$ actions (known as $C\Box C$ and $C^3$) exist and that only a specific linear combination of them (up to overall scaling) admits a $\cN=(2,0)$ uplift, or completion. Hence, the same is true for the Bach tensors --- so our $\cN = (2,0)$ Bach tensor must truncate into this unique $\cN = (1,0)$ Bach tensor.

We also performed some more checks of the $\cN = (2,0)$ result. First, we comment that $K_{a} B^{ij,kl} = 0$ gives only $b_5 = - b_6$, the rest of the terms are inherently conformal primaries. 
We checked the flat case where $\nabla_a \to \partial_a$ and $\nabla_\a^i \to D_\a^i$ take their flat superspace values. In the flat case, only linear terms survive. The only linear term is that of $\partial^2 Y^{i j, k l}$. In this case, it is annihilated by $S_i^\a$ and it obeys the conservation equation \eqref{ConservationEquation}. It is relatively easy to see the conservation equation holds since, in flat $(2,0)$ superspace, $\partial_a$ commutes with $D_\a^i$ and the expansion of $D_\a^i Y^{jk,l i_1}$ given by \eqref{QYijkl} can be applied immediately and it matches perfectly with \eqref{ConservationEquation}. Alternatively, equation \eqref{QYijkl} itself expresses that the contribution $\ytableausetup{smalltableaux}\ydiagram{3,2}^{\,{\rm trcls}}$ is set to zero, and that $D_\a^i Y^{jk,l i_1}$ is only valued in $\ydiagram{2,1}^{\,{\rm trcls}}$.

The general 6D $\cN=(1,0)$ Bach tensor can be seen in equation (6.2) of \cite{Butter:2016}. We also give a brief review here. 
We start by taking the 9-parameter ansatz for the $(1,0)$ Bach tensor used in \cite{Butter:2016}, which is originally written in the Yang--Mills frame. Next, we move from the Yang--Mills to the truncated frame. In the notation of Section~\ref{sect:truncationto6DN=(1,0)2}, the general 6D $\cN=(1,0)$ Bach tensor is then given by
\begin{align} \label{(1,0)eq1}
    B &= c_1 \hnabla^2 Y
	+ (c_2 + 3c_1) Y^2
	+ \ri \, (c_3 -3c_1) \,X^{\alpha i} \hnabla_{\alpha \beta} X^\beta_i
	+ \ri \,c_4 \,X_\alpha^i{}^{\beta \gamma} \hnabla_{\gamma \delta} X_{\beta i}{}^{\alpha \delta}
	\nonumber\\&\quad
    + c_5 \,Y_\alpha{}^\beta{}^{ij} Y_{\beta}{}^{\alpha}{}_{ij}
	+ c_6 \,Y_{\alpha \beta}{}^{\gamma \delta} Y_{\gamma \delta}{}^{\alpha \beta}
	+ c_7 \,W^{\a\g} \hnabla_{\a\b} \hnabla_{\g\d} W^{\d\b}
	+ c_8 \hnabla_{\b\a}W^{\a\g}  \hnabla_{\g\d} W^{\d\b} 
	\nonumber\\&\quad
	+ c_9 \eps_{\a_1 \a_2 \a_3 \a_4} \eps_{\b_1 \b_2 \b_3 \b_4} W^{\a_1 \b_1} W^{\a_2 \b_2} W^{\a_3 \b_3} W^{\a_4 \b_4}
    \,,
\end{align}
where $\hnabla_{a}$ is the $\cN=(1,0)$ truncated frame covariant derivative, $\hnabla^2 = \hnabla^a \hnabla_a$ and $c_i$ are real coefficients. We see that changing from the Yang--Mills to the truncated frame only mixes a few terms of the original ansatz of \cite{Butter:2016} --- $c_1$ makes a contribution to the terms that originally had only $c_2$ and $c_3$ respectively.
In 6D $\cN = (1,0)$ we still have $\bbD B = 4 B$ and $S_i^\a B = 0$, but the conservation equation is different. It is instead given by\footnote{This equation is independent of any deformations of the $\cN=(1,0)$ gauged algebra induced by the change of covariant derivatives in equation \eqref{(1,0)generalframedeformation}.}
\begin{align} \label{(1,0)eq2}
    \hnabla_{[\a}^{(i} \hnabla_{\b}^j \hnabla_{\g]}^{k)} B = 0
    \,.
\end{align}
 After implementing these constraints, \cite{Butter:2016} showed the solution is a 2-parameter family
\begin{alignat}{3} \label{(1,0)eq3}
c_3 &= -\frac{8}{3} c_2 - 5 c_1 \,, &\quad
c_4 &= - \frac{32}{15} c_2 - 16 c_1 \,, &\quad
c_5 &= \frac{2}{15} c_2 + \frac{6}{5} c_1 \,, 
\nonumber\\
c_6 &=  \frac{2}{45} c_2 + \frac{1}{3} c_1\,, &\quad
c_7 &= -\frac{2}{15} c_2 - \frac{1}{5} c_1\,, &\quad
c_8 &= \frac{1}{2} c_7 = -\frac{1}{15} c_2 - \frac{1}{10} c_1\,,
\nonumber\\
c_9 &= 0 \,.
\end{alignat}
The $\cN=(2,0)$ upliftable version simply corresponds to turning off contributions we do not see in the $\cN=(2,0)$ case.\footnote{This is readily inferred from the truncation procedure: i) use rules \eqref{truncationrules3}, ii) replace $(2,0)$ covariant derivatives with $(1,0)$ truncated frame covariant derivatives $\hnabla_a$. Hence, there is no overall change in the structure that could enable different terms to talk to eachother.} In particular, it corresponds to $c_2 = -(15/2) c_1$ which sets $c_4=0=c_6$ and gives a 1-parameter family, which makes it unique to up scaling.

We give a brief justification of all of the contributions to our Bach tensor ansatz \eqref{(2,0)BachTensor}. First, note that $B^{ij,kl}$'s $\Sp(4,\bbC)$ irrep corresponds to the traceless Young diagram
\begin{equation}
\ytableausetup{nosmalltableaux}
    B^{ij,kl} \equiv \begin{ytableau}
        i&k\\
        j&l
    \end{ytableau}^{\,{\rm trcls}}
    \,.
\end{equation}
We then require all terms in the ansatz to be superconformally covariant, which restricts to having only objects constructed out of $W_{abc}{}^{kl}$, its independent superconformal descendants defined in \eqref{XYDefs}, and $\nabla_a$ derivatives acting on all superfields.
The first term $\nabla^2 Y^{ij,kl}$ is an obvious contribution because it is dimension 4 and $Y^{ij,kl}$ lives in the same $\Sp(4,\bbC)$ irrep.
Next, $Y^{pq,ij} Y_{pq}{}^{kl}$ is the unique trace of $Y^{ij,kl}Y^{i_1j_1,k_1l_1}$ with four free indices by the multi-term symmetries of $Y^{ij,kl}$. Moreover, $Y^{pqij} Y_{pq}{}^{kl}$ is almost inherently in the Young diagram. It is antisymmetric in $i,j$ and $k,l$ and symmetric in the pairs. But, antisymmetrising $i,j,k$ does not make it vanish. However,
\begin{align}
Y^{pq,[ij}Y_{pq}{}^{k]l} = \frac{1}{3} Y^{pq,ij}Y_{pq}{}^{kl} + \frac{2}{3} Y^{pq,i[k} Y_{pq}{}^{l]j}
\end{align}
is antisymmetric in $i,j$ and $k,l$ and symmetric in the pairs and can be used to minus this away to get into the Young diagram. All that remains is to minus away the traces.

Regarding contributions from fermions $X_{\a}{}^{\b \g i}$ and $X^{\a i,jk}$, naively on dimension grounds, we expect possible contributions from terms of the form $XW^{\a\b ij}X$, $X\nabla_a X$. However, $\SL(4,\bbC) \times \Sp(4,\bbC)$ representation theory guarantees that $X^{p,ij} \slashed{\nabla} X_p{}^{kl}$ projected onto $B^{ij,kl}$'s irrep is the only contribution. Indeed, $X^{p,ij} \slashed{\nabla} X_p{}^{kl}$ is the unique trace of $X^{i,jk} \slashed{\nabla} X^{l, i_1j_1}$ by the multi-term symmetries of $X^{i,jk}$; also, we can project $X^{p,ij} \slashed{\nabla} X_p{}^{kl}$ into the Young diagram and minus away the traces.

Now, naively on dimension grounds we can expect contributions from terms with a $Y^{ij,kl}$, $Y_\a{}^{\b ij}$ or $Y_{\a\b}{}^{\g\d}$ of the form $Y^2$, $\nabla_a\nabla_b Y$, $W^{\a\b ij} \nabla_a Y$, $Y \nabla_a W^{\a\b ij}$ and $W^{\a\b ij} W^{\g\d kl} Y$. If one $Y$ is required to be $Y^{ij,kl}$, the only possibilities are the two terms discussed above; the $\SL(4,\bbC)$-scalar requirement is too restrictive for the others. If one $Y$ is required to be $Y_{\a}{}^{\b ij}$, the only term which contributes is $Y_\a{}^{\b k[i} Y_\b{}^{\a j]l}$; satisfying the $\Sp(4,\bbC)$ irrep and the $\SL(4,\bbC)$-scalar is too restrictive for others. In particular, we need only minus away traces from this term to put it in the correct $\Sp(4,\bbC)$ irrep. If one $Y$ is required to be $Y_{\a\b}{}^{\g\d}$, then no contributions are found; the only possibilities with four $\Sp(4,\bbC)$ indices are $W^{\a\b ij} W^{\g\d kl} Y_{\e\z}{}^{\k\t}$ and $Y^{ij,kl} Y_{\e\z}{}^{\k\t}$. However, no $\SL(4,\bbC)$-scalar can be made out of such terms.

The remaining terms must be built using only vector derivatives $\nabla_a$ and $W^{\a\b ij}$. Representation theory guarantees that terms schematically of the form $(\nabla_aW)(\nabla_bW)$, $W\nabla_a\nabla_b W$ and $W^4$ will contribute. In particular, the $\SL(4,\bbC)$ scalars of such terms are unique. Also, the $\Sp(4,\bbC)$ four-index traceless irrep is also unique; this is quite easily seen for the first two terms, but the $W^4$ is not as obvious at first glance. 
The restriction that guarantees that $W^4$ generates the desired $\Sp(4,\bbC)$ irrep in a unique way comes from the $\Sp(4,\bbC)$ identity 
\begin{equation}
 W^{\g\q l[i} W^{\d\eta j]}{}_l = \frac{1}{4} \O^{ij} W^{\g\q kl} W^{\d\eta}{}_{kl}
\,.
\end{equation}
In particular, the unique $\SL(4,\bbC)$-scalar part of $W^4$ means we should contract this with a Levi-Civita symbol to get
\begin{equation}
\ve_{\a\b\q\eta} W^{\g\q l[i} W^{\d\eta j]}{}_l = \frac{1}{4} \O^{ij} \ve_{\a\b\q\eta} W^{\g\q kl} W^{\d\eta}{}_{kl}
\,.
\end{equation}
Note that antisymmetrising in $i,j$ is equivalent to antisymmetrising in $\g,\d$ for the LHS of the above equation.
Using condensed notation and noting that all appropriate pairs of $\SL(4,\bbC)$ indices, which we drop for simplicity below, are antisymmetrised (so that all appropriate pairs of $\Sp(4,\bbC)$ indices are antisymmetrised), then
\begin{align}
W^{i}{}_p W^{jp} W^{k}{}_q W^{lq} &\equiv \O^{ij} \O^{kl} (W_{pq} W^{pq})^2
\,,\\
W^{pq} W^{i}{}_p W^{j}{}_q W^{kl} &\equiv W^{ij} W^{kl} W_{pq} W^{pq}
\,.
\end{align}
Hence, when projecting onto the $\Sp(4,\bbC)$ irrep, we only see one independent contribution from $W^{ij}W^{kl} W_{pq} W^{pq}$.

\subsection{\texorpdfstring{Truncation to 6D $\cN = (1,0)$}{Truncation to 6D N=(1,0)}}\label{sect:truncationto6DN=(1,0)}
The truncation results for the standard Weyl multiplet in components are all in Section~\ref{sect:truncationto6DN=(1,0)2}. Here we will apply these results to analyse the truncation of the Bach tensor. The basic procedure is simple. We first project our $\cN = (2,0)$ Bach tensor to components, then use the truncation rules \eqref{truncationrules3}, with the additional understanding that the truncated $(2,0)$ covariant derivatives $\nabla_a$ are replaced with the $(1,0)$ truncated frame covariant derivatives $\hnabla_a$. We again abuse notation and denote the component field $B^{ij,kl}|$ as $B^{ij,kl}$ and the same for the other covariant fields.

We stress that this procedure is a truncation, with key objects the $(2,0)$ and $(1,0)$ Weyl multiplets, and not just a multiplet reduction, as many parts of various $(2,0)$ fields have been set to zero. Correspondingly, the truncation of a given $(2,0)$ field is only defined if it can be expressed in terms of the $(2,0)$ Weyl multiplet. The $(2,0)$ Bach tensor $B^{ij,kl}$ has a truncation determined by our ansatz \eqref{(2,0)BachTensor}. In particular, if we consider the RHS, we can see that only zero, two or four indices may have a primed index. If there is an odd number of primed indices, then it immediately vanishes due to the truncation rules  \eqref{truncationrules3}. Hence, the same must hold for $B^{ij,kl}$ itself. Up to some irrelevant overall scaling, we find that it must have the truncation rule
\begin{subequations}\label{BachTensorTruncation}
\begin{alignat}{2}
B^{ij}{}_{kl} &= - \eps^{ij} \eps_{kl} B \,, &\quad
B^{ij}{}_{k'l'} &= \eps^{ij} \eps_{k'l'} B \,, \\
B^{ij'}{}_{kl'} &= - \hf \d^i_k \d^{j'}_{l'} B  \,,&\quad
B^{i'j'}{}_{k'l'} &= - \eps^{i'j'} \eps_{k'l'} B \,.
\end{alignat}
\end{subequations}
Indeed, the truncation enforces $B^{ij,kl} = \ve^{ij} \ve^{kl} B$ up to scaling. The rest are fixed by symmetries as we discussed in Section~\ref{sect:truncationto6DN=(1,0)2}, e.g.\ $B^{ij,k'l'}$ is fixed by $\USp(4)$-tracelessness and the block-diagonal decomposition of $\O^{ij}$ into $\ve^{ij}$ and $\ve^{i'j'}$ in \eqref{omegatruncation}.

With this in mind, to compare to the $(1,0)$ Bach tensor in \eqref{(1,0)eq1} one can take the $(2,0)$ Bach tensor \eqref{(2,0)BachTensor} and use \eqref{BachTensorTruncation}. Since $\O_{i j} \O_{k l} B^{ij,kl} = 0$ vanishes, there is no point computing this. We instead take \eqref{(2,0)BachTensor} and project the four $\USp(4)$ indices $i,j,k,l=1,2,3,4$ onto just $i,j,k,l=1,2$ and use the truncation formula $B^{ij,kl} = \ve^{ij} \ve^{kl} B$.\footnote{This is a choice, but any other projection will amount to the same $(1,0)$ truncation of the equation. Indeed, the rest of the truncation rules of $B^{ij,kl}$ are determined by the first rule $B^{ij,kl} = \ve^{ij} \ve^{kl} B$ by using symmetries of $B^{ij,kl}$.} We then contract both sides with $(1/4) \ve_{ij} \ve_{kl}$ and the LHS becomes $B$ whereas the RHS needs to be evaluated using the truncation rules of the Weyl multiplet \eqref{truncationrules3} and the $(1,0)$ truncated frame covariant derivatives $\hnabla_a$.

In order to complete the computations, one needs the exact expansion of the traces being minused away in the $(2,0)$ Bach tensor ansatz \eqref{(2,0)BachTensor}. This can be done in a combined way. First, it is useful to note the $\Sp(4,\bbC)$ decomposition
\begin{align}
\ytableausetup{smalltableaux} 
\ydiagram{2,2} = \ydiagram{2,2}^{\, {\rm trcls}} + \ydiagram{1,1}
\,.
\end{align}
This means that after projecting into the Young tableau of $B^{ij,kl}$, we will need to subtract traces determined by a single 2-form. Let $A^{ij,kl}$ represent such a term that has been projected, then it has a 2-form trace $A^{ij} = A^{ij,l}{}_l$ and also a scalar trace $A = A^i{}_i$. To find the traceless irrep of $A^{ij,kl}$ that actually appears in the ansatz of $B^{ij,kl}$ \eqref{(2,0)BachTensor}, we use the projector of $\ydiagram{2,2}$ on the `pure trace' terms $\O^{i j} A^{k l}$ and $\O^{i j} \O^{k l} A$ and add them together with two undetermined coefficients.\footnote{ Recall this projector is proportional to symmetrising over $i,k$, symmetrising over $j,l$, then antisymmetrising over $i,j$ and antisymmetrising over $k,l$.} The coefficients are fixed by setting the traces to zero --- it is sufficient to set the trace over $i,j$ to zero. We find that contributions to $B^{ij,kl}$ take the form
\begin{align}
    \ytableausetup{nosmalltableaux}
\begin{ytableau}
    i&k\\
    j&l
\end{ytableau}^{\,{\rm trcls}}
    &= A^{ij,kl} + \frac{1}{6} ( \O^{ij} A^{kl} + \O^{kl} A^{ij} + \O^{l[i} A^{j]k} - \O^{k[i} A^{j]l} ) 
    \nonumber\\&\quad
    + \frac{1}{30} ( \O^{ij} \O^{kl} - \O^{k[i} \O^{j]l} ) A
    \,.
\end{align}
Consider the terms of \eqref{(2,0)BachTensor} with coefficients $b_2,b_5,b_7$, we can describe them using $T^{ij}$, a traceless 2-form. The projection of $T^{ij}T^{kl}$ onto the Young diagram of $B^{ij,kl}$ is proportional to
\begin{equation}
A^{ij,kl} = T^{ij}T^{kl} - T^{k[i}T^{j]l}
\,.
\end{equation}
We have $A^{ij} = T^{il} T^{j}{}_l$ and $A = T^{ij} T_{ij}$.
Consider the terms of \eqref{(2,0)BachTensor} with coefficients $b_3,b_6$, we can describe them using $H^{ij}T^{kl}$ with both $H^{ij}$ and $T^{ij}$ traceless 2-forms. The Young diagram projection is proportional to
\begin{equation}
A^{ij,kl} = H^{ij}T^{kl} + H^{kl}T^{ij} + H^{l[i} T^{j]k} - H^{k[i} T^{j]l}
\,.
\end{equation}
We have $A^{ij} = 2 H^{[i|l|} T^{j]}{}_l$ and $A = 2 H^{ij} T_{ij}$. 
Consider the term of \eqref{(2,0)BachTensor} with coefficient $b_4$, we can describe it using $G^{ij}$, a symmetric tensor. The projection of $G^{ij}G^{kl}$ onto the Young diagram is proportional to
\begin{equation}
A^{ij,kl} = G^{k[i}G^{j]l}
\,.
\end{equation}
We have $A^{ij} = G^{il} G^{j}{}_l$ and $A = G^{ij} G_{ij}$. These three cases encapsulate all the terms in the ansatz aside from $b_1$, which can readily be dealt with.
In the truncation, we are essentially computing 
\begin{align}
    &\frac{1}{4} \ve_{ij} \ve_{kl}\bigg(  A^{ij,kl} + \frac{1}{6} ( \O^{ij} A^{kl} + \O^{kl} A^{ij} + \O^{l[i} A^{j]k} - \O^{k[i} A^{j]l} ) 
    \nonumber\\&
    + \frac{1}{30} ( \O^{ij} \O^{kl} - \O^{k[i} \O^{j]l} ) A\bigg)
=
\frac{1}{4} \ve_{ij} \ve_{kl} A^{ij,kl} - \frac{1}{4} \ve_{kl} A^{kl} + \frac{1}{20} A
\,,
\end{align}
with the understanding that the $\USp(4)$ indices $i,j,k,l=1,2,3,4$ have been projected to $i,j,k,l=1,2$ in the same way we projected the $(2,0)$ Bach tensor. One just needs to substitute the appropriate expansions of $A^{ij,kl}$, $A^{ij}$ and $A$ in terms of the Weyl multiplet that appear in the $(2,0)$ Bach tensor ansatz \eqref{(2,0)BachTensor}

In the end, our truncation gives the result
\begin{align}
B &= 
80b_1 \hnabla^2 Y
+ \frac{9}{2}(80^2) b_2 Y^2
+ \frac{9}{4}(40^2) \ri\, b_3   X^i \slashed{\hnabla} X_i
+ \frac{1}{20} (96^2) b_4 Y_{\a}{}^{\b i j} Y_{\b}{}^{\a}{}_{ij}
  \nonumber\\&\quad
 - \frac{6}{5} (4^2) b_5 \hnabla_{\a \b} W^{\a \g}  \hnabla_{\g \d} W^{\b \d}
  -\frac{12}{5} (4^2) b_6 W^{\a \g} \hnabla_{\a \b}\hnabla_{\g \d} W^{\b \d }
 \nonumber\\&\quad
 +  \frac{24}{5}(4^4) b_7  \ve_{\a_1 \a_2 \a_3 \a_4} \ve_{\b_1 \b_2 \b_3 \b_4}  W^{\a_1 \b_1} W^{\a_2 \b_2} W^{\a_3 \b_3} W^{\a_4 \b_4}
 \,.
\end{align}
Comparing the previous equation with the 6D $\cN=(1,0)$ Bach tensor \eqref{(1,0)eq1} and its coefficients \eqref{(1,0)eq3}, one uniquely obtains the component 6D $\cN=(2,0)$ Bach tensor \eqref{(2,0)BachTensor} with coefficients (\ref{Bach-constraints},~\ref{Bach-constraints2}). This readily lifts back to superspace.

\subsection*{Acknowledgments}
We are grateful to Lorenzo Casarin, Sergei Kuzenko and Emmanouil Raptakis for discussions. 
We are also grateful to Gregory Gold and Saurish Khandelwal for discussions and for their open-source \emph{Cadabra} codes \cite{Gold:2024git,Gold:2024nbw}. 
Many of the calculations presented in this paper have been carried out by using \emph{Cadabra} \cite{Peeters:2006,Peeters:2007,Peeters:2018}.
 We acknowledge the kind hospitality and financial support at the MATRIX Program “New Deformations of Quantum Field and Gravity Theories,” which took place in Creswick (Australia) between
22 January and 2 February 2024, and we thank the participants of the workshop for stimulating discussions related to this paper.
CK and GT-M acknowledge the kind hospitality and financial support extended to them at the 2025 ANZAMP Meeting, which took place in Bendigo (Australia) between 11 and 13 Feb 2025.
CK is supported by a postgraduate scholarship at the University of Queensland.
GT-M has been supported by the Australian Research Council (ARC) Future Fellowship FT180100353, ARC Discovery Project DP240101409, the Capacity Building Package at the University of Queensland and a faculty start-up funding of UQ's School of Mathematics and Physics.

\appendix
\section{Notation and conventions}
\label{appendix:notationsandconventions}

\subsection{Gamma matrix and symplectic algebra}\label{appendix:gammamatrixandsymplecticalgebra}
Our flat metric is $\eta_{ab} = {\rm diag}(-1,1,1,1,1,1)$ and our flat Levi-Civita symbols $\ve_{abcdef}$ are $\ve_{012345} = 1 = -\ve^{012345}$.
All of our conventions mostly follow \cite{Butter:2016}, but we include some useful formulae of theirs. 
The only difference in conventions is that we use $\USp(4)$ instead of $\SU(2)$. In particular, our $\USp(4)$ symplectic form can be taken without loss of generality\footnote{In all of our pure 6D $\cN=(2,0)$ calculations, the actual form of $\Omega^{ij}$ is not constrained. This particular form of $\Omega^{ij}$ is convenient to implement the symplectic Majorana--Weyl condition and readily realise $\SU(2) \times \SU(2)$ as a subgroup, making truncation to $(1,0)$ easier with its $\SU(2)$ R-symmetry. We comment that just enforcing the symplectic Majorana--Weyl condition with $\USp(4)$ or $\SU(2)$ only requires that the matrix representation of the symplectic form obeys $\Omega\Omega^* = -1$ or $\ve\ve^* = -1$ respectively.} to be
\begin{equation}\label{SympleticFormDef}
\Omega^{ij} = \begin{pmatrix}
\varepsilon^{ij} & 0 \\
0 & \varepsilon^{i'j'}
\end{pmatrix},
\qquad
\Omega_{ij} = \begin{pmatrix}
\varepsilon_{ij} & 0 \\
0 & \varepsilon_{i'j'}
\end{pmatrix},
\qquad
\O_{i j} \O^{j k} = \d_{i}^{k}
\,,
\end{equation}
where we use the convention of \cite{Butter:2016} that the $\SU(2)$ symplectic form satifies $\varepsilon^{12} = -\varepsilon_{12} = 1$. This makes it easy to see that our $\cN=(2,0)$ superconformal algebra reduces immediately to the $\cN=(1,0)$ superconformal algebra of \cite{Butter:2016} upon replacing $\O^{ij},\O_{ij}$ with $\e^{ij},\e_{ij}$. Our raising and lowering conventions also match \cite{Butter:2016}
\begin{align}
    v^i = \O^{ij} v_j
    \,,\qquad
    v_i = \O_{ij} v^j
    \,.
\end{align}

In 6D, Weyl spinors live in $\bbC^4$ and can be endowed with a Lorentz covariant charge conjugation. This charge conjugation is given by a Lorentz covariant quaternionic structure on $\bbC^4$. In particular, one can view this in terms of a conjugate-linear map $J$, with its matrix representation satisfying $JJ^* = -1$. This $J$ is inherently built into the Spin group
\begin{align}
    \Spin(1,5) \cong \SL(2,\bbH) \cong \SU^*(4) = \{ U \in {\rm Mat}_4(\bbC) \mid U^*J = J U \text{ and } \det(U)=1\}
    \,.
\end{align}
The map $J$ can essentially be identified with the matrix $B$ in Appendix~A of \cite{Butter:2016} which obeys $\Gamma_a = B (\Gamma_a)^* B^{-1}$ and $BB^* = -1$.\footnote{A technicality is that $B$ is a bilinear form, but $J$ is a conjugate-linear map. They are essentially related by $B(\psi,\chi) = \langle J(\psi),\chi\rangle$ where $\langle -,-\rangle$ is a complex inner product. Note that one needs to extend the action of $J$ from Weyl spinors to Dirac spinors.} Here, $\Gamma_a$ denotes the Dirac gamma matrices. This is not immediate from the definition of $\SU^*(4)$ and instead needs to be developed from realising an embedding $\SU^*(4) \to {\rm Cl}(1,5)$ into the Clifford algebra as the Spin group $\Spin(1,5) \subseteq {\rm Cl}(1,5)$. The standard analysis of gamma matrices in Appendix~A of \cite{Butter:2016} effectively does this.

The main point is that we take charge conjugation conventions matching \cite{Butter:2016}, and this amounts to having $\overline{\Gamma_a} = \Gamma_a$ and the same for the Weyl gamma matrices. The other charge conjugation convention taken in \cite{Butter:2016} is given by using $\SU(2)$ indices and setting
\begin{align}\label{SMConditionSU2}
    (\psi^{\a}_i)^c = \overline{\psi^{\a}_i} = - \ve^{i j} \psi_j^\a =- \psi^{\a i}
    \,,\qquad
    (\chi_{\a i})^c = \overline{\chi_{\a i}} = \ve^{i j} \chi_{\a j} = \chi_\a^i
    \,.
\end{align}
One can view this as embedding a Weyl spinor as $\psi^{\a}_i = (\overline{\psi^{\a}},\psi^\a)$, $\chi_{\a i} = (\chi_\a, \overline{\chi_{\a}})$, which itself is a freedom.\footnote{That is, one could of course take $\psi^{\a}_i = (\psi^{\a},\overline{\psi^\a})$ and flip the sign above, but this will change reality conditions --- like those of the superconformal algebra. We are not interested in doing this, so we stick to the above convention.
}
This readily extends to the $\USp(4)$ case since $\O^{ij}$ is built by $\ve^{ij}$ in diagonal blocks. We simply insert one more independent Weyl spinor, e.g.\ $\psi^{\a}_i = (\overline{\psi^{\a}},\psi^\a,\overline{\psi'^{\a}},\psi'^\a)$ and have
\begin{align}\label{SMCondition}
    (\psi^{\a}_i)^c = \overline{\psi^{\a}_i} = - \O^{i j} \psi_j^\a =- \psi^{\a i}
    \,,\qquad
    (\chi_{\a i})^c = \overline{\chi_{\a i}} = \O^{i j} \chi_{\a j} = \chi_\a^i
    \,.
\end{align}
We can evaluate $\psi^{\a}_i = (\overline{\psi^{\a}},\psi^\a,\overline{\psi'^{\a}},\psi'^\a)$ and obtain $\overline{\overline{\psi^{\a}_i}} = - \psi_i^\a$. This enforces that our matrix representation of $\O$ must obey $\O\O^* = -1$ or $\sum_j \O^{ij} (\O^{jk})^* = \d^{ik}$.
Our choice satisfies this constraint and has
\begin{align}
    (\O^{ij})^* = \O^{ij} = -\O_{ij}
    \,.
\end{align}
This is again matched by $\SU(2)$ since $(\ve^{ij})^* = \ve^{ij} = -\ve_{ij}$.

For products of spinors one needs to consider conjugation of the (odd parity) Grassmann numbers that spinors are implicitly valued in.\footnote{This is true both for spinor superfields on $\mathcal{M}^{6|16}$ and spinor fields on $\mathcal{M}^6$. One understands that spinor superfields have Grassmann contributions from the fermionic coordinates $\theta$ of $\mathcal{M}^{6|16}$, but also an independent set of Grassmann numbers, say $\theta'$. We then require that only odd products of $\theta$ and $\theta'$ contribute, to fix the overall parity of a spinor to be odd. Spinor fields simply have odd $\theta'$ contributions.} We find that \eqref{SMCondition} then implies that the product of two Weyl spinors satisfies
\begin{subequations}
    \begin{align}
            \overline{\psi^\a_i \chi_{\b j}} &= \psi^{\a i} \chi_\b^j
    \,,\\
        \overline{\psi_{\a i} \chi_{\b j}} &=  - \psi_{\a}^i \chi_{\b}^j
    \,,\\
        \overline{\psi^{\a}_{i} \chi^{\b}_{j}} &=  - \psi^{\a i} \chi^{\b j}
    \,.
    \end{align}
\end{subequations}
Hence, we have the special cases
\begin{equation}
    \overline{\psi^{\a i} \chi_{\a i}} = \psi^{\a i} \chi_{\a i}
    \,,\qquad
    \overline{\psi^{\a}_i \chi_{\a j}} = \psi^{\a i} \chi_{\a}^j
    \,.
\end{equation}
In particular, for bosonic quantities built from spinors, we will in general have (with potential spinor indices omitted)
\begin{align}
    \overline{B_{ij}} = \pm B^{ij}
    \,,
\end{align}
depending on how the spinors contribute. For bosonic quantities not built from spinors, consistency requires the same reality conditions. Moreover, all the reality conditions are enforced by conjugate-linear maps, hence multiplication by $\ri$ will flip the sign of a given reality condition. An important point is that these reality conditions for bosonic fields represent a real structure --- applying it twice gives the identity. For objects with no spinor indices, it is genuine complex conjugation so that $\overline{B_{ij}} = (B_{ij})^*$. This holds, for example, with $\overline{\O_{ij}} = (\O_{ij})^*$ and $\overline{\O^{ij}} = (\O^{ij})^*$. The same thing holds for the $\SU(2)$ case with $\ve^{ij}$. We now list the reality conditions of the (gauged) superconformal algebra generators and the important superfields we make use of.
For the algebra, we have
\begin{subequations}
    \begin{alignat}{4}
    \overline{M_{ab}} &= M_{ab}
    \,,&\qquad
    \overline{\nabla_a} &= \nabla_a
    \,,&\qquad
    \overline{\nabla_\a^i} &= -\nabla_{\a i}
    \,,&\qquad
    \overline{\bbD} &= \bbD
    \,,\\
    \overline{J^{ij}} &= J_{ij}
    \,,&\qquad
    \overline{S_i^\a} &= - S^{\a i}
    \,,&\qquad
    \overline{K_a} &= K_a
    \,.
\end{alignat}
\end{subequations}
For the superfields, we have
\begin{subequations}
    \begin{alignat}{3}
    \overline{W_{abc}{}^{ij}} &= - W_{abc \, ij}
    \,,&\qquad
    \overline{X^{\a i,jk}} &= X^{\a}{}_{i,jk}
    \,,&\qquad
    \overline{X_{\a}{}^{\b\g i}} &= X_{\a}{}^{\b\g}{}_i
    \,,\\
    \overline{Y_{\a\b}{}^{\g\d}} &= Y_{\a\b}{}^{\g\d}
    \,,&\qquad
    \overline{Y_{\a}{}^{\b ij}} &= Y_{\a}{}^{\b}{}_{ij}
    \,,&\qquad
    \overline{Y^{ij,kl}} &= Y_{ij,kl}
    \,.
\end{alignat}
\end{subequations}

We now detail some algebraic facts about our $4 \times 4$ Weyl gamma matrices $(\gamma^a)_{\alpha\beta}$ 
and $(\tilde\gamma^a)^{\alpha\beta}$. They are antisymmetric and related by
\begin{align}
(\tilde\gamma^a)^{\alpha\beta} = \frac{1}{2} \ve^{\a\b\g\d} (\gamma^a)_{\g\d}
\,,\qquad
(\gamma^a)^* = \tilde\gamma_a
\,,\qquad
\overline{\gamma^a} = \gamma^a
\,,\qquad
\overline{\tilde{\gamma}^a} = \tilde{\gamma}^a
\,.
\end{align}
where $\ve^{\a\b\g\d}$ is the Levi-Civita symbol of $\Spin(1,5) \cong \SU^*(4)$ whose $4 \times 4$ matrices have determinant one and hence preserve top forms. 
They obey
\begin{subequations}
\begin{align}
(\g^a)_{\a\b} (\tilde{\g}^b)^{\b\g}
+ (\g^b)_{\a\b} (\tilde{\g}^a)^{\b\g} &= - 2 \eta^{ab} \d^\g_\a \ , \\
(\tilde{\g}^a)^{\a\b} (\g^b)_{\b\g}
+ (\tilde{\g}^b)^{\a\b} (\g^a)_{\b\g} 
&= - 2 \eta^{ab} \d^\a_\g \ .
\end{align}
\end{subequations}
The completeness relations are
\begin{subequations}
\begin{align}
(\gamma^a)_{\a\b} (\tilde\gamma_{a})^{\g\d} &= 4\, \delta_{[\a}{}^\g \delta_{\b]}{}^{\d}\,, \\
(\gamma^{ab})_\a{}^\b (\gamma_{ab})_\g{}^\d &= - 8\,\delta_{\a}{}^\d \delta_{\g}{}^{\b}
	+ 2\, \delta_{\a}{}^\b \delta_{\g}{}^{\d}\,, \\
(\gamma^{abc})_{\a\b} (\tilde\gamma_{abc})^{\g\d} &= 48\, \delta_{(\a}{}^\g \delta_{\b)}{}^{\d}\,, \\
(\gamma^{abc})_{\a\b} (\tilde\gamma_{abc})_{\g\d} &= (\gamma^{abc})^{\a\b} (\tilde\gamma_{abc})^{\g\d} = 0\,.
\end{align}
\end{subequations}
Using these, we can establish natural isomorphisms between tensors of $\SO(1,5)$ built from $\bbC^6$ and those of $\SU^*(4)$ built from $\bbC^4$.
Vectors $V^a$ and antisymmetric matrices $V_{\a\b} = - V_{\b\a}$ 
\begin{equation}
V_{\a\b} := (\g^a)_{\a\b} V_a \quad \Longleftrightarrow  \quad V_a = \frac{1}{4} (\tilde{\g}_a)^{\a\b} V_{\a\b} \ .
\end{equation}
2-forms $F_{ab}$ are related to traceless matrices $F_\a{}^\b$ 
via
\begin{equation}
F_\a{}^\b := - \frac{1}{4} (\g^{ab})_\a{}^\b F_{ab} \ , \quad F_\a{}^\a = 0 \quad
\Longleftrightarrow 
\quad F_{ab} = \hf (\g_{ab})_\b{}^\a F_\a{}^\b = - F_{ba} \ .
\end{equation}
Self-dual and anti-self-dual 3-forms $T^{(\pm)}_{abc}$,
\begin{equation}
(*T^{(\pm)})^{abc} = \frac{1}{3!} \eps^{abcdef} T_{def}^{(\pm)}  = \pm T^{(\pm)abc} \ ,
\end{equation}
are related to symmetric matrices $T_{\a\b}$ and $T^{\a\b}$ 
via
\begin{subequations}
\begin{align}
T_{\a\b} &:= \frac{1}{3!} (\g^{abc})_{\a\b} T_{abc} = T_{\b\a} \quad \Longleftrightarrow \quad 
T_{abc}^{(+)} = \frac{1}{8} (\tilde{\g}_{abc})^{\a\b} T_{\a\b} \ , \\
T^{\a\b} &:= \frac{1}{3!} (\tilde{\g}^{abc})^{\a\b} T_{abc} = T^{\b\a} \quad 
\Longleftrightarrow \quad
T^{(-)}_{abc} = \frac{1}{8} (\g_{abc})_{\a\b} T^{\a\b} \ .
\end{align}
\end{subequations}
Further irreps of $\Spin(1,5)$ take particularly
simple forms when written with $\bbC^4$ spinor indices. For example, a gamma-traceless
left-handed spinor two-form $\Psi_{ab}{}^{\g}$ is related to a symmetric traceless
$\Psi_\a{}^{\b\g}$, 
\begin{align}
\Psi_\a{}^{\b \g} &:= - \frac{1}{4} (\g^{ab})_\a{}^\b \Psi_{ab}{}^\g
	= \Psi_\a{}^{\g \b}\ , \quad \Psi_\a{}^{\a\g} = 0 \quad
\Longleftrightarrow \nonumber\\
\Psi_{ab}{}^\g &= \hf (\g_{ab})_\b{}^\a \Psi_\a{}^{\b \g} \ ,\quad
(\gamma^b)_{\d\g} \Psi_{ab}{}^\g = 0\,,
\end{align}
and a rank-four tensor $C_{abcd}$ with the symmetries of the Weyl
tensor is related to a symmetric traceless $C_{\a\g}{}^{\b\d}$ 
via
\begin{align}
C_{\a\g}{}^{\b\d} &:= \frac{1}{16} (\g^{ab})_\a{}^\b (\g^{cd})_\g{}^\d \, C_{abcd}
	= C_{(\a\g)}{}^{(\b\d)}\,, \qquad C_{\a\g}{}^{\b\g} = 0
	\quad \Longleftrightarrow \nonumber\\
C_{abcd} &= \frac{1}{4} (\g_{ab})_\b{}^\a (\g_{cd})_\d{}^\g\, C_{\a\g}{}^{\b\d}
	= C_{[cd] [ab]}\,, \qquad
	C_{[abc]d} = 0~.
\end{align}

\subsection{Superforms}\label{appendix:superforms}
We work only with even superforms, i.e.\ the superform $J$ itself is even, but its components are both even and odd dependent on the basis vectors $dz^M$. Exceptions only occur if $J$ is vector-valued, at which point one must be careful of this grading. For example, $T^A J^{\ub C} =(-1)^{A(\ub + C)} J^{\ub C} T^A$, where we abuse notation and use the index itself to denote the grading --- all objects are assumed to have grading determined by their indices, and if no indices are present, then they are even-graded. 
None of the below conventions depend upon whether the form is vector-valued or not. Recall our local coordinates are $z^M = (x^m, \theta_i^\mu)$.
Given a frame/supervielbein $E_M{}^A$ and its inverse $E_A{}^M$ we have 1-forms and vector fields given by
\begin{align}
    E^A = dz^M E_M{}^A
    \,,\quad
    E_A = E_A{}^M \partial_M
    \,.
\end{align}

\subsubsection{Superform components}
The coordinate and in-frame components of an $n$-form are defined by
\begin{align}
J &= \frac{1}{n!} dz^{M_n} \wedge \dots \wedge dz^{M_1} J_{M_1 \dots M_n}
\\
&=
\frac{1}{n!} E^{A_n} \wedge \dots \wedge E^{A_1} J_{A_1 \dots A_n}
\,.
\end{align}
A key point is that, for 1-forms we naturally have
\begin{equation}
J = E^A J_A = dz^M E_M{}^A J_A = dz^M J_M
\end{equation}
so that $J_M = E_M{}^A J_A$. For $n$-forms with $n>1$
\begin{equation}
J_{M_1 \dots M_n} = (-1)^\rho E_{M_n}{}^{A_n} \dots E_{M_1}{}^{A_1} J_{A_1 \dots A_n} 
\end{equation}
for a non-trivial grading sign $\rho = \sum_{i=1}^n \rho_i$ with $\rho_1 = 0$ and
\begin{equation}
\rho_i = (M_i + A_i) \sum_{j=1}^{n-1} M_j
\,.
\end{equation}

\subsubsection{Exterior derivatives}
The exterior derivative is defined in coordinates in the usual way, but it takes into account the rank of a superform via a sign. That is, on components of $n$-forms
\begin{equation}
d = (-1)^n dz^M \partial_M = (-1)^n E^A E_A
\,.
\end{equation}
We take this convention to ensure that its action on $n$-forms is natural, i.e.\,
\begin{align}
dJ 
&=
\frac{1}{(n+1)!} dz^{M_{n+1}} \wedge \dots \wedge dz^{M_2} \wedge dz^{M_1} \left[ (n+1) \partial_{M_1} J_{M_2 \dots M_{n+1}} \right]
\,,
\end{align}
so that the components of $dJ$ have no signs 
\begin{align}
(dJ)_{M_1 \dots M_{n+1}} = (n+1) \partial_{[M_1} J_{M_2 \dots M_{n+1}]}
\,.
\end{align}
We still have $d^2=0$ as $\partial_{[M} \partial_{N]} = 0$. In this convention, the product rule takes the form
\begin{equation}\label{dproductrule}
d(J\wedge J') = J \wedge d J' + (-1)^{\text{rank}(J')} dJ \wedge J'
\,.
\end{equation} 
If an exterior covariant derivative $\nabla$ with connection 1-form $\omega = \omega^{\ua} X_{\ua}$ is used (with any structure group), some care is necessary. When acting on $n$-forms, the operator takes the form
\begin{align}
\nabla &= (-1)^n dz^M \nabla_M
=
 (-1)^n dz^M ( \partial_M - \omega_M)
=
d - (-1)^n \omega
\,. 
\end{align}
The same rules that held for $d$ will now hold for $\nabla$. In particular, in the $\nabla$ case, we require the forms to be covariant in the sense that $X_{\ua} J$ is well-defined. The wedge product of vector-valued forms is given by the usual wedge product along with a tensor product of the vector spaces they are valued in.

\subsubsection{Interior products}
One of the nice results about differential forms is that their Lie derivatives satisfy Cartan's formula
\begin{equation}
\mathcal{L}_X J = (d\iota_X + \iota_Xd)J
\,.
\end{equation}
Given that we have adjusted the typical convention of exterior derivatives and now have the rule \eqref{dproductrule}, we have to employ the same convention for the interior product, otherwise Cartan's formula will fail. We define for 1-forms
\begin{equation}
\iota_X J = X(J) = X^M \partial_M (dz^N J_N) = X^M J_M
\,,
\end{equation}
and we extend to $n$-forms via
\begin{equation}
\iota_{X}(J\wedge J') = J \wedge \iota_X J' + (-1)^{\text{rank}(J')} \iota_X J \wedge J'
\,.
\end{equation}
In particular, for a 2-form
\begin{equation}
\iota_X J = - X^M dz^N J_{NM} = - X^A E^B J_{BA}
\,.
\end{equation}

\subsection{Conversion of conventions}
\label{appendix:conversionofconventions}

First it should be noted that we and \cite{Bergshoeff:1999} have the same flat metric and the same flat Levi-Civita symbols, $\eta_{ab} = {\rm diag}(-1,1,1,1,1,1)$ and $\ve_{012345} = 1 = -\ve^{012345}$, respectively. Our gamma matrices differ by a factor of $\pm\ri$.  We both (anti)symmetrise with unit weight, e.g.\ $2 T_{[ab]} = T_{ab} - T_{ba}$.

We take our Dirac gamma matrices $\Gamma_a$, our charge conjugation matrix\footnote{Note that the charge conjugation matrix we use is $C = \begin{pmatrix}
    0&I\\
    I&0
\end{pmatrix}$, matching \cite{Butter:2017}. As mentioned earlier in Appendix~\ref{appendix:gammamatrixandsymplecticalgebra} $C$, all our conventions match theirs.} and our symplectic form $\O^{ij}$ and compare them to \cite{Bergshoeff:1999}'s. Here we denote theirs as $\Gamma'_{a}$, $C'$ and $\O'^{ij}$.  Due to the use of the symplectic form to employ Dirac conjugation and charge conjugation, we ignore these in favour of dealing directly with the symplectic form. The only detectable change to $\Gamma_a$, $C$, $\O^{ij}$ are the rescalings below
\begin{align}
\G_a = \ri \e_\G \G'_a
\,,\quad
C= \e^{\ri}_C C'
\,,\quad
\O^{ij} = \e_\O \O'^{ij}
\,,\quad
\O_{ij} = -\e_\O \O'_{ij}
\,,
\end{align}
where $\e\equiv\pm1$ and $\e^\ri\equiv\pm1,\pm\ri$ have 2 and 4 possibilities respectively. The lower index symplectic form needs an extra minus sign because we lower $\USp(4)$ indices oppositely to \cite{Bergshoeff:1999}. We now do the same to our superconformal algebra. We need to be careful about $\USp(4)$ because $\O^{ij} = \e_\O \O'^{ij}$, e.g.~if $T_i = \l_T T'_i$, then $T^i = \l_T \e_\O T'^i$. Thus we fix our USp(4) indices in the positions $X_{\ua} = (\dots,Q_\a^i,S^{\a i}, J_{ij})$ before comparison. We then allow complex rescalings\footnote{For consistency with reality conditions each of these will inevitably be real or imaginary.}
\begin{align}
P_A = \l_{P_A} P'_A
\,,\quad
X_{\ua} = \l_{X_{\ua}} X'_{\ua}
\,.
\end{align}
The only thing we will fix is $\l_{P} = 1$ because $e_m{}^a P_a = \partial_m$ generates the usual partial derivative, and the vielbeins should be the same because our flat metrics are the same. With this fixed, one can compare covariant derivatives and curvatures too.
We also allow our matter fields to undergo complex rescaling as
\begin{subequations}
\begin{align}
W_{abc}{}^{ij} | = W_{abc}{}^{ij} = \lambda_W T_{abc}{}^{ij} 
\,,\\
X^{\a i,jk} | = X^{\a i,jk} = \lambda_X\chi^{\a i,jk}
\,,\\
Y^{ij,kl} | = Y^{ij,kl} = \lambda_YD^{ij,kl}
\,.
\end{align}
\end{subequations}
After comparison with \cite{Bergshoeff:1999}, one has four distinct choices which will result in the same superconformal algebra.\footnote{Once the algebra conversion is determined, there are really no other constraints one needs to impose. Any remaining freedom is a genuine freedom.} All choices stem from choosing $\e_\G,\e_\O,\e_Q$ such that
\begin{equation}
\e_\G \e_\O = -1
\,.
\end{equation}
The quantities depending on this choice are
\begin{align}
\l_Q &= 2 \e_Q
\,,\quad
\l_S = 2 \e_\G \e_Q
\,,\quad
\l_J = 2\e_\O
\,,\\
\l_W &= -2\ri \e_\O
\,,\quad
\l_X = - \frac{16}{3} \e_\G \e_\O \e_Q
\,.
\end{align}
The rescalings not written are fixed already, and they can be seen in Table~\ref{table:conversionofconventions}, where we also make the choice corresponding to the table in Appendix B of \cite{Butter:2017}
\begin{equation}
\e_\O =  1
\,,\quad
\e_\G = \e_Q = -1
\,.
\end{equation}
In Table~\ref{table:conversionofconventions}, we also replace the primed objects with their actual notations in \cite{Bergshoeff:1999}.
\begin{table*}[!htbp]
\centering
\renewcommand{\arraystretch}{1.4}
\resizebox{14cm}{!}{
\begin{tabular}{@{}cc@{}} \toprule
Our notation&  Bergshoeff et al. \cite{Bergshoeff:1999}  \\ \midrule
$\eta^{ab}$, 
$\ve^{abcdef}$, $\G^a$, $\G^{a_1\cdots a_n}$, $\G_*$, $C$, $\O^{ij}$, $\O_{i j}$ $\cdots$
&~~~~~~
$\eta^{ab}$,
$\ve^{abcdef}$, $-\ri \gamma^a$, $(-\ri)^n\g^{a_1\cdots a_n}$, $\g_7$, $C$, $\O^{ij}$, $-\O_{i j}$ $\cdots$
\\
\midrule
$P_a$, $K^a$, $Q^i$,  $S_i$ 
&~~~~~~
$P_a$, $K^a$, $-2Q^i$,  $2S_i$ 
 \\
$M_{ab}$,
$J_{ij}$,
$\mathbb D$
&~~~~~~
$-2M_{ab}$,
$2U_{ij}$,
$D$
  \\
\midrule
$e_m{}^a$,
$\hat{\mathfrak{f}}_m{}^a$,
$\psi_m{}^i$,
$\hat\phi_m{}^i$ 
& ~~~~~~
 $e_\mu{}^a$,
$f_\mu{}^a$,
 $\psi_\mu{}^i$,
 $\phi_\mu{}^i $  
 \\
$\hat\omega_m{}^{ab}$,
$\cV_m{}^{ij}$, 
$b_m$
&~~~~~~
$-\omega_\mu{}^{ab}$,
$\frac{1}{2} V_\mu{}^{ij}$,
$b_\mu$
 \\
\midrule
$\hat R(P)_{ab}{}^{c}$,
$\hat R(K)_{ab}{}_c$,
$\hat R(Q)_{ab}{}^i$,
$\hat R(S)_{ab}{}_i$ 
&~~~~~~
$R(P)_{ab}{}^{c}$,
$R(K)_{ab}{}_c$,
$\frac{1}{2} \hat R(Q)_{ab}{}^i$, 
$\frac{1}{2} \hat R(S)_{ab}{}_i$ 
\\
$\hat R(M)_{ab}{}^{cd}$,
$\hat R(J)_{ab}{}^{ij}$,
$\hat R(\mathbb D)_{ab}$,
&~~~~~~
$-R(M)_{ab}{}^{cd}$,
$\frac{1}{2} R(J)_{ab}{}^{ij}$,
$R(D)_{ab}$
\\
\midrule
$\x^a$,
$\l^a$,
$\x_i$,
$\eta^i$
&~~~~~~
$\x^a$,
$\L_K^a$,
$\frac{1}{2}\ve_i$,
$\frac{1}{2}\eta^i$
\\
$\l^{ab}$,
$\l^{ij}$,
$\s$
&~~~~~~
$-\ve^{ab}$,
$\frac{1}{2}\L^{ij}$,
$\L_D$
\\
\midrule
$W_{abc}{}^{ij}$,
$X^{i,jk}$,
$Y^{ij,kl}$
&~~~~~~
$-2\ri\,T_{abc}{}^{ij}$,
$- \frac{16}{3} \chi^{i,jk}$,
$\frac{32}{3} D^{ij,kl}$
\\
\bottomrule
\end{tabular}}
\caption{Translation of notation and conventions.}
\label{table:conversionofconventions}
\end{table*}

\newpage
\section{Representation theory}\label{appendix:reptheory}
We discuss some aspects of the representation theory of $\SL(n+1,\bbC)$ and $\Sp(2n,\bbC)$.

\subsection{Young diagrams to dimensions and highest weights}\label{appendix:reptheory-dimensions&highestweights}
To enable relatively easy conversion to other ways of denoting irreps, we give a method to convert the (traceless) Young diagrams into their respective highest weights of $\SL(n+1,\bbC)$ and $\Sp(2n,\bbC)$. 

We first give the dimensions and highest weights for $\SL(n+1,\bbC)$, then we give the formula for the scaling factor associated with Young symmetrisers for $\SL(n+1,\bbC)$. Then we finish by giving the dimensions and highest weights for $\Sp(2n,\bbC)$.

\subsubsection{\texorpdfstring{$\SL(n+1,\bbC)$ dimensions and highest weights}{SL(n+1,C) dimensions and highest weights}}
Let $\lambda=(\lambda_1,\dots,\lambda_n)$ denote a $\SL(n+1,\bbC)$-highest weight. The complex dimension of the complex $\SL(n+1,\bbC)$ irrep $V_\lambda$ furnished by a Young diagram is
\begin{equation}
\dim V_\lambda = \prod_{\text{boxes}} \frac{(n + 1) + d_{\text{box}}}{h_{\text{box}}}
\,,
\end{equation}
where $d_{\text{box}}$ is given by the following way of filling a diagram
\begin{equation}
\ytableausetup{nosmalltableaux}
\begin{ytableau}
~ & ~ & ~ \\
~ & ~ & ~ \\
~
\end{ytableau}
\longrightarrow
\begin{ytableau}
0 & 1 & 2 \\
-1 & 0 & 1 \\
-2
\end{ytableau}
\,,
\end{equation}
i.e.\ start with $0$ then add 1 for each box along a row, and minus 1 for each box down a column. The quantity $h_{\text{box}}$ is called the hook-length and is defined for each box by
\begin{equation}
h_{\text{box}} = \#(\text{boxes below}) + \#(\text{boxes to the right}) + 1
\,.
\end{equation}
The corresponding $\SL(n+1,\bbC)$-highest weight $\l = (\l_1,\dots,\l_n)$ of $V_\lambda$ depends upon its Young diagram in the following way\footnote{Note that a Young diagram for $\SL(n+1,\bbC)$ never has more than $(n+1)$-boxes in a column because an $(n+2)$-form vanishes for a $(n+1)$-dimensional vector space. Moreover, any diagram with columns of length $n+1$ can have these columns discarded, it is equivalent to a tensor product of the unique scalar representation with the remaining part of the diagram --- one way to see this is to note that $\SL(n+1,\bbC)$ preserves top forms.}
\begin{equation}
\l_i = \#(\text{columns of length $i$}) 
\,.
\end{equation}
For example, the defining representation of $\SL(n+1,\bbC)$ on $\bbC^{n+1}$ has weight $\l = (1,0,\dots,0)$. The dual representation $\L^n(\bbC^{n+1}) \cong (\bbC^{n+1})^*$ has weight $(0,\dots,0,1)$.

\subsubsection{The scaling factor of the Young symmetriser}
We now find a formula for the scaling factor associated with Young symmetrisers seen in the examples \eqref{YoungSymmetrisersExample}. The key is to explicitly leverage the role of the symmetric group in Young symmetrisers. This train of thought leads to Schur--Weyl duality. Prior to this we need to introduce the notion of integer partitions and a particular filling of the Young diagrams.

There is a bijection between Young diagrams with $k$ boxes and integer partitions of $k$, in the sense that the length of each row is an integer, and
\begin{align}
    k = \sum_{\text{rows}} (\text{length of row})
    \,.
\end{align}
For example, $\ytableausetup{smalltableaux}\ydiagram{2,1}$ is the integer partition $3 = 2 + 1$. The $\SL(n+1,\bbC)$-highest weights $\lambda$ also uniquely determine Young diagrams by telling us how many columns of each length appear, but in this case, we allow the number of boxes to vary. If we consider only diagrams with $k$ boxes, then we can equivalently denote it by the highest weight $\lambda=(\lambda_1,\dots,\lambda_n)$ or its integer partition $\mu = (\mu_1,\dots,\mu_n)$, $k = \sum_{i=1}^n \mu_i$. Young diagrams that have columns with length greater than $n+1$ naturally do not have an associated highest weight. Given this, we will now use the notation
\begin{equation}
    V_\mu = \begin{cases}
        V_\lambda\,,\quad&\text{if $\mu$ gives a valid diagram}
        \,,\\
        0\,,\quad&\text{otherwise}
        \,.
    \end{cases}
\end{equation}
We also introduce a filling of a diagram by integers in the following way
\begin{align}\label{FilledYoungDiagram}
\ytableausetup{nosmalltableaux}
    \ydiagram{3,3,1} \longrightarrow 
\begin{ytableau}
1 & 2 & 3 \\
4 & 5 & 6 \\
7
\end{ytableau}
\,.
\end{align}
 We now leverage Schur--Weyl duality and the symmetric group to find the scaling factor for Young symmetrisers. A simplified form of Schur--Weyl duality of $\SL(n+1,\bbC)$ and the symmetric group $S_k$ is a special fusion rule given by
\begin{align}
        (\bbC^{n+1})^{\otimes k} = \sum_\mu S^\mu \otimes V_\mu
    \,,
\end{align}
where the sum is over all integer partitions $\mu$ of $k$ and $S^\mu$ is a Specht module of $S_k$.

The Specht modules $S^\mu$ of the symmetric group $S_k$ are precisely its finite-dimensional complex irreps.
The Specht modules can be generated using Young symmetrisers on the group algebra $\bbC[S_k]$ of $S_k$ --- we now have to be more careful about whether the Young symmetriser is acting on $\bbC[S_k]$ or $(\bbC^{n+1})^{\otimes k}$ and the implicit conventions in each case (scaling, left or right action).
We will work with left actions only and deal with the scaling soon.

We note that the (complex) group algebra $\bbC[G]$ of a finite group $G$ is defined by setting the basis to be given by vectors $e_g$ for $g \in G$. The multiplication is defined by $e_ge_{g'} = e_{gg'}$.

When acting on $\bbC[S_k]$, the Young symmetriser is just an element of $\bbC[S_k]$ given by $c_\mu = b_\mu a_\mu$. To define $a_\mu$ and $b_\mu$, we need to use $\mu$ to fix subgroups of $S_k$. Recall that the symmetric group acts on $1,\dots,k$ by permutations and consider the filling of the Young diagram $\mu$ described in \eqref{FilledYoungDiagram}. Define the subgroups $P_\mu,Q_\mu$ of $S_k$ by\footnote{By preservation, we mean that $\sigma \in P_\mu$ will only permute the numbers in a row with themselves, and this holds for all rows. In particular, $P_\mu \cong S_{\mu_1} \times \dots \times S_{\mu_n}$.}
\begin{subequations}
    \begin{align}
    P_\mu &= \{ \sigma \in S_k \mid \sigma\text{ preserves each row of $\mu$}\}
    \,,\\
    Q_\mu &= \{ \sigma \in S_k \mid \sigma\text{ preserves each column of $\mu$}\}
    \,.
\end{align}
\end{subequations}
Define $a_\mu,b_\mu \in \bbC[S_k]$ by
\begin{align}
    a_\mu = \sum_{\sigma \in P_\mu} e_\sigma
    \,,\qquad
    b_\mu = \sum_{\sigma \in Q_\mu} \sgn(\sigma) e_\sigma 
    \,.
\end{align}
This construction precisely makes $c_\mu = b_\mu a_\mu$ symmetrise over the rows, then antisymmetrise over the columns of the Young diagram. The Specht module $S^\mu$ is then given by $c_\mu \bbC[S_k]$. Similarly, the corresponding $\SL(n+1,\bbC)$ irrep $V_\mu$ is given by $c_\mu (\bbC^{n+1})^{\otimes k}$. The action of $c_\mu$ can be defined by linearly extending the action of each $e_\sigma$ comprising it. The action of $e_\sigma$ is to permute the vectors of a given $(\bbC^{n+1})^{\otimes k}$-tensor as follows
\begin{align}
    e_\sigma(v_1 \otimes \dots \otimes v_k) = \sigma(v_1 \otimes \dots \otimes v_k) = v_{\sigma(1)} \otimes \dots v_{\sigma(k)}
    \,.
\end{align}
The key point here is that $c_\mu$, when acted on $(\bbC^{n+1})^{\otimes k}$, is proportional to the explicit projectors in the examples \eqref{YoungSymmetrisersExample}. 
However, since we constructed $c_\mu$ in a way explicitly emphasizing the symmetric group $S_k$, then the symmetric group's properties determine that\footnote{See Section 4 of Fulton and Harris \cite{Fulton:1991}.}
\begin{align}
    c_\mu^2 = \alpha_\mu c_\mu
    \,,\qquad
    \alpha_\mu = \frac{k!}{\dim S^\mu}
    \,.
\end{align}
The dimension of a complex Specht module is given by
\begin{align}
    \dim S^\mu = \frac{k!}{\prod_{\text{boxes}} h_{\text{box}}}
    \,.
\end{align}
So, the projector, or idempotent associated with a Young symmetriser is $p_\mu = c_\mu/\alpha_\mu$ with $\alpha_\lambda$ given by the product of all hook-lengths.

When considering our use case of $\SL(n+1,\bbC)$ irreps, note that we use conventions where $T_{[ab]} = (1/2!)(T_{ab} - T_{ab})$ etc.\ Hence, the scaled version of $c_\mu$  we act on a tensor $T_{\a_1\dots\a_k} \, e^{\a_1} \otimes \dots \otimes e^{\a_k} \in (\bbC^{n+1})^{\otimes k}$, with $e^\a$ a basis of $\bbC^{n+1}$, is really $c_\mu$ divided by such factorials. Thus, the scaling factor that we need to use in equations like \eqref{YoungSymmetrisersExample} is given by
\begin{align}\label{scalingfactor}
    (\text{scaling factor}) = \frac{\prod(\text{factorials})}{\prod_{\text{boxes}} h_{\text{box}}}
    \,,\qquad
    \prod(\text{factorials}) = \left( \prod_i \mu_i!\right) \left( \prod_j \nu_j! \right)
    \,,
\end{align}
where $\mu_i = (\text{length of row $i$})$ and $\nu_i = (\text{length of column $i$})$. For example, the scaling factor of the diagram $\ytableausetup{smalltableaux}\ydiagram{2,2}$ is
\begin{equation}
    \frac{(2!)^4}{(3)(2)(2)} = \frac{4}{3}
    \,.
\end{equation}
The scaling factor of $\ydiagram{3,1}$ is
\begin{eqnarray}
    \frac{(3!)(2!)}{(4)(2)} = \frac{3}{2}
    \,.
\end{eqnarray}

\subsubsection{Garnir relations}
We give a brief explanation of multi-term symmeries, or Garnir relations that we introduce in \eqref{GarnirDiagramsExample} and \eqref{GarnirTensorsExample} using some of the formalism developed above about the symmetric group. We follow Section 2.6 of Sagan \cite{Sagan:1991} and Section 7 of James \cite{James:1978}.

Take any filled Young diagram $\mu$ with Young symmetriser $c_\mu = b_\mu a_\mu$ and consider picking two adjacent columns, say column $i$ and $i+1$. As in the diagram \eqref{GarnirABDiagram} below, take portions of the these columns $A$ and $B$ so that $A$ starts from the bottom, $B$ starts from the top, and $A$ and $B$ share at least one row.\footnote{Taking $A$ and $B$ to be as small as possible with just one row of overlap is inherently more interesting. Otherwise, as we saw earlier when we introduced Garnir relations in \eqref{GarnirDiagramsExample} and \eqref{GarnirTensorsExample}, we will be antisymmetrising over more indices than needed --- we could still get zero by antisymmetrising over less of them.}
\begin{equation}\label{GarnirABDiagram}
\begin{tikzpicture}[scale=0.65, baseline={(current bounding box.center)}]
  % Outer part of diagram
  \draw[thick] (0,0) -- (1,0) -- (1,1) -- (4,1) -- (4,4) -- (0,4) -- (0,0);
  
  % Filled rectangles for A and B
  \fill[myblack] (1,1) rectangle (2,3);  % A's box
  \fill[myblack] (2,2) rectangle (3,4);  % B's box

  % Inner part of diagram (borders over fill)
  \draw[thick] (1,1) -- (1,3) -- (2,3) -- (2,1);
  \draw[thick] (2,4) -- (2,2) -- (3,2) -- (3,4);

  % Labels (white)
  \node at (1.5,2) {\textcolor{white}{$A$}};
  \node at (2.5,3) {\textcolor{white}{$B$}};
\end{tikzpicture}
\end{equation}
As we saw in \eqref{GarnirDiagramsExample} and \eqref{GarnirTensorsExample}, we expect antisymmetrisation over the combined boxes $A \cup B$ should give zero on a tensor $T \in V_\mu$ associated with the Young diagram \eqref{GarnirABDiagram}. We formalise this below.

Using the filling of the Young diagram $\mu$ with integers $\{1,\dots,k\}$ as in \eqref{FilledYoungDiagram}, redefine $A$ and $B$ as subsets of $\{1,\dots,k\}$ corresponding to the integers contained within them in the diagram \eqref{GarnirABDiagram}. Consider the group $S_A \times S_B$ where $S_A$ is the symmetric group permuting elements of $A$ and the same for $S_B$. This is a subgroup of $S_{A\cup B}$ which also allows permutations between $A$ and $B$. Of course, these are all subgroups of $S_k$ which allows permutations between any boxes in the Young diagram. 
Define $g_A,g_B,g_{A\cup B} \in \bbC[S_k]$ by
\begin{align}
    g_{A} = \sum_{\sigma \in S_A} \sgn(\sigma) e_{\sigma}
    \,,\qquad
    g_{B} = \sum_{\sigma \in S_B} \sgn(\sigma) e_{\sigma}
    \,,\qquad
    g_{A\cup B} = \sum_{\sigma \in S_{A\cup B}} \sgn(\sigma) e_{\sigma}
    \,.
\end{align}
The action of $g_A$ on a tensor $T \in V_\mu$ associated with the Young diagram is to antisymmetrise over indices in contained within the boxes $A$ covers. The same holds for $g_B$ and $g_{A\cup B}$. The Garnir relation is $g_{A\cup B} c_\mu = 0$, so that the Young symmetriser is annihiliated by $g_{A\cup B}$. Hence, also $g_{A\cup B} T = 0$ for tensors $T \in V_\mu$ --- giving their multi-term symmetry like in \eqref{GarnirDiagramsExample} and \eqref{GarnirTensorsExample}. 
This can be stated in a way closer to the mathematical literature which uses a minimal $g_{A,B}$ instead of $g_{A\cup B}$.\footnote{To give some concreteness, if we act on $T_{\a\b,\g}$ associated with $\ydiagram{2,1}$ (there is only one choice of $A,B$). The explicit action of $g_{A\cup B}$ is $g_{A\cup B} T_{\a\b,\g} = T_{\a\b,\g} - T_{\b\a,\g} + T_{\g\a,\b} - T_{\a\g,\b} + T_{\b\g,\a} - T_{\g\b,\a}$. Whereas, this minimal $g_{A,B}$ has explicit action $g_{A,B} T_{\a\b,\g} = T_{\a\b,\g} + T_{\g\a,\b} + T_{\b\g,\a}$. In both cases, we are antisymmetrising over $\a,\b,\g$, but $g_{A,B}$ incorporates that $T_{\a\b,\g}$ is already antisymmetric in the first two indices.}

Allow the permutations $\sigma_1,\dots,\sigma_l \in S_{A\cup B}$ to be a choice of coset representatives of $S_A \times S_B$ in $S_{A\cup B}$.  A \emph{Garnir element} is an element $g_{A,B} \in \bbC[S_k]$ is constructed as
\begin{align}
    g_{A,B} = \sum_{i=1}^l \sgn(\sigma_i) e_{\sigma_i}
    \,.
\end{align}
Precisely because these $\sigma_i$ are cosets of $S_A \times S_B$ in $S_{A\cup B}$, one finds the factorisation $g_{A,B} g_A g_B = g_{A \cup B}$. 
The Garnir relations can be restated as $g_{A,B} c_\lambda = 0$ since $g_{A \cup B} c_\lambda = |A|! |B|! g_{A,B} c_\lambda$ where $|A|$, $|B|$ denote the number of elements in the sets $A$, $B$, or the number of boxes they contain, respectively. Indeed, $g_A,g_B$ antisymmetrise over $A$ and $B$, but $b_\mu$ (in $c_\mu = b_\mu a_\mu$) already antisymmetrises over all columns. In particular, it already antisymmetrises over the column containing $A$ and the column containing $B$. Thus, $g_Ag_B b_\mu = |A|! |B|! b_\mu$.

It is interesting to note that, along with the inherent antisymmetry between indices in columns of the Young diagram, the Garnir relations completely characterise the linear dependence between basis vectors $e^{\alpha_1} \otimes \dots \otimes e^{\alpha_k}$ of $(\bbC^{n+1})^{\otimes k}$ after they are projected to the $\SL(n+1,\bbC)$ irrep $V_\mu = c_\mu (\bbC^{n+1})^{\otimes k}$. Put another way, all symmetries of a tensor $T_{\alpha_1\dots\alpha_k}$ in an irrep $V_\mu$ are described by the inherent antisymmetry between indices in columns of the Young diagram and the Garnir relations.

To give further context, we mention that this is also closely related to the polytabloid construction of the Specht modules $S^\mu$. It is arguably easier to prove an equivalent statement to $g_{A,B} c_\mu = 0$ in this setting, and this is what Sagan \cite{Sagan:1991} and James \cite{James:1978} do. In this setting, one uses Garnir relations to find linear dependence between polytabloids and determine a basis given by standard polytabloids for $S^\mu$. In particular, Garnir elements provide an iterative algorithm to remove row descents one at a time --- typically called a \emph{straightening algorithm}. Iterating allows one write all polytabloids in terms of standard polytabloids.

\subsubsection{\texorpdfstring{$\Sp(2n,\bbC)$ dimensions and highest weights}{Sp(2n,C) dimensions and highest weights}}

The $\Sp(2n,\bbC)$-highest weight $\l = (\l_1,\dots,\l_n)$ of a complex irrep $V_\lambda$ depends upon its traceless Young diagram in the following way. First, the highest weights of the defining representation on $\bbC^{2n}$ and its traceless exterior products $\L^k(\bbC^{2n})^{\,{\rm trcls}}$ for $k=2,\dots,n$ are given by
\begin{align}\ytableausetup{smalltableaux}
(1,0,\dots,0) \equiv \begin{ytableau}
~
\end{ytableau}
\,,\quad
(1,1,0,\dots,0) \equiv \begin{ytableau}
~\\
~
\end{ytableau}^{\,{\rm trcls}}
\,,\dots\quad
(1,\dots,1) \equiv \begin{ytableau}
~\\
\none\\
\none[\svdots]\\
~
\end{ytableau}^{\,{\rm trcls}}
\,.
\end{align}
Given a traceless Young diagram with columns of length less than or equal to $n$, its $\Sp(2n,\bbC)$ highest weight $\l = (\l_1,\dots,\l_n)$ is obtained by summing the weights associated with the columns as above. That is, if $m_i = \#(\text{columns of length $i$})$, then $\l_i = \sum_{j=i}^{n} m_j$. Diagrams with columns of length greater than $n$ are not considered because $\Sp(2n,\bbC)$ has an invariant bilinear form $\O$ and a natural top form $\L^{2n}(\Omega)$. Hence, there are natural morphisms $\L^{2n-k}(\bbC^{2n}) \cong \L^{k}(\bbC^{2n})$ which lead to redundancy. In particular, using this morphism, Young diagrams with columns of length greater than $n$ can always be rewritten as a sum of those with column lengths less than or equal to $n$ --- this is an application of the branching rules. In the $n=2$ case, see our earlier example of a diagram with one column of length 3 and one column of length 1. 
For completeness, we mention that the Weyl dimension formula for $\Sp(4,\bbC)$ is
\begin{equation}
\dim V_{(\l_1,\l_2)} = \frac{1}{6} (\l_1-\l_2 + 1)(\l_1 + 2)(\l_2 + 1)(\l_1+\l_2 + 3)
\,.
\end{equation}

\subsection{\texorpdfstring{Irrep decompositions of $Q$-actions on dimension 3/2 fields}{Irrep decompositions of Q-actions on dimension 3/2 fields}}\label{appendix:reptheory-irrepQX}
We find that $\nabla_\a^i X^{\b j, kl}$ decomposes into the following $\SL(4,\bbC)$ and $\Sp(4,\bbC)$ irreps:
\begin{align}
\ytableausetup{smalltableaux}
\SL(4,\bbC)\,:& \quad
\begin{ytableau}
~
\end{ytableau}
\otimes
\begin{ytableau}
~\\
~\\
~
\end{ytableau}
=
\begin{ytableau}
~&~\\
~\\
~
\end{ytableau}
+
\bullet
\,,\\
\Sp(4,\bbC)\,:& \quad
\begin{ytableau}
~
\end{ytableau}
\otimes
\begin{ytableau}
~&~\\
~
\end{ytableau}^{\,{\rm trcls}}
=
\begin{ytableau}
~&~\\
~&~
\end{ytableau}^{\,{\rm trcls}}
+
\begin{ytableau}
~&~&~\\
~
\end{ytableau}^{\,{\rm trcls}}
+
\begin{ytableau}
~\\
~
\end{ytableau}^{\,{\rm trcls}}
+
\begin{ytableau}
~&~
\end{ytableau}
\,.
\end{align}
It is worth mentioning that the above $\Sp(4,\bbC)$ calculation is easier when one leverages that $\ve_{ijkl} = 3\O_{[ij} \O_{kl]}$ is a top form and also virtual representations.
We find $\nabla_\a^i X_\b{}^{\g\d j}$ decomposes into the following $\SL(4,\bbC)$ and $\Sp(4,\bbC)$ irreps:
\begin{align}
\SL(4,\bbC)\,:& \quad
\begin{ytableau}
~
\end{ytableau}
\otimes
\begin{ytableau}
~&~&~\\
~&~\\
~&~
\end{ytableau}
=
\begin{ytableau}
~&~\\
~\\
~
\end{ytableau}
+
\begin{ytableau}
~&~&~\\
~&~&~\\
~&~
\end{ytableau}
+
\begin{ytableau}
~&~&~&~\\
~&~\\
~&~
\end{ytableau}
\,,\\
\Sp(4,\bbC)\,:& \quad
\begin{ytableau}
~
\end{ytableau}
\otimes
\begin{ytableau}
~
\end{ytableau}
=
\begin{ytableau}
~\\
~
\end{ytableau}^{\,{\rm trcls}}
+
\begin{ytableau}
~&~
\end{ytableau}
+
\bullet
\,.
\end{align}
Using these decompositions, along with the information about which irreps are set to zero \eqref{QXconstraints1}, we find (\ref{QX_YExpansion1},~\ref{QX_YExpansion2}).

\subsection{\texorpdfstring{Irrep decompositions of $Q$-actions on dimension 2 fields}{Irrep decompositions of Q-actions on dimension 2 fields}}\label{appendix:reptheory-irrepQY}
We find that $\nabla_\a^i Y^{jk,l i_1}$ decomposes into the following $\Sp(4,\bbC)$
\begin{align}
\Sp(4,\bbC)\,:& \quad
\begin{ytableau}
~
\end{ytableau}
\otimes
\begin{ytableau}
~&~\\
~&~
\end{ytableau}^{\,{\rm trcls}}
=
\begin{ytableau}
~&~&~\\
~&~
\end{ytableau}^{\,{\rm trcls}}
+
\begin{ytableau}
~&~\\
~
\end{ytableau}^{\,{\rm trcls}}
\,.
\end{align}
We find that $\nabla_\a^i Y_{\b\g}{}^{\d\e}$ decomposes into the following $\SL(4,\bbC)$ irreps
\begin{align}
\SL(4,\bbC)\,:& \quad
\begin{ytableau}
~
\end{ytableau}
\otimes
\begin{ytableau}
~&~&~&~\\
~&~\\
~&~
\end{ytableau}
=
\begin{ytableau}
~&~&~&~\\
~&~&~\\
~&~
\end{ytableau}
+
\begin{ytableau}
~&~&~&~&~\\
~&~\\
~&~
\end{ytableau}
+
\begin{ytableau}
~&~&~\\
~\\
~
\end{ytableau}
\,.
\end{align}
We find that $\n_{\a}^{i} Y_{\b}{}^{\g jk}$ decomposes into the following $\SL(4,\bbC)$ and $\Sp(4,\bbC)$ irreps
\begin{align}
\SL(4,\bbC)\,:& \quad
\begin{ytableau}
~
\end{ytableau}
\otimes
\begin{ytableau}
~&~\\
~\\
~
\end{ytableau}
=
\begin{ytableau}
~&~\\
~&~\\
~
\end{ytableau}
+
\begin{ytableau}
~&~&~\\
~\\
~
\end{ytableau}
+
\begin{ytableau}
~
\end{ytableau}
\,,\\
\Sp(4,\bbC)\,:& \quad
\begin{ytableau}
~
\end{ytableau}
\otimes
\begin{ytableau}
~&~
\end{ytableau}
=
\begin{ytableau}
~&~\\
~
\end{ytableau}^{\,{\rm trcls}}
+
\begin{ytableau}
~&~&~
\end{ytableau}
+
\begin{ytableau}
~
\end{ytableau}
\,.
\end{align}
Using these decompositions, along with the information about which irreps are set to zero \eqref{QYConstraints1}, we find \eqref{QY_ZExpansions}.

\bibliography{biblio} 

\providecommand{\href}[2]{#2}\begingroup\raggedright\begin{thebibliography}{10}

\bibitem{Deser:1970hs}
S.~Deser, ``{Scale invariance and gravitational coupling},''
  \href{http://dx.doi.org/10.1016/0003-4916(70)90402-1}{{\em Annals Phys.}
  {\bfseries 59} (1970) 248--253}.

\bibitem{Zumino:1970tu}
B.~Zumino, ``Effective lagrangians and broken symmetries,'' in {\em Lectures on
  Elementary Particles and Quantum Field Theory, Vol. 2}, S.~Deser, M.~Grisaru,
  and H.~Pendleton, eds., pp.~437--500.
\newblock MIT Press, Cambridge, Mass., 1970.

\bibitem{Freedman:2012zz}
D.~Z. Freedman and A.~Van~Proeyen,
  \href{http://dx.doi.org/10.1017/CBO9781139026833}{{\em {Supergravity}}}.
\newblock Cambridge Univ. Press, Cambridge, UK, 5, 2012.

\bibitem{Lauria:2020rhc}
E.~Lauria and A.~Van~Proeyen,
  \href{http://dx.doi.org/10.1007/978-3-030-33757-5}{{\em {${\cal N}=2$
  Supergravity in $D=4,5,6$ Dimensions}}}, vol.~966.
\newblock 3, 2020.
\newblock \href{http://arxiv.org/abs/2004.11433}{{\ttfamily arXiv:2004.11433
  [hep-th]}}.

\bibitem{Gates:1983nr}
S.~J. Gates, M.~T. Grisaru, M.~Rocek, and W.~Siegel, {\em {Superspace Or One
  Thousand and One Lessons in Supersymmetry}}, vol.~58 of {\em Frontiers in
  Physics}.
\newblock 1983.
\newblock \href{http://arxiv.org/abs/hep-th/0108200}{{\ttfamily
  arXiv:hep-th/0108200}}.

\bibitem{Buchbinder:1998twe}
I.~L. Buchbinder and S.~M. Kuzenko,
  \href{http://dx.doi.org/10.1201/9780367802530}{{\em {Ideas and Methods of
  Supersymmetry and Supergravity or A Walk Through Superspace: A Walk Through
  Superspace}}}.
\newblock Taylor \& Francis, 1998.

\bibitem{Kuzenko:2022skv}
S.~M. Kuzenko, E.~S.~N. Raptakis, and G.~Tartaglino-Mazzucchelli, {\em
  {Superspace Approaches to $\mathscr {N} = \text{1}$ Supergravity}}.
\newblock 2023.
\newblock \href{http://arxiv.org/abs/2210.17088}{{\ttfamily arXiv:2210.17088
  [hep-th]}}.

\bibitem{Kuzenko:2022ajd}
S.~M. Kuzenko, E.~S.~N. Raptakis, and G.~Tartaglino-Mazzucchelli, {\em
  {Covariant Superspace Approaches to $\mathscr {N}=\text{2}$ Supergravity}}.
\newblock 2023.
\newblock \href{http://arxiv.org/abs/2211.11162}{{\ttfamily arXiv:2211.11162
  [hep-th]}}.

\bibitem{deWit:1979dzm}
B.~de~Wit, J.~W. van Holten, and A.~Van~Proeyen, ``{Transformation Rules of N=2
  Supergravity Multiplets},''
  \href{http://dx.doi.org/10.1016/0550-3213(80)90125-X}{{\em Nucl. Phys. B}
  {\bfseries 167} (1980) 186}.

\bibitem{Bergshoeff:1980is}
E.~Bergshoeff, M.~de~Roo, and B.~de~Wit, ``{Extended Conformal Supergravity},''
  \href{http://dx.doi.org/10.1016/0550-3213(81)90465-X}{{\em Nucl. Phys. B}
  {\bfseries 182} (1981) 173--204}.

\bibitem{Siegel:1978mj}
W.~Siegel and S.~J. Gates, Jr., ``{Superfield Supergravity},''
  \href{http://dx.doi.org/10.1016/0550-3213(79)90416-4}{{\em Nucl. Phys. B}
  {\bfseries 147} (1979) 77--104}.

\bibitem{Howe:1981gz}
P.~S. Howe, ``{Supergravity in Superspace},''
  \href{http://dx.doi.org/10.1016/0550-3213(82)90349-2}{{\em Nucl. Phys. B}
  {\bfseries 199} (1982) 309--364}.

\bibitem{Fradkin:1982xc}
E.~S. Fradkin and A.~A. Tseytlin, ``{ASYMPTOTIC FREEDOM IN EXTENDED CONFORMAL
  SUPERGRAVITIES},'' \href{http://dx.doi.org/10.1016/0370-2693(82)91018-8}{{\em
  Phys. Lett. B} {\bfseries 110} (1982) 117--122}. [Erratum: Phys.Lett.B 126,
  (1983)].

\bibitem{Bergshoeff:1985mz}
E.~Bergshoeff, E.~Sezgin, and A.~Van~Proeyen, ``{Superconformal Tensor Calculus
  and Matter Couplings in Six-dimensions},''
  \href{http://dx.doi.org/10.1016/0550-3213(86)90503-1}{{\em Nucl. Phys. B}
  {\bfseries 264} (1986) 653}. [Erratum: Nucl.Phys.B 598, 667 (2001)].

\bibitem{Bergshoeff:1999}
E.~Bergshoeff, E.~Sezgin, and A.~Van~Proeyen, ``{(2,0) tensor multiplets and
  conformal supergravity in D = 6},''
  \href{http://dx.doi.org/10.1088/0264-9381/16/10/311}{{\em Class. Quant.
  Grav.} {\bfseries 16} (1999) 3193--3206},
  \href{http://arxiv.org/abs/hep-th/9904085}{{\ttfamily arXiv:hep-th/9904085}}.

\bibitem{Kugo:2000hn}
T.~Kugo and K.~Ohashi, ``{Supergravity tensor calculus in 5-D from 6-D},''
  \href{http://dx.doi.org/10.1143/PTP.104.835}{{\em Prog. Theor. Phys.}
  {\bfseries 104} (2000) 835--865},
  \href{http://arxiv.org/abs/hep-ph/0006231}{{\ttfamily arXiv:hep-ph/0006231}}.

\bibitem{Bergshoeff:2001hc}
E.~Bergshoeff, T.~de~Wit, R.~Halbersma, S.~Cucu, M.~Derix, and A.~Van~Proeyen,
  ``{Weyl multiplets of N=2 conformal supergravity in five-dimensions},''
  \href{http://dx.doi.org/10.1088/1126-6708/2001/06/051}{{\em JHEP} {\bfseries
  06} (2001) 051}, \href{http://arxiv.org/abs/hep-th/0104113}{{\ttfamily
  arXiv:hep-th/0104113}}.

\bibitem{Butter:2009cp}
D.~Butter, ``{N=1 Conformal Superspace in Four Dimensions},''
  \href{http://dx.doi.org/10.1016/j.aop.2009.09.010}{{\em Annals Phys.}
  {\bfseries 325} (2010) 1026--1080},
  \href{http://arxiv.org/abs/0906.4399}{{\ttfamily arXiv:0906.4399 [hep-th]}}.

\bibitem{Butter:2011sr}
D.~Butter, ``{N=2 Conformal Superspace in Four Dimensions},''
  \href{http://dx.doi.org/10.1007/JHEP10(2011)030}{{\em JHEP} {\bfseries 10}
  (2011) 030}, \href{http://arxiv.org/abs/1103.5914}{{\ttfamily arXiv:1103.5914
  [hep-th]}}.

\bibitem{Kuzenko:2011xg}
S.~M. Kuzenko, U.~Lindstrom, and G.~Tartaglino-Mazzucchelli, ``{Off-shell
  supergravity-matter couplings in three dimensions},''
  \href{http://dx.doi.org/10.1007/JHEP03(2011)120}{{\em JHEP} {\bfseries 03}
  (2011) 120}, \href{http://arxiv.org/abs/1101.4013}{{\ttfamily arXiv:1101.4013
  [hep-th]}}.

\bibitem{Butter:2013goa}
D.~Butter, S.~M. Kuzenko, J.~Novak, and G.~Tartaglino-Mazzucchelli,
  ``{Conformal supergravity in three dimensions: New off-shell formulation},''
  \href{http://dx.doi.org/10.1007/JHEP09(2013)072}{{\em JHEP} {\bfseries 09}
  (2013) 072}, \href{http://arxiv.org/abs/1305.3132}{{\ttfamily arXiv:1305.3132
  [hep-th]}}.

\bibitem{Butter:2013rba}
D.~Butter, S.~M. Kuzenko, J.~Novak, and G.~Tartaglino-Mazzucchelli,
  ``{Conformal supergravity in three dimensions: Off-shell actions},''
  \href{http://dx.doi.org/10.1007/JHEP10(2013)073}{{\em JHEP} {\bfseries 10}
  (2013) 073}, \href{http://arxiv.org/abs/1306.1205}{{\ttfamily arXiv:1306.1205
  [hep-th]}}.

\bibitem{Butter:2014xxa}
D.~Butter, S.~M. Kuzenko, J.~Novak, and G.~Tartaglino-Mazzucchelli,
  ``{Conformal supergravity in five dimensions: New approach and
  applications},'' \href{http://dx.doi.org/10.1007/JHEP02(2015)111}{{\em JHEP}
  {\bfseries 02} (2015) 111}, \href{http://arxiv.org/abs/1410.8682}{{\ttfamily
  arXiv:1410.8682 [hep-th]}}.

\bibitem{Butter:2016}
D.~Butter, S.~M. Kuzenko, J.~Novak, and S.~Theisen, ``{Invariants for minimal
  conformal supergravity in six dimensions},''
  \href{http://dx.doi.org/10.1007/JHEP12(2016)072}{{\em JHEP} {\bfseries 12}
  (2016) 072}, \href{http://arxiv.org/abs/1606.02921}{{\ttfamily
  arXiv:1606.02921 [hep-th]}}.

\bibitem{Butter:2017pbp}
D.~Butter, S.~Hegde, I.~Lodato, and B.~Sahoo, ``{$N=2$ dilaton Weyl multiplet
  in 4D supergravity},'' \href{http://dx.doi.org/10.1007/JHEP03(2018)154}{{\em
  JHEP} {\bfseries 03} (2018) 154},
  \href{http://arxiv.org/abs/1712.05365}{{\ttfamily arXiv:1712.05365
  [hep-th]}}.

\bibitem{Butter:2019edc}
D.~Butter, F.~Ciceri, and B.~Sahoo, ``{$N=4$ conformal supergravity: the
  complete actions},'' \href{http://dx.doi.org/10.1007/JHEP01(2020)029}{{\em
  JHEP} {\bfseries 01} (2020) 029},
  \href{http://arxiv.org/abs/1910.11874}{{\ttfamily arXiv:1910.11874
  [hep-th]}}.

\bibitem{Howe:2020xrg}
P.~S. Howe and U.~Lindstr\"om, ``{Local supertwistors and conformal
  supergravity in six dimensions},''
  \href{http://dx.doi.org/10.1098/rspa.2020.0683}{{\em Proc. Roy. Soc. Lond. A}
  {\bfseries 476} no.~2243, (2020) 20200683},
  \href{http://arxiv.org/abs/2008.10302}{{\ttfamily arXiv:2008.10302
  [hep-th]}}.

\bibitem{Howe:2020hxi}
P.~S. Howe and U.~Lindstr\"om, ``{Superconformal geometries and local
  twistors},'' \href{http://dx.doi.org/10.1007/JHEP04(2021)140}{{\em JHEP}
  {\bfseries 04} (2021) 140}, \href{http://arxiv.org/abs/2012.03282}{{\ttfamily
  arXiv:2012.03282 [hep-th]}}.

\bibitem{Hutomo:2022hdi}
J.~Hutomo, S.~Khandelwal, G.~Tartaglino-Mazzucchelli, and J.~Woods,
  ``{Hyperdilaton Weyl multiplets of 5D and 6D minimal conformal
  supergravity},'' \href{http://dx.doi.org/10.1103/PhysRevD.107.046009}{{\em
  Phys. Rev. D} {\bfseries 107} no.~4, (2023) 046009},
  \href{http://arxiv.org/abs/2209.05748}{{\ttfamily arXiv:2209.05748
  [hep-th]}}.

\bibitem{Kuzenko:2022qnb}
S.~M. Kuzenko and E.~S.~N. Raptakis, ``{Conformal (p, q) supergeometries in two
  dimensions},'' \href{http://dx.doi.org/10.1007/JHEP02(2023)166}{{\em JHEP}
  {\bfseries 02} (2023) 166}, \href{http://arxiv.org/abs/2211.16169}{{\ttfamily
  arXiv:2211.16169 [hep-th]}}.

\bibitem{Kuzenko:2023qkg}
S.~M. Kuzenko and E.~S.~N. Raptakis, ``{$ \mathcal{N} $ = 3 conformal
  superspace in four dimensions},''
  \href{http://dx.doi.org/10.1007/JHEP03(2024)026}{{\em JHEP} {\bfseries 03}
  (2024) 026}, \href{http://arxiv.org/abs/2312.07242}{{\ttfamily
  arXiv:2312.07242 [hep-th]}}.

\bibitem{Adhikari:2023tzi}
S.~Adhikari and B.~Sahoo, ``{$ \mathcal{N} $ = 2 conformal supergravity in five
  dimensions},'' \href{http://dx.doi.org/10.1007/JHEP07(2024)028}{{\em JHEP}
  {\bfseries 07} (2024) 028}, \href{http://arxiv.org/abs/2312.01879}{{\ttfamily
  arXiv:2312.01879 [hep-th]}}.

\bibitem{Adhikari:2024qxg}
S.~Adhikari, A.~Aikot, M.~Mishra, and B.~Sahoo, ``{Dilaton Weyl multiplets for
  $ \mathcal{N} $ = 3 conformal supergravity in four dimensions},''
  \href{http://dx.doi.org/10.1007/JHEP04(2025)062}{{\em JHEP} {\bfseries 04}
  (2025) 062}, \href{http://arxiv.org/abs/2412.14874}{{\ttfamily
  arXiv:2412.14874 [hep-th]}}.

\bibitem{Adhikari:2024esl}
S.~Adhikari and B.~Sahoo, ``{SU(2) \texttimes{} SU(2) dilaton Weyl multiplets
  for maximal conformal supergravity in four, five, and six dimensions},''
  \href{http://dx.doi.org/10.1007/JHEP02(2025)059}{{\em JHEP} {\bfseries 02}
  (2025) 059}, \href{http://arxiv.org/abs/2411.16322}{{\ttfamily
  arXiv:2411.16322 [hep-th]}}.

\bibitem{Adhikari:2025wwb}
S.~Adhikari, A.~Aikot, B.~Sahoo, and M.~Mishra, ``{Variant dilaton Weyl
  Multiplet for N=3 conformal supergravity in four dimensions},''
  \href{http://arxiv.org/abs/2502.13683}{{\ttfamily arXiv:2502.13683
  [hep-th]}}.

\bibitem{Kuzenko:2025bud}
S.~M. Kuzenko and E.~S.~N. Raptakis, ``{${\mathcal N}=3$ nonlinear multiplet
  and supergravity},'' \href{http://arxiv.org/abs/2501.11339}{{\ttfamily
  arXiv:2501.11339 [hep-th]}}.

\bibitem{Pestun:2016zxk}
V.~Pestun {\em et~al.}, ``{Localization techniques in quantum field
  theories},'' \href{http://dx.doi.org/10.1088/1751-8121/aa63c1}{{\em J. Phys.
  A} {\bfseries 50} no.~44, (2017) 440301},
  \href{http://arxiv.org/abs/1608.02952}{{\ttfamily arXiv:1608.02952
  [hep-th]}}.

\bibitem{Baggio:2014hua}
M.~Baggio, N.~Halmagyi, D.~R. Mayerson, D.~Robbins, and B.~Wecht, ``{Higher
  Derivative Corrections and Central Charges from Wrapped M5-branes},''
  \href{http://dx.doi.org/10.1007/JHEP12(2014)042}{{\em JHEP} {\bfseries 12}
  (2014) 042}, \href{http://arxiv.org/abs/1408.2538}{{\ttfamily arXiv:1408.2538
  [hep-th]}}.

\bibitem{Butter:2018wss}
D.~Butter, J.~Novak, M.~Ozkan, Y.~Pang, and G.~Tartaglino-Mazzucchelli,
  ``{Curvature squared invariants in six-dimensional ${\cal N} = (1,0)$
  supergravity},'' \href{http://dx.doi.org/10.1007/JHEP04(2019)013}{{\em JHEP}
  {\bfseries 04} (2019) 013}, \href{http://arxiv.org/abs/1808.00459}{{\ttfamily
  arXiv:1808.00459 [hep-th]}}.

\bibitem{Bobev:2020egg}
N.~Bobev, A.~M. Charles, K.~Hristov, and V.~Reys, ``{The Unreasonable
  Effectiveness of Higher-Derivative Supergravity in AdS$_4$ Holography},''
  \href{http://dx.doi.org/10.1103/PhysRevLett.125.131601}{{\em Phys. Rev.
  Lett.} {\bfseries 125} no.~13, (2020) 131601},
  \href{http://arxiv.org/abs/2006.09390}{{\ttfamily arXiv:2006.09390
  [hep-th]}}.

\bibitem{Bobev:2021oku}
N.~Bobev, A.~M. Charles, K.~Hristov, and V.~Reys, ``{Higher-derivative
  supergravity, AdS$_{4}$ holography, and black holes},''
  \href{http://dx.doi.org/10.1007/JHEP08(2021)173}{{\em JHEP} {\bfseries 08}
  (2021) 173}, \href{http://arxiv.org/abs/2106.04581}{{\ttfamily
  arXiv:2106.04581 [hep-th]}}.

\bibitem{Bobev:2021qxx}
N.~Bobev, K.~Hristov, and V.~Reys, ``{AdS$_{5}$ holography and
  higher-derivative supergravity},''
  \href{http://dx.doi.org/10.1007/JHEP04(2022)088}{{\em JHEP} {\bfseries 04}
  (2022) 088}, \href{http://arxiv.org/abs/2112.06961}{{\ttfamily
  arXiv:2112.06961 [hep-th]}}.

\bibitem{Liu:2022sew}
J.~T. Liu and R.~J. Saskowski, ``{Four-derivative corrections to minimal gauged
  supergravity in five dimensions},''
  \href{http://dx.doi.org/10.1007/JHEP05(2022)171}{{\em JHEP} {\bfseries 05}
  (2022) 171}, \href{http://arxiv.org/abs/2201.04690}{{\ttfamily
  arXiv:2201.04690 [hep-th]}}.

\bibitem{Hristov:2022lcw}
K.~Hristov, ``{ABJM at finite N via 4d supergravity},''
  \href{http://dx.doi.org/10.1007/JHEP10(2022)190}{{\em JHEP} {\bfseries 10}
  (2022) 190}, \href{http://arxiv.org/abs/2204.02992}{{\ttfamily
  arXiv:2204.02992 [hep-th]}}.

\bibitem{Cassani:2022lrk}
D.~Cassani, A.~Ruip\'erez, and E.~Turetta, ``{Corrections to AdS$_{5}$ black
  hole thermodynamics from higher-derivative supergravity},''
  \href{http://dx.doi.org/10.1007/JHEP11(2022)059}{{\em JHEP} {\bfseries 11}
  (2022) 059}, \href{http://arxiv.org/abs/2208.01007}{{\ttfamily
  arXiv:2208.01007 [hep-th]}}.

\bibitem{Gold:2023ymc}
G.~Gold, J.~Hutomo, S.~Khandelwal, M.~Ozkan, Y.~Pang, and
  G.~Tartaglino-Mazzucchelli, ``{All Gauged Curvature-Squared Supergravities in
  Five Dimensions},''
  \href{http://dx.doi.org/10.1103/PhysRevLett.131.251603}{{\em Phys. Rev.
  Lett.} {\bfseries 131} no.~25, (2023) 251603},
  \href{http://arxiv.org/abs/2309.07637}{{\ttfamily arXiv:2309.07637
  [hep-th]}}.

\bibitem{Cassani:2024tvk}
D.~Cassani, A.~Ruip\'erez, and E.~Turetta, ``{Higher-derivative corrections to
  flavoured BPS black hole thermodynamics and holography},''
  \href{http://dx.doi.org/10.1007/JHEP05(2024)276}{{\em JHEP} {\bfseries 05}
  (2024) 276}, \href{http://arxiv.org/abs/2403.02410}{{\ttfamily
  arXiv:2403.02410 [hep-th]}}.

\bibitem{Casarin:2024qdn}
L.~Casarin, C.~Kennedy, and G.~Tartaglino-Mazzucchelli, ``{Conformal anomalies
  for (maximal) 6d conformal supergravity},''
  \href{http://dx.doi.org/10.1007/JHEP10(2024)227}{{\em JHEP} {\bfseries 10}
  (2024) 227}, \href{http://arxiv.org/abs/2403.07509}{{\ttfamily
  arXiv:2403.07509 [hep-th]}}.

\bibitem{Ma:2024ynp}
L.~Ma, P.-J. Hu, Y.~Pang, and H.~Lu, ``{Effectiveness of Weyl gravity in
  probing quantum corrections to AdS black holes},''
  \href{http://dx.doi.org/10.1103/PhysRevD.110.L021901}{{\em Phys. Rev. D}
  {\bfseries 110} no.~2, (2024) L021901},
  \href{http://arxiv.org/abs/2403.12131}{{\ttfamily arXiv:2403.12131
  [hep-th]}}.

\bibitem{Saskowski:2024otc}
R.~J. Saskowski, {\em {Explorations in Precision Holography and
  Higher-derivative Supergravity}}.
\newblock PhD thesis, Michigan U., 2024.
\newblock \href{http://arxiv.org/abs/2404.04134}{{\ttfamily arXiv:2404.04134
  [hep-th]}}.

\bibitem{Hristov:2024cgj}
K.~Hristov, ``{Equivariant localization and gluing rules in 4d $\mathcal{N}=2$
  higher derivative supergravity},''
\newblock 6, 2024.
\newblock \href{http://arxiv.org/abs/2406.18648}{{\ttfamily arXiv:2406.18648
  [hep-th]}}.

\bibitem{Cassani:2024kjn}
D.~Cassani, A.~Ruip\'erez, and E.~Turetta, ``{Localization of the 5D
  supergravity action and Euclidean saddles for the black hole index},''
  \href{http://dx.doi.org/10.1007/JHEP12(2024)086}{{\em JHEP} {\bfseries 12}
  (2024) 086}, \href{http://arxiv.org/abs/2409.01332}{{\ttfamily
  arXiv:2409.01332 [hep-th]}}.

\bibitem{Ozkan:2024euj}
M.~Ozkan, Y.~Pang, and E.~Sezgin, ``{Higher derivative supergravities in
  diverse dimensions},''
  \href{http://dx.doi.org/10.1016/j.physrep.2024.07.002}{{\em Phys. Rept.}
  {\bfseries 1086} (2024) 1--95},
  \href{http://arxiv.org/abs/2401.08945}{{\ttfamily arXiv:2401.08945
  [hep-th]}}.

\bibitem{Osborn:1993cr}
H.~Osborn and A.~C. Petkou, ``{Implications of conformal invariance in field
  theories for general dimensions},''
  \href{http://dx.doi.org/10.1006/aphy.1994.1045}{{\em Annals Phys.} {\bfseries
  231} (1994) 311--362}, \href{http://arxiv.org/abs/hep-th/9307010}{{\ttfamily
  arXiv:hep-th/9307010}}.

\bibitem{Duff:1980qv}
M.~J. Duff and P.~van Nieuwenhuizen, ``{Quantum Inequivalence of Different
  Field Representations},''
  \href{http://dx.doi.org/10.1016/0370-2693(80)90852-7}{{\em Phys. Lett. B}
  {\bfseries 94} (1980) 179--182}.

\bibitem{Duff:1993wm}
M.~J. Duff, ``{Twenty years of the Weyl anomaly},''
  \href{http://dx.doi.org/10.1088/0264-9381/11/6/004}{{\em Class. Quant. Grav.}
  {\bfseries 11} (1994) 1387--1404},
  \href{http://arxiv.org/abs/hep-th/9308075}{{\ttfamily arXiv:hep-th/9308075}}.

\bibitem{Bonora:1985cq}
L.~Bonora, P.~Pasti, and M.~Bregola, ``{WEYL COCYCLES},''
  \href{http://dx.doi.org/10.1088/0264-9381/3/4/018}{{\em Class. Quant. Grav.}
  {\bfseries 3} (1986) 635}.

\bibitem{Buchbinder:1986im}
I.~L. Buchbinder and S.~M. Kuzenko, ``{Matter Superfields in External
  Supergravity: Green Functions, Effective Action and Superconformal
  Anomalies},'' \href{http://dx.doi.org/10.1016/0550-3213(86)90532-8}{{\em
  Nucl. Phys. B} {\bfseries 274} (1986) 653--684}.

\bibitem{Kuzenko:2013gva}
S.~M. Kuzenko, ``{Super-Weyl anomalies in N=2 supergravity and (non)local
  effective actions},'' \href{http://dx.doi.org/10.1007/JHEP10(2013)151}{{\em
  JHEP} {\bfseries 10} (2013) 151},
  \href{http://arxiv.org/abs/1307.7586}{{\ttfamily arXiv:1307.7586 [hep-th]}}.

\bibitem{Gomis:2015yaa}
J.~Gomis, P.-S. Hsin, Z.~Komargodski, A.~Schwimmer, N.~Seiberg, and S.~Theisen,
  ``{Anomalies, Conformal Manifolds, and Spheres},''
  \href{http://dx.doi.org/10.1007/JHEP03(2016)022}{{\em JHEP} {\bfseries 03}
  (2016) 022}, \href{http://arxiv.org/abs/1509.08511}{{\ttfamily
  arXiv:1509.08511 [hep-th]}}.

\bibitem{Cordova:2015fha}
C.~Cordova, T.~T. Dumitrescu, and K.~Intriligator, ``{Anomalies,
  renormalization group flows, and the a-theorem in six-dimensional (1, 0)
  theories},'' \href{http://dx.doi.org/10.1007/JHEP10(2016)080}{{\em JHEP}
  {\bfseries 10} (2016) 080}, \href{http://arxiv.org/abs/1506.03807}{{\ttfamily
  arXiv:1506.03807 [hep-th]}}.

\bibitem{Cordova:2015vwa}
C.~Cordova, T.~T. Dumitrescu, and X.~Yin, ``{Higher derivative terms, toroidal
  compactification, and Weyl anomalies in six-dimensional (2, 0) theories},''
  \href{http://dx.doi.org/10.1007/JHEP10(2019)128}{{\em JHEP} {\bfseries 10}
  (2019) 128}, \href{http://arxiv.org/abs/1505.03850}{{\ttfamily
  arXiv:1505.03850 [hep-th]}}.

\bibitem{Herzog:2013ed}
C.~P. Herzog and K.-W. Huang, ``{Stress Tensors from Trace Anomalies in
  Conformal Field Theories},''
  \href{http://dx.doi.org/10.1103/PhysRevD.87.081901}{{\em Phys. Rev. D}
  {\bfseries 87} (2013) 081901},
  \href{http://arxiv.org/abs/1301.5002}{{\ttfamily arXiv:1301.5002 [hep-th]}}.

\bibitem{Bastianelli:2000hi}
F.~Bastianelli, S.~Frolov, and A.~A. Tseytlin, ``{Conformal anomaly of (2,0)
  tensor multiplet in six-dimensions and AdS / CFT correspondence},''
  \href{http://dx.doi.org/10.1088/1126-6708/2000/02/013}{{\em JHEP} {\bfseries
  02} (2000) 013}, \href{http://arxiv.org/abs/hep-th/0001041}{{\ttfamily
  arXiv:hep-th/0001041}}.

\bibitem{Deser:1993yx}
S.~Deser and A.~Schwimmer, ``{Geometric classification of conformal anomalies
  in arbitrary dimensions},''
  \href{http://dx.doi.org/10.1016/0370-2693(93)90934-A}{{\em Phys. Lett. B}
  {\bfseries 309} (1993) 279--284},
  \href{http://arxiv.org/abs/hep-th/9302047}{{\ttfamily arXiv:hep-th/9302047}}.

\bibitem{Nahm:1977tg}
W.~Nahm, ``{Supersymmetries and Their Representations},''
  \href{http://dx.doi.org/10.1201/9781482268737-2}{{\em Nucl. Phys. B}
  {\bfseries 135} (1978) 149}.

\bibitem{Kulaxizi:2009pz}
M.~Kulaxizi and A.~Parnachev, ``{Supersymmetry Constraints in Holographic
  Gravities},'' \href{http://dx.doi.org/10.1103/PhysRevD.82.066001}{{\em Phys.
  Rev. D} {\bfseries 82} (2010) 066001},
  \href{http://arxiv.org/abs/0912.4244}{{\ttfamily arXiv:0912.4244 [hep-th]}}.

\bibitem{Beccaria:2015uta}
M.~Beccaria and A.~A. Tseytlin, ``{Conformal a-anomaly of some non-unitary 6d
  superconformal theories},''
  \href{http://dx.doi.org/10.1007/JHEP09(2015)017}{{\em JHEP} {\bfseries 09}
  (2015) 017}, \href{http://arxiv.org/abs/1506.08727}{{\ttfamily
  arXiv:1506.08727 [hep-th]}}.

\bibitem{Butter:2017}
D.~Butter, J.~Novak, and G.~Tartaglino-Mazzucchelli, ``{The component structure
  of conformal supergravity invariants in six dimensions},''
  \href{http://dx.doi.org/10.1007/JHEP05(2017)133}{{\em JHEP} {\bfseries 05}
  (2017) 133}, \href{http://arxiv.org/abs/1701.08163}{{\ttfamily
  arXiv:1701.08163 [hep-th]}}.

\bibitem{Casarin:2023ifl}
L.~Casarin, ``{Conformal anomalies in 6D four-derivative theories: A
  heat-kernel analysis},''
  \href{http://dx.doi.org/10.1103/PhysRevD.108.025014}{{\em Phys. Rev. D}
  {\bfseries 108} no.~2, (2023) 025014},
  \href{http://arxiv.org/abs/2306.05944}{{\ttfamily arXiv:2306.05944
  [hep-th]}}.

\bibitem{Castellani:1991eu}
L.~Castellani, R.~D'Auria, and P.~Fr{\`e}, {\em {Supergravity and Superstrings:
  A Geometric Perspective. Vol. 2: Supergravity}}.
\newblock {World Scientific}, {Singapore}, {1991}.

\bibitem{Gates:1997ag}
S.~J. Gates, Jr., M.~T. Grisaru, M.~E. Knutt-Wehlau, and W.~Siegel,
  ``{Component actions from curved superspace: Normal coordinates and
  ectoplasm},'' \href{http://dx.doi.org/10.1016/S0370-2693(97)01557-8}{{\em
  Phys. Lett. B} {\bfseries 421} (1998) 203--210},
  \href{http://arxiv.org/abs/hep-th/9711151}{{\ttfamily arXiv:hep-th/9711151}}.

\bibitem{Gates:1998hy}
S.~J. Gates, Jr., ``{Ectoplasm has no topology},''
  \href{http://dx.doi.org/10.1016/S0550-3213(98)00819-0}{{\em Nucl. Phys. B}
  {\bfseries 541} (1999) 615--650},
  \href{http://arxiv.org/abs/hep-th/9809056}{{\ttfamily arXiv:hep-th/9809056}}.

\bibitem{Kuzenko:2013vha}
S.~M. Kuzenko, J.~Novak, and G.~Tartaglino-Mazzucchelli, ``{N=6 superconformal
  gravity in three dimensions from superspace},''
  \href{http://dx.doi.org/10.1007/JHEP01(2014)121}{{\em JHEP} {\bfseries 01}
  (2014) 121}, \href{http://arxiv.org/abs/1308.5552}{{\ttfamily arXiv:1308.5552
  [hep-th]}}.

\bibitem{Butter:2016mtk}
D.~Butter, F.~Ciceri, B.~de~Wit, and B.~Sahoo, ``{Construction of all N=4
  conformal supergravities},''
  \href{http://dx.doi.org/10.1103/PhysRevLett.118.081602}{{\em Phys. Rev.
  Lett.} {\bfseries 118} no.~8, (2017) 081602},
  \href{http://arxiv.org/abs/1609.09083}{{\ttfamily arXiv:1609.09083
  [hep-th]}}.

\bibitem{Novak:2017wqc}
J.~Novak, M.~Ozkan, Y.~Pang, and G.~Tartaglino-Mazzucchelli, ``{Gauss-Bonnet
  supergravity in six dimensions},''
  \href{http://dx.doi.org/10.1103/PhysRevLett.119.111602}{{\em Phys. Rev.
  Lett.} {\bfseries 119} no.~11, (2017) 111602},
  \href{http://arxiv.org/abs/1706.09330}{{\ttfamily arXiv:1706.09330
  [hep-th]}}.

\bibitem{Gold:2023dfe}
G.~Gold, J.~Hutomo, S.~Khandelwal, and G.~Tartaglino-Mazzucchelli,
  ``{Curvature-squared invariants of minimal five-dimensional supergravity from
  superspace},'' \href{http://dx.doi.org/10.1103/PhysRevD.107.106013}{{\em
  Phys. Rev. D} {\bfseries 107} no.~10, (2023) 106013},
  \href{http://arxiv.org/abs/2302.14295}{{\ttfamily arXiv:2302.14295
  [hep-th]}}.

\bibitem{Gold:2023ykx}
G.~Gold, J.~Hutomo, S.~Khandelwal, and G.~Tartaglino-Mazzucchelli,
  ``{Components of curvature-squared invariants of minimal supergravity in five
  dimensions},'' \href{http://dx.doi.org/10.1007/JHEP07(2024)221}{{\em JHEP}
  {\bfseries 07} (2024) 221}, \href{http://arxiv.org/abs/2311.00679}{{\ttfamily
  arXiv:2311.00679 [hep-th]}}.

\bibitem{Butter:2013lta}
D.~Butter, B.~de~Wit, S.~M. Kuzenko, and I.~Lodato, ``{New higher-derivative
  invariants in N=2 supergravity and the Gauss-Bonnet term},''
  \href{http://dx.doi.org/10.1007/JHEP12(2013)062}{{\em JHEP} {\bfseries 12}
  (2013) 062}, \href{http://arxiv.org/abs/1307.6546}{{\ttfamily arXiv:1307.6546
  [hep-th]}}.

\bibitem{Kuzenko:2014jra}
S.~M. Kuzenko and J.~Novak, ``{Supergravity-matter actions in three dimensions
  and Chern-Simons terms},''
  \href{http://dx.doi.org/10.1007/JHEP05(2014)093}{{\em JHEP} {\bfseries 05}
  (2014) 093}, \href{http://arxiv.org/abs/1401.2307}{{\ttfamily arXiv:1401.2307
  [hep-th]}}.

\bibitem{Kuzenko:2015jda}
S.~M. Kuzenko, J.~Novak, and G.~Tartaglino-Mazzucchelli, ``{Higher derivative
  couplings and massive supergravity in three dimensions},''
  \href{http://dx.doi.org/10.1007/JHEP09(2015)081}{{\em JHEP} {\bfseries 09}
  (2015) 081}, \href{http://arxiv.org/abs/1506.09063}{{\ttfamily
  arXiv:1506.09063 [hep-th]}}.

\bibitem{Kuzenko:2015jxa}
S.~M. Kuzenko and J.~Novak, ``{On curvature squared terms in N=2
  supergravity},'' \href{http://dx.doi.org/10.1103/PhysRevD.92.085033}{{\em
  Phys. Rev. D} {\bfseries 92} no.~8, (2015) 085033},
  \href{http://arxiv.org/abs/1507.04922}{{\ttfamily arXiv:1507.04922
  [hep-th]}}.

\bibitem{Kuzenko:2016tfz}
S.~M. Kuzenko, J.~Novak, and I.~Sachs, ``{Minimal $ \mathcal{N}=4 $
  topologically massive supergravity},''
  \href{http://dx.doi.org/10.1007/JHEP03(2017)109}{{\em JHEP} {\bfseries 03}
  (2017) 109}, \href{http://arxiv.org/abs/1610.09895}{{\ttfamily
  arXiv:1610.09895 [hep-th]}}.

\bibitem{Kowalski:2014}
E.~Kowalski, {\em {An Introduction to the Representation Theory of Groups}}.
\newblock {American Mathematical Society}, 2014.

\bibitem{Fulton:1991}
W.~Fulton and J.~Harris, {\em {Representation Theory: A First Course}}.
\newblock {Springer New York}, 1991.

\bibitem{Peeters:2006}
K.~Peeters, ``{A Field-theory motivated approach to symbolic computer
  algebra},'' \href{http://dx.doi.org/10.1016/j.cpc.2007.01.003}{{\em Comput.
  Phys. Commun.} {\bfseries 176} (2007) 550--558},
  \href{http://arxiv.org/abs/cs/0608005}{{\ttfamily arXiv:cs/0608005}}.

\bibitem{Peeters:2007}
K.~Peeters, ``{Introducing Cadabra: A Symbolic computer algebra system for
  field theory problems},''
  \href{http://arxiv.org/abs/hep-th/0701238}{{\ttfamily arXiv:hep-th/0701238}}.

\bibitem{Peeters:2018}
K.~Peeters, ``{Cadabra2: computer algebra for field theory revisited},''
  \href{http://dx.doi.org/10.21105/joss.01118}{{\em J. Open Source Softw.}
  {\bfseries 3} no.~32, (2018) 1118}.

\bibitem{Gold:2024nbw}
G.~Gold, S.~Khandelwal, and G.~Tartaglino-Mazzucchelli, ``{Supergravity
  Component Reduction with Computer Algebra},''
\newblock 6, 2024.
\newblock \href{http://arxiv.org/abs/2406.19687}{{\ttfamily arXiv:2406.19687
  [hep-th]}}.

\bibitem{Gold:2024git}
G.~Gold, S.~Khandelwal, and G.~Tartaglino-Mazzucchelli, ``{SUGRA Component
  Reduction},'' 2024.
\newblock \url{{https://github.com/gregory-gold/sugra-component-reduction}}.

\bibitem{Kuzenko:2017zsw}
S.~M. Kuzenko, J.~Novak, and S.~Theisen, ``{Non-conformal supercurrents in six
  dimensions},'' \href{http://dx.doi.org/10.1007/JHEP02(2018)030}{{\em JHEP}
  {\bfseries 02} (2018) 030}, \href{http://arxiv.org/abs/1709.09892}{{\ttfamily
  arXiv:1709.09892 [hep-th]}}.

\bibitem{Sagan:1991}
B.~E. Sagan, {\em {The Symmetric Group: Representations, Combinatorial
  Algorithms, and Symmetric Functions}}.
\newblock {Springer New York}, 2001.

\bibitem{James:1978}
J.~G.~D., {\em {The Representation Theory of the Symmetric Groups}}.
\newblock {Springer Berlin, Heidelberg}, 1978.

\end{thebibliography}\endgroup
\bibliographystyle{utphys}

\end{document}